\documentclass[12pt]{article}%
\usepackage{amsmath, amsthm,comment}
\usepackage{amsfonts}
\usepackage{amssymb,graphics,psfrag}
\usepackage{array,epsfig,stmaryrd,graphicx}
\usepackage{heck}
\usepackage{cite}
\usepackage{graphicx}
\usepackage{makeidx}
\usepackage{multicol}
\usepackage{geometry}
\usepackage{amsfonts}
\usepackage{mathrsfs}
\usepackage{amssymb}
\usepackage{amsmath}
\usepackage{slashed}%
\providecommand{\U}[1]{\protect\rule{.1in}{.1in}}

\newcommand{\beq}{\begin{equation}}
\newcommand{\eeq}{\end{equation}}
\newcommand{\be}{\begin{equation}}
\newcommand{\ee}{\end{equation}}
\newcommand{\bea}{\begin{eqnarray}}
\newcommand{\eea}{\end{eqnarray}}
\newcommand{\ben}{\begin{eqnarray*}}
\newcommand{\een}{\end{eqnarray*}}
\newcommand{\ba}{\begin{aligned}}
\newcommand{\ea}{\end{aligned}}
\newcommand{\bt}{\begin{tabular}}
\newcommand{\et}{\end{tabular}}
\newcommand{\bc}{\begin{center}}
\newcommand{\ec}{\end{center}}

\newcommand{\cref}{{\bf [check ref]}}

\newcommand{\bs}{\begin{subarray}{c}}
\newcommand{\es}{\end{subarray}}

\begin{document}

\date{December, 2008}
\title{Cosmology of F-theory GUTs}

\preprint{arXiv:0812.3155}

\institution{HarvardU}{Jefferson Physical Laboratory, Harvard University, Cambridge,
MA 02138, USA}

\authors{Jonathan J. Heckman\footnote{e-mail: {\tt jheckman@fas.harvard.edu}},
Alireza Tavanfar\footnote{e-mail: {\tt tavanfar@physics.harvard.edu}}
and
Cumrun Vafa\footnote{e-mail: {\tt vafa@physics.harvard.edu}}}

\abstract{In this paper we study the interplay between the recently
proposed F-theory GUTs and cosmology. Despite the fact that the parameter range
for F-theory GUT models is very narrow, we find that F-theory GUTs beautifully
satisfy most cosmological constraints without any further restrictions. The viability
of the scenario hinges on the interplay between various components of the axion supermultiplet,
which in F-theory GUTs is also responsible for breaking supersymmetry. In these models,
the gravitino is the LSP and develops a mass by eating the axino mode. The radial component
of the axion supermultiplet known as the saxion typically begins to oscillate in the
early Universe, eventually coming to dominate the energy density. Its decay reheats the
Universe to a temperature of $\sim 1$ GeV, igniting BBN and diluting all thermal relics such as the
gravitino by a factor of $\sim 10^{-4} - 10^{-5}$ such that gravitinos contribute a sizable component
of the dark matter. In certain cases, non-thermally
produced relics such as the axion, or gravitinos generated from the decay of the saxion can also
contribute to the abundance of dark matter. Remarkably enough,
this cosmological scenario turns out to be independent of the initial reheating temperature of
the Universe. This is due to the fact that the initial oscillation temperature of the saxion
coincides with the freeze out temperature for gravitinos in F-theory GUTs. We also find that saxion dilution
is compatible with generating the desired baryon asymmetry from standard leptogenesis.
Finally, the gravitino mass range in F-theory
GUTs is $10-100$ MeV,  which interestingly coincides with the window of values
required for the decay of the NLSP to solve the problem of $^7Li$ over-production.}%

\maketitle

\tableofcontents

\section{Introduction}

Two prominent triumphs of modern theoretical physics are the Standard Models
of particle physics and cosmology. Moreover, the interplay between particle
physics and astrophysics/early Universe cosmology has already proven to be a
fruitful arena of investigation for both fields. On the astrophysics side,
this interplay has led to the very successful predictions of Big Bang
Nucleosynthesis\ (BBN), which accounts for the abundance of light elements. In
addition, ideas from particle physics have provided several plausible
mechanisms such as baryogenesis or leptogenesis which could generate the
observed baryon asymmetry. Finally, many extensions of the Standard Model
include dark matter candidates.

On the particle physics side, constraints from astrophysics have led to novel,
and sometimes quite stringent conditions on possible extensions of the
Standard Model. These constraints can translate into important bounds on
parameters of a candidate model which may be inaccessible from other avenues
of investigation, and which can have repercussions beyond their immediate
astrophysics applications. For example, compatibility with the successful
predictions of BBN imposes important restrictions such as the
requirement -- spectacularly confirmed by LEP -- that essentially there are at most three generations of
light neutrinos! This is in amazing accord with the Standard Model of particle
physics, and indicates a deep link between these two fields. Other
cosmological considerations such as over-production of gravitinos in
supersymmetric models, or a deficit in the observed baryon asymmetry provide
additional constraints. Satisfying all of these constraints is often a
non-trivial task for a given model, but can also point the way to novel
mechanisms which may not be available in the standard cosmology.

At a more refined level, the interrelations between particle physics and
cosmology roughly bifurcate into issues where gravity itself plays a central
role, and questions where gravity plays
only a supporting role in addressing more detailed features of a given
particle physics model. For example, issues connected to the cosmological
constant, or the homogeneity of the early Universe fall in the former
category, whereas particle physics oriented issues such as the origin of dark
matter or the overall baryon asymmetry fit most naturally in the latter
category. Due to the vast array of observational data from probes wholly
separate from cosmology, issues more closely tied to particle physics appear
to at present be more tractable. In this regard, it is therefore quite
natural, as is often done in the particle physics literature, to exclusively
focus on the \textquotedblleft low energy\textquotedblright\ aspects
associated with the cosmology of a given particle physics model,
parameterizing our ignorance of what occurs in the very early Universe in
terms of an \textquotedblleft initial reheating temperature\textquotedblright%
\ for the Universe, $T_{RH}^{0}$. For the purposes of particle physics
considerations, the cosmology of the Universe effectively begins at this
temperature. In many cases, this reheating temperature ends up being smaller
than the Planck, or GUT\ scale, and models with $T_{RH}^{0}$ as low as
$10-100$ TeV have also been discussed.

It is interesting that this division in cosmology between gravity and particle
physics issues parallels recent developments in the string theory literature.
It is clear that a vast landscape of consistent string theory vacua exists.
While this makes the string theory paradigm a very rich and flexible physical
model, at present it also lacks predictive power because it does not lead to
any particularly distinguished vacua! As a principle which can be used to
limit the search for semi-realistic vacua, in
\cite{BHVI,BHVII,HVGMSB,HVLHC,HVFLAV} (see also
\cite{WijnholtDonagi,WatariTATARHETF,IbanezSUSYFTHEORY,DonagiWijnholtBreak,HMSSNV,MarsanoGMSB,MarsanoToolbox,Wijnholt:2008db,Jiang:2008yf,Blumenhagen:2008zz,Blumenhagen:2008aw,Font:2008id}%
), it was shown that in the context of F-theory based models, demanding the
existence of a limit where the Planck scale (and thus associated gravitational
questions) decouples, in tandem with some qualitative particle physics
considerations such as the existence of a GUT, leads to a remarkably limited,
and predictive framework. In fact, without adding any additional ingredients,
traditionally vexing particle physics issues related to flavor hierarchies,
the doublet-triplet splitting problem, the $\mu/B\mu$-problem, and undesirable
GUT\ mass relations all find natural solutions in F-theory GUT\ models. In
addition, natural estimates for the overall magnitude of the Yukawa couplings,
axion decay constant, $\mu$ parameter and MSSM\ soft mass terms all fall in an
acceptable phenomenological range.

The aim of the present paper is to examine the cosmology of the F-theory
GUT\ scenario as a model of particle physics. We find that these models
naturally satisfy the typically stringent constraints derived from
compatibility with BBN as well as bounds on the overall relic abundance
of dark matter candidates, such as the gravitino. Perhaps the
most striking feature of this analysis is the \textit{absence} of any major
problem, namely, that \textit{consistency with cosmology imposes almost no
constraint at all on the initial reheating temperature} $T_{RH}^{0} $.

That this is the case is a highly non-trivial consequence of the parameter
range found for F-theory GUTs. Indeed, one potentially significant source of
tension could in principle have originated from the fact that as a model of
gauge mediated supersymmetry breaking with a relatively high mass for the
gravitino \cite{HVGMSB}:%
\begin{equation}
m_{3/2}=\frac{1}{\sqrt{3}}\frac{F}{M_{PL}}\sim10-100\text{ MeV,}%
\end{equation}
the relic abundance of gravitinos, which is the LSP, can potentially overclose
the Universe. In the above, as throughout this paper, $M_{PL}$ denotes the
reduced Planck mass $M_{PL}=2.4\times10^{18}$\ GeV, and $F$ denotes the
component of the GUT\ group singlet chiral superfield $X$ responsible for
supersymmetry breaking:%
\begin{equation}
\left\langle X\right\rangle =x+\theta^{2}F
\end{equation}
where as shown in \cite{HVGMSB}, simultaneously solving the $\mu$ problem and
generating viable soft mass terms in a minimal gauge mediation scenario
requires:%
\begin{align}
F  &  \sim10^{17}\text{ GeV}^{2}\label{OURF}\\
x  &  \sim10^{12}\text{ GeV.}%
\end{align}
We note that the Goldstino mode corresponding to the fermionic component of
$X$ is eaten by the gravitino.

In many gauge mediation models, the scale of supersymmetry breaking is
significantly lower. From the perspective of cosmology, the relic abundance of
gravitinos would at first appear to provide strong motivation for lowering the
value of $m_{3/2}$. Indeed, in the most straightforward approximation, it is
quite natural to take $T_{RH}^{0}$ as high as the GUT\ scale. In this case, a
well known estimate for the relic abundance of gravitinos in many
supersymmetric models requires:%
\begin{equation}
\frac{m_{3/2}}{2\text{ keV}}\leq0.1
\end{equation}
to avoid an overabundance of gravitinos in the present Universe. In
particular, this would appear to suggest an upper bound for $F$ of order:%
\begin{equation}
F\lesssim7\times10^{11}\text{ GeV}^{2}%
\end{equation}
which is significantly lower than equation (\ref{OURF})! In the gravitino
cosmology literature, it is common to take $T_{RH}^{0}\ll M_{GUT}\sim
3\times10^{16}$ GeV to truncate the production of thermally produced
gravitinos, thus avoiding precisely these types of issues. At a conceptual
level, though, it is somewhat distressing that particle physics considerations
demand such a stringent upper bound on $T_{RH}^{0}$. Indeed, insofar as the
value of $T_{RH}^{0}$ is dictated by gravitational issues which are a priori
wholly separate from details of a particular particle physics models, this
type of tuning of parameters is quite puzzling.

In F-theory GUTs, the resolution of this apparent dilemma again resides in the
chiral superfield $X$. The essential point is that $X$ plays a dual role in
F-theory GUTs because its bosonic component breaks the anomalous global $U(1)$
Peccei-Quinn symmetry of the low energy theory. As such, the associated
Goldstone mode is the QCD\ axion, with decay constant:%
\begin{equation}
f_{a}\sim\sqrt{2}x\sim10^{12}\text{ GeV,}%
\end{equation}
solving the strong CP\ problem.\ In the context of supersymmetric theories,
however, the axion corresponds to one of two real degrees of freedom
associated with the bosonic component of the corresponding axion
supermultiplet. The other degree of freedom, known as the saxion is exactly
massless in the limit where supersymmetry is restored, and in the present
context has a mass of order $100$ GeV. An exciting feature of the F-theory
GUT\ is that the mass of the saxion is controlled by UV sensitive details of
the compactification, such as the mass of the anomalous $U(1)_{PQ}$ gauge
boson. As such, cosmological constraints for the saxion provide a window into
the high scale dynamics of the model.

Because the potential of the saxion is nearly flat, it is easily displaced
from its minimum, and will generically begin to oscillate as the early
Universe cools from the initial reheating temperature $T_{RH}^{0}$ until the
saxion decays. In a generic situation, the initial amplitude for $s_{0}$ is
sufficiently large that its vacuum energy density comes to dominate the energy
density of the Universe. The value of $s_{0}$ is on the order of the
characteristic Kaluza-Klein scale for the $X$ field:%
\begin{equation}
s_{0}\sim M_{X}\sim10^{15.5}\text{ GeV.}%
\end{equation}
The decay of the saxion releases a large amount of entropy into the Universe,
effectively diluting the overall relic abundance of all particle species. As a
consequence, we find that rather neatly, one component of the axion
supermultiplet, the saxion, counteracts the potentially dangerous features of
the fermionic gravitino component (which includes the axino as the
longitudinal degree of freedom)! We note that the decay of the saxion or some
other cosmological modulus as a means to dilute the relic abundance of a
particle species to acceptable levels has certainly been discussed in the
literature before, for example in \cite{BanksDineGraesser}. The primary
novelty here is that without any additional assumptions, \textit{F-theory GUT
models automatically resolve the most problematic features of gravitino
cosmology.} The requisite era of saxion dominance occurs provided the initial
reheating temperature satisfies:%
\begin{equation}
T_{RH}^{0} \gtrsim 10^{6}\text{ GeV.}%
\end{equation}
In this case, the dilution of the saxion is such that the gravitino could
naturally make up a prominent component of the dark matter density. Depending
on whether the axion begins to oscillate before or after an era of saxion domination,
the axion can also provide a component of the dark matter. We are currently investigating
whether F-theory GUTs provide additional dark matter candidates
\cite{FGUTSDARK}. Finally, we also find that lower values of the reheating
temperature are also possible, and are compatible with a regime where the
saxion has a smaller initial amplitude. In this case, the gravitino and axion
can potentially both contribute to the dark matter density.

But while the evolution of the saxion neatly solves the gravitino
\textquotedblleft problem\textquotedblright, it can in principle introduce
additional constraints. For example, the decay of the saxion, with its
significant entropy release might disrupt the start of BBN. Most
conservatively, this requires that the reheating temperature for the saxion
remain above the starting temperature for BBN. Quite remarkably, this
requirement is again satisfied quite naturally in F-theory GUT\ models. We
find that the reheating temperature of the saxion, $T_{RH}^{s}$ satisfies:%
\begin{equation}
T_{RH}^{s} \sim 1\text{ GeV} > T_{BBN}\sim 2\text{ MeV.}%
\end{equation}
Compatibility with BBN also imposes strong constraints on the decay
products of the saxion. Using the well known bound (which we shall review) on
the branching ratio of the saxion to relativistic species such as the axion,
we find that we must either posit the existence of an additional decay product
not found in the MSSM, or that the saxion must have sufficient mass to decay
dominantly to two Higgs fields so that:%
\begin{equation}
m_{sax}>2m_{h^{0}}\sim230\text{ GeV.}%
\end{equation}

A sufficient amount of baryon asymmetry must be generated at high temperatures
in order for BBN\ to produce the observed abundances of light elements. In the
standard solution to the gravitino problem, lowering the initial reheating
temperature $T_{RH}^{0}$ has the deleterious consequence of removing some of
the more efficient mechanisms for generating such an asymmetry, such as
GUT\ scale baryogenesis, or standard leptogenesis. Rather, in this context it
is common to invoke a mechanism where a large baryon asymmetry can be
generated by the coherent oscillation of a field which carries either
non-trivial baryon, or lepton number, such as in the Affleck-Dine scenario. In
the present context, however, the dilution of the saxion completely relaxes
any upper bound on $T_{RH}^{0}$. Thus, the only condition left to check is
whether any of the available mechanisms are capable of generating a
sufficiently large initial baryon asymmetry which can survive the effects of
dilution.\ Performing this analysis, we find that standard leptogenesis
indeed produces an appropriate baryon asymmetry. In other words, the range of
parameters suggested by F-theory GUTs naturally fall within the small range of
dilution factors compatible with diluting the gravitinos to an amount where
they do not overclose the Universe, but which nevertheless preserve a
sufficient baryon asymmetry.\footnote{See \cite{Kumar:2008vs} for other recent
work on moduli stabilization, leptogenesis, and dark matter in the
context of gauge and gravity mediated scenarios.}

The organization of this paper is as follows. In section \ref{CosmoReview} we
review the main interrelations between cosmology and particle physics which
will occupy a central role in the present paper. In section \ref{FGUTAXION} we
study the axion supermultiplet and its interactions in the context of F-theory
GUTs. Section \ref{FCOSMO} forms the main body of our paper in which we
present the particular cosmological scenario suggested by F-theory GUTs,
paying special attention to the role of the saxion. In this same section, we
analyze the gravitino and axion relic abundance, study constraints from BBN,
and determine the overall baryon asymmetry generated by standard
leptogenesis. Finally, in section \ref{CONC} we briefly discuss future
avenues of investigation in the cosmology of F-theory GUTs.

\section{Cosmology and Particle Physics\label{CosmoReview}}

One of the primary aims of the present paper is to determine how cosmological
considerations constrain F-theory GUTs. In this section we review the main
issues which must be addressed in a viable cosmological scenario. This section
is entirely review, and can safely be skipped by the reader familiar with this material.

We first describe the main concepts which will figure prominently in the
estimates to follow. After this, we proceed to a more detailed review of the
features which will be especially important for analyzing the cosmology of
F-theory GUTs. To this end, in the next subsection we provide a brief
introduction to the FRW\ Universe, next describing the history of the standard
cosmology as the Universe evolves in such a Universe. As will be apparent in
later sections, the cosmology of the gravitino plays an especially important
role in assessing the viability of a supersymmetric model. For this reason, we
next review the computation of relic abundances, and in particular, describe
in some detail the usual \textquotedblleft gravitino problem\textquotedblright%
. While this is indeed a significant constraint on many models, in certain
situations the dilution due to a late decaying cosmological modulus can
significantly alter this analysis. With this in mind, we next review the primary
consequences of late-decaying cosmological moduli. We next review the
constraint on the number of thermalized relativistic species derived from BBN.
Finally, we conclude this section with a discussion of more refined features
of BBN\ which are present in models such as F-theory GUTs which possess a
late-decaying NLSP.

\subsection{FRW\ Universe}

Below the scale of compactification, and before the present epoch, to good
approximation, the early Universe is described by the four-dimensional
Robertson-Walker (RW) metric:%
\begin{equation}
ds^{2}=dt^{2}-a^{2}(t)\left(  \frac{dr^{2}}{1-kr^{2}}+r^{2}\left(  d\theta
^{2}+\sin^{2}\theta d\phi^{2}\right)  \right)  \text{,}%
\end{equation}
where in the above, we have introduced the scale factor $a(t)$, as well as the
curvature constant $k$, which after a suitable rescaling takes the values
$k=+1,0,-1$ for a respectively closed, flat, or open Universe. This describes
the evolution of an isotropic Universe with homogeneous energy distribution.
The overall expansion rate of the Universe is measured by the Hubble
parameter:%
\begin{equation}
H\equiv\frac{\dot{a}}{a}\text{.}%
\end{equation}
It is also common to introduce a dimensionless variant of $H$, called $h$
defined by the equation:%
\begin{equation}
H=100h\text{ km Mpc}^{-1}\text{ }\sec^{-1}\text{.} \label{Hubblenow}%
\end{equation}
The present value of $h$ is given by \cite{Dunkley:2008ie}:%
\begin{equation}
h_{0}\sim0.7\text{.}%
\end{equation}
When the context is clear, we will often drop this subscript to avoid
cluttering various equations.

The background energy density determines the expansion rate of the Universe
via the Friedmann equations:%
\begin{align}
\ddot{a}  &  =-\frac{4\pi}{3}G_{N}(\rho+3p)a\\
H^{2}  &  =\frac{8\pi G_{N}}{3}\rho-\frac{k}{a^{2}} \label{Friedtwo}%
\end{align}
where in the above, $p$ denotes the pressure of the given \textquotedblleft
fluid\textquotedblright, $G_{N}$ denotes the four-dimensional Newton's
constant, and $\rho$ corresponds to the total energy density of the Universe.
The critical density $\rho_{c}$ is defined as the value of the total energy
density for which $k=0$, so that:%
\begin{equation}
\rho_{c}=\frac{3H^{2}}{8\pi G_{N}}\text{.}%
\end{equation}
Note that the critical density is a non-trivial function of $t$.

The total energy density will in general receive various types of
contributions, so that $\rho$ is given by a sum of the form:%
\begin{equation}
\rho=\underset{i}{\sum}\rho_{i}\text{.}%
\end{equation}
Plugging this expression into equation (\ref{Friedtwo}) now yields:%
\begin{equation}
\Omega_{\text{tot}}=\underset{i}{\sum}\Omega_{i}=1+\frac{k}{H^{2}a^{2}%
}=1+\frac{k}{\dot{a}^{2}}%
\end{equation}
where we have introduced the parameter:%
\begin{equation}
\Omega_{i}\equiv\frac{\rho_{i}}{\rho_{c}}\text{.}%
\end{equation}
The sign of $k$ correlates with the magnitude of $\Omega_{\text{tot}}$:%
\begin{align}
\Omega_{\text{tot}}  &  >1:k=+1\\
\Omega_{\text{tot}}  &  =1:k=0\\
\Omega_{\text{tot}}  &  <1:k=-1\text{.}%
\end{align}
Due to the overall dependence on $H^{2}$ in $\rho_{c}$, it is sometimes
convenient to introduce the quantity $\Omega_{i}h^{2}$, with $h=\kappa H$ with
$\kappa$ the constant implicitly defined by equation (\ref{Hubblenow}). The
overall dependence on $H^{2}$ factors out, yielding:%
\begin{equation}
\Omega_{i}h^{2}\equiv\frac{\rho_{i}}{\rho_{c}}\cdot\left(  \kappa H\right)
^{2}=\frac{8\pi G_{N}\kappa^{2}}{3}\cdot\rho_{i}\text{.}%
\end{equation}

The energy density of the $i^{th}$ species also evolves with scale. In the
approximation where the energy density $\rho_{i}$ is proportional to the
pressure $p_{i}$, the equation of state is given by:%
\begin{equation}
p_{i}=w_{i}\rho_{i}\text{,}%
\end{equation}
for some constant $w_{i}$. The scaling behavior of $\rho$ for three common
choices is:%
\begin{align}
\rho_{r}  &  \propto a^{-4}:w_{r}=1/3\label{radscale}\\
\rho_{m}  &  \propto a^{-3}:w_{m}=0\label{mattscale}\\
\rho_{\Lambda}  &  \propto\text{const}:w_{\Lambda}=-1
\end{align}
where $r$, $m$ and $\Lambda$ respectively denote \textquotedblleft
radiation\textquotedblright\ or relativistic matter, non-relativistic matter,
and vacuum energy density.

Observation indicates that the Universe has recently transitioned from an era
of matter domination to one where a background vacuum energy density plays a
dominant role, with \cite{Dunkley:2008ie}:%
\begin{align}
\Omega_{\Lambda}  &  \sim0.7\\
\Omega_{m}  &  \sim0.3\text{.}%
\end{align}
The matter content $\Omega_{m}$ further subdivides into a subdominant
``visible'' matter contribution with $\Omega_{\text{visible}} \sim 0.05$) with the
rest being comprised of the so-called dark matter $\Omega_{DM} \sim 0.25$, which
by definition interacts weakly with the Standard Model degrees of freedom. The
existence of such a large additional component of matter indicates that
the Standard Model must be extended in some fashion. An important feature
of this is that the overall energy density is quite close to the critical value:%
\begin{equation}
\Omega_{\text{tot}}=\Omega_{\Lambda}+\Omega_{m}\sim1\text{.}%
\end{equation}

\subsection{Timeline of the Standard Cosmology\label{Timeline}}

Having reviewed the main concepts which we shall use throughout this paper, we
now describe in reverse chronological order the timeline for the standard
cosmology. At various points we also indicate where deviations from this
trajectory are possible. This material can be found in many standard
textbooks, such as \cite{KT,Mukhanov}, for example.

\subparagraph{Era of Structure Formation: $T\sim1$ meV$\rightarrow100$ meV}

At present, the Universe is roughly $10^{18}$ sec old and the background
photon radiation is at a temperature of $2.7$ K $\sim1$ meV, which is
presently characterized as an era where large scale structures have formed.
\ In addition, the energy density of the Universe is composed of roughly $5\%$
visible baryonic matter, $25\%$ non-baryonic matter, or \textquotedblleft dark
matter\textquotedblright, and $70\%$ of some other type of energy density
which appears to share the characteristics of a vacuum energy density, or
cosmological constant. \ At this point, the abundances of all of the light
elements have been produced via cosmological processes earlier in the history
of the universe, as well as other, more astrophysical processes having to do
with the birth and death of stars, for example. \ At somewhat earlier
timescales around $10^{16}$ sec described by a temperature of a few hundred
meV, the present galaxies begin to form.

\subparagraph{Era of BBN and Atom Formation: $T\sim1$ eV$\rightarrow5$ MeV}

Prior to the start of structure formation, radiation and matter decouple at
roughly $10^{12}$ sec and a temperature of $\sim1$ eV. \ At this point, all of
the present atoms have formed and will soon begin to form large scale
structure. \ At only slightly earlier times, or higher temperatures, the
universe is sufficiently hot to overcome the binding energy of atoms. \ At
this point, nucleosynthesis has finished, and an appropriate abundance of the
ions such as $H^{+}$, $D^{+}$, $T^{+}$, $^{3}He^{++}$, $^{4}He^{++}$,
$^{7}Li^{+++}$ have formed.\footnote{As reviewed in \cite{KT}, for example,
astrophysical processes also account for a fraction of the $^{4}He$
abundance.} \ These ions will soon combine with electrons to form neutral
atoms. \ Heavier elements are produced later as a result of astrophysical
phenomena. \ Other than $^{7}Li$, the observed abundances of these elements
agree quite well with theoretical computations.\ As we shall later review,
this information is sufficiently accurate to provide a strong bound on the
number of additional relativistic species in thermal equilibrium.\ BBN\ begins
at a temperature $T\sim1-5$ MeV. \ Generating the appropriate light element
abundances requires a sufficient number of baryons. \ Letting $n_{B}$ and
$n_{\overline{B}}$ denote the respective number densities of baryons and
anti-baryons in a comoving volume, BBN\ requires that the ratio between the
net baryon number density and photon number density, $n_{\gamma}$ is:%
\begin{equation}
\eta_{B}\equiv\frac{n_{B}-n_{\overline{B}}}{n_{\gamma}}\sim6\times
10^{-10}\text{.}%
\end{equation}

\subparagraph{Era of Radiation/Coherent Oscillation Domination: $T\sim1$
GeV$\rightarrow T_{B}$}

Before the start of BBN, the temperature of the universe is sufficiently hot
that the particles of the Standard Model exist in a plasma. \ \ In standard
cosmological scenarios, it is assumed that some initial amount of baryon
asymmetry necessary for BBN has been generated at a temperature $T_{B}$. \ At
this point, the particles generated during baryogenesis are in thermal
equilibrium in a hot plasma. \ As the Universe cools and expands, a particle
species may fall out of thermal equilibrium.

The abundance of these species is particularly important for cosmology. \ The
essential point is that each frozen out species will contribute to the present
matter abundance $\Omega_{M}h^{2}$. \ In particular, because the observed dark
matter abundance is given by $\Omega_{DM}h^{2}\sim0.1$, each species must
satisfy the bound:%
\begin{equation}
\Omega_{i}h^{2}\leq0.1\text{.}%
\end{equation}
When this inequality is not satisfied, it is common to say that the species
\textquotedblleft overcloses\textquotedblright\ the Universe. \ As explained
in \cite{KT}, the more precise statement is that when $\Omega_{i}h^{2}$ is too
large, the predicted value of $h$ would not be in accord with observation.

In extensions of the standard cosmology, additional dynamics beyond those
associated with the freeze out of various species may also play an important
role. \ For example, at temperatures above the start of BBN, scalar fields
undergoing coherent oscillations can come to dominate the energy density of
the Universe.\ When such particles decay, they can release significant amounts
of entropy into the Universe. \ This has the effect of diluting the abundance
of various frozen out species. \ In addition, the decay products of such
fields can also increase the relic abundance of certain species.

\subparagraph{Era of Baryon Asymmetry Creation: $T\sim T_{B}$ $\rightarrow
T_{RH}^{0}$}

At a temperature $T_{B}$, the baryon asymmetry requisite for successful
BBN\ is assumed to be created.\ As found by Sakharov, generating an
appropriate baryon asymmetry requires that first, the accidental $U(1)_{B}$
baryon symmetry of the Standard Model must be broken, and further, that both C
and CP must be violated. In addition, the baryon asymmetry must be generated
due to some out of equilibrium process. In order to achieve the required value of
$\eta_{B}$, there must be an additional source of CP violation beyond that
which is present in the Standard Model. In scenarios where $T_{B}$ is close to
the GUT\ scale, off diagonal gauge bosons of the GUT\ group can potentially
generate the required values of $\eta_{B}$. \ Lower temperatures for $T_{B}$
are also possible depending on the particular mechanism in question. \ For
example, in leptogenesis, an asymmetry in lepton number is converted to an
asymmetry in baryon number via sphaleron processes. \ In these examples, heavy
right-handed Majorana neutrinos decay to Standard Model particles, generating
the initial lepton number asymmetry so that the associated
temperature $T_{B}$ is roughly given by the Majorana mass of the right-handed
neutrinos. An important consequence of this is that the initial reheating
temperature $T_{RH}^{0}$ must satisfy the inequality $T_{RH}^{0}\gtrsim
\min(M_{maj})$, where $M_{maj}$ is shorthand for the Majorana masses of the
heavy right-handed neutrinos. As we will review later, this can lead to a
certain amount of tension in many supersymmetric models which typically aim to
lower $T_{RH}^{0}$ to avoid overproduction of gravitinos. Lower values for
$T_{B}$ are also possible in less standard leptogenesis scenarios, as well as
in models where the coherent oscillation of a field generates a large lepton
or baryon number. This latter scenario, known as the Affleck-Dine scenario is
particularly attractive in models where other cosmological constraints require
$T_{B}$ to be relatively low compared to the value required for other baryon
asymmetry scenarios.

\subparagraph{Era of Speculation: $T$ $\sim T_{RH}^{0}\rightarrow M_{PL}$}

Above temperatures where the baryon asymmetry is created, there is some
initial temperature $T_{RH}^{0}$ corresponding to the \textquotedblleft
reheating\textquotedblright\ (RH) of the Universe. Below this temperature,
radiation domination commences. This temperature is typically associated with
the end of some era of where some high scale dynamics generates the required
density perturbations and flatness of the present Universe. We stress that
below the temperature $T_{RH}^{0}$, the primary mechanism which sets these
initial conditions, which could originate from a mechanism such as inflation
or string gas cosmology is inconsequential for the analysis of the paper. In
this way, in adhering to the philosophy of
\cite{BHVI,BHVII,HVGMSB,HVLHC,HVFLAV} these issues can be deferred to a later
stage of analysis.

\subsection{Thermodynamics in the Early Universe}

In the previous subsection we provided a rough sketch for the evolution of the
standard cosmology. We now discuss in greater detail the
thermodynamics of the Universe, and in particular, review the computation of
the relic abundances for hot and cold relics.

\subsubsection{Equilibrium Thermodynamics}

The expanding Universe corresponds to the stage on which the interactions of a
given particle physics model will play out. We now summarize some basic
features of equilibrium thermodynamics in the early Universe. Much of the
following discussion is explained in lucid detail in Chapter 3 of \cite{KT}.
Our aim here is to give a rough intuitive summary of the various formulae
which will be important in later discussions.

Assuming that some high scale dynamics sets the initial temperature of the
Universe, which we denote by $T_{RH}^{0}$, we can follow the subsequent
evolution of a cosmological model. At sufficiently high temperatures, various
particle species will be in thermal equilibrium. The corresponding interaction
rates $\Gamma_{\text{int}}$ within the thermal bath are specified by the
collision time for the particle species in question so that:%
\begin{equation}
\Gamma_{\text{int}}\sim n_{i}\left\langle \sigma_{i}v_{i}\right\rangle
\end{equation}
where $\left\langle \sigma_{i}v_{i}\right\rangle$ denotes the thermally averaged cross section
for the species and $n_{i}$ denotes its number density. In the limit where the
temperature is much greater than the chemical potential, the
number density, and energy density of a relativistic species $(m\gg T)$ are
given by:%
\begin{align}
n_{\text{rel}}  &  \sim T^{3}\label{nrel}\\
\rho_{\text{rel}}  &  \sim T^{4} \label{rhorel}%
\end{align}
for a relativistic species. In the other limit where the chemical potential
$\mu$ dominates, the above formula holds with $T~$replaced by $\mu$. For a
non-relativistic species of mass $m$, the number and energy density are:%
\begin{align}
n_{\text{n-rel}}  &  \sim\left(  mT\right)  ^{3/2}\exp\left(  -\frac{m-\mu}%
{T}\right)  \text{,}\label{boltzmannsupp}\\
\rho_{\text{n-rel}}  &  \sim mn_{\text{n-rel}}\text{.} \label{rhonrel}%
\end{align}
Finally, the entropy density $s$ of the thermal bath is primarily determined
by the interactions of the relativistic species: and is given by:%
\begin{equation}
s\sim g_{\ast S}(T)T^{3}\text{,} \label{sdense}%
\end{equation}
where here,
\begin{equation}
g_{\ast S}(T)\equiv\underset{\text{bose}}{\sum}g_{i}\left(  \frac{T_{i}}%
{T}\right)  ^{3}+\frac{7}{8}\underset{\text{fermi}}{\sum}g_{i}\left(
\frac{T_{i}}{T}\right)  ^{3}%
\end{equation}
with $g_{i}$ the number of internal degrees of freedom associated to a given
species, so that for example, an electron and positron both have $g_{e^{+}%
}=g_{e^{-}}=2$. Further, $T_{i}$ denotes the actual temperature of the given
species, which in general may differ from $T$. Nevertheless, this distinction
is largely unimportant when a given species is in equilibrium with the
background bath. It is also convenient to introduce a count of the total
number of relativistic species defined as:%
\begin{equation}
g_{\ast}(T)\equiv\underset{\text{bose}}{\sum}g_{i}\left(  \frac{T_{i}}%
{T}\right)  ^{4}+\frac{7}{8}\underset{\text{fermi}}{\sum}g_{i}\left(
\frac{T_{i}}{T}\right)  ^{4}\text{.} \label{gstdef}%
\end{equation}
We note that at high temperatures, $T_{i}\sim T$, and $g_{\ast}(T)$ and
$g_{\ast S}(T)$ may be used interchangeably. In the high temperature limit
where all degrees of freedom are relativistic, the value of $g_{\ast}$ in the
Standard Model and MSSM are respectively:%
\begin{align}
g_{\ast}(SM)  &  =106.75\\
g_{\ast}(MSSM)  &  =228.75\text{.}%
\end{align}
In the present era, the net energy and entropy density are given, for example, in Appendix A
of \cite{KT}:%
\begin{align}
\rho_{c,0}  &  \sim(8.1\times10^{-47})\;h^{2}\;\mathrm{GeV}^{4}\label{rhopres}%
\\
s_{0}  &  \sim2.3\times10^{-38}\;\mathrm{GeV}^{3} \label{spres}%
\end{align}
where in the above, the subscript $0$ reflects the evaluation of this quantity
at present times.

The connection between the expansion of the Universe and the temperature of
the thermal bath follows from the fact that the total entropy in a co-moving
volume:%
\begin{equation}
S\propto\left(  g_{\ast S}T^{3}\right)  \cdot a^{3}=\text{constant.}
\label{entropyconstant}%
\end{equation}
As a result, the temperature of the universe evolves as:%
\begin{equation}
T\propto g_{\ast S}^{-1/3}a^{-1}\text{.} \label{Tbathevolve}%
\end{equation}
Finally, in an era of radiation domination, the fact that the energy density
scales as $\rho\propto g_{\ast}T^{4}$, in tandem with the second Friedmann
equation (\ref{Friedtwo}), implies that the Hubble parameter is related to the
temperature as:%
\begin{equation}
H^{2}\sim g_{\ast}\frac{T^{4}}{M_{PL}^{2}}\text{.} \label{raddom}%
\end{equation}
This relation will be quite important when we discuss the decoupling of
thermal relics during an era of radiation domination.

\subsubsection{Relics and Decoupling\label{SUBSEC:RELDEC}}

In the above, we have implicitly assumed that all species in question remain
in thermal equilibrium. In this subsection we review what happens when a
species decouples from this thermal bath. Much of the material of this
subsection is reviewed in greater detail in Chapter 5 of \cite{KT}, and we
refer the interested reader there for further discussion.

As the Universe expands, the thermal bath cools, and a given species may
decouple. This occurs when the associated comoving volume $a^{3}$ becomes too
large to allow efficient interactions, so that the $i^{th}$ species
\textquotedblleft freezes out\textquotedblright\ at the temperature\ $T_{i}%
^{f}$ implicitly defined by:%
\begin{equation}
H(T_{i}^{f})\sim n_{i}\left\langle \sigma_{i}v_{i}\right\rangle \text{.}
\label{freezeout}%
\end{equation}
More precisely, the evolution of the number density $n_{i}$ as a
function of $t$ is given by the Boltzmann equations in the presence of a
dissipation term which accounts for the overall expansion of the Universe:%
\begin{equation}
\frac{dn_{i}}{dt}+3Hn_{i}=C_{i}\text{,} \label{Boltzequation}%
\end{equation}
The left hand side corresponds to the time evolution of the number density,
and $C_{i}$ is determined by the reaction rates of the thermal bath which can
generate the $i^{th}$ species. The principle of detailed balance implies that
$C_{i}$ is given by:%
\begin{equation}
C_{i}=\left\langle \sigma_{i}v_{i}\right\rangle \left(  n_{i,\text{EQ}}%
^{2}-n_{i}^{2}\right)  \text{,} \label{CDEF}%
\end{equation}
where $n_{i,\text{EQ}}$ denotes the equilibrium number density of the $i^{th}$ species.

After a species is no longer in contact with the thermal bath, its number
density redshifts with the expansion of the Universe, scaling as $a^{-3}$. On
the other hand, returning to equation (\ref{entropyconstant}), it also follows
that when the entropy remains constant within a comoving volume, that the
entropy density $s$ will also scale as $a^{-3}$. In determining the relic
abundance associated with a frozen out species, it is therefore convenient to
introduce the yield:
\begin{equation}
Y_{i}\equiv\frac{n_{i}}{s}\text{,}%
\end{equation}
which modulo subtleties connected to changes in the entropy, remains constant
after the $i^{th}$ species has frozen out. During an era of radiation
domination $t\propto T^{-2}$, and the Boltzmann equation for the yield
attains the form:%
\begin{equation}
\frac{dY_{i}}{dx_{i}}=\frac{s\left\langle \sigma_{i}v_{i}\right\rangle
}{H(m_{i})}\left(  Y_{i,\text{EQ}}^{2}-Y_{i}^{2}\right)  \text{,}
\label{BoltzYield}%
\end{equation}
where in the above, we have introduced the parameter:%
\begin{equation}
x_{i}=\frac{m_{i}}{T}\text{.}%
\end{equation}
After a species has frozen out, the left hand side of equation
(\ref{BoltzYield}) is to leading order, negligible.\footnote{See Chapter 5 of
\cite{KT} for a more precise discussion based on integrating the Boltzmann
equations.} Hence, the yield at the time of freeze out $Y_{i,\infty}$ is given
by $Y_{i,\text{EQ}}$ evaluated at the freeze out temperature:%
\begin{equation}
Y_{i,\infty}=Y_{i,\text{EQ}}(x_{i}^{f})\text{.}%
\end{equation}

Using the yield, we can determine the relic abundance of the $i^{th}$ species.
The key point is that because the yield does not change after freeze out, the
number density at present times is given by:%
\begin{equation}
n_{i,0}=s_{0}\cdot Y_{i,\infty}\text{.}%
\end{equation}
As a consequence, the relic abundance is:%
\begin{equation}
\Omega_{i}h^{2}=\frac{\rho_{i,0}h^{2}}{\rho_{c,0}}=\frac{m_{i}n_{i,0}h^{2}%
}{\rho_{c,0}}=\frac{s_{0}h^{2}}{\rho_{c,0}}m_{i}Y_{i,\infty}\sim
2.8\times10^{8}\text{ GeV}^{-1}\cdot m_{i}Y_{i,\infty}\text{,}
\label{relicyield}%
\end{equation}
where in the final equality, we have used the explicit values of $\rho_{c,0}$
and $s_{0}$ given by equations (\ref{rhopres}) and (\ref{spres}).

The actual yield of the $i^{th}$ species strongly depends on whether it is
relativistic, or only semi-relativistic at the time of freeze out. In the
former case, the freeze out temperature is far above the mass of the given
particle, so that it is appropriate to use the number density of equation
(\ref{nrel}). In the latter case, the mass may be comparable to, or larger
than the freeze out temperature, in which case the number density is given by
equation (\ref{boltzmannsupp}). Restoring all numerical factors, and using the
value of the entropy density in equation (\ref{sdense}), the yield at the time
of freeze out for a relativistic species is:%
\begin{equation}
Y_{i,\infty}^{\text{rel}}=\frac{n_{i,\infty}^{\text{rel}}}{s(T_{i}^{f})}%
\sim\frac{g_{\text{eff}}}{g_{\ast S}(x_{i}^{f})}\text{,}%
\end{equation}
where in the above, $g_{\text{eff}}=g$ for bosons and $g_{\text{eff}}=3g/4$
for fermions, with $g$ the number of degrees of freedom associated with the
given species.

The evaluation of the yield in the case of a species which is at most
semi-relativistic at freeze out is somewhat more involved. Nevertheless, for
our present purposes, the main point is that in a rough approximation, the
number density is given by evaluating the equilibrium number density at the
temperature of decoupling. Returning to the freeze out condition of equation
(\ref{freezeout}), the number density at freeze out is:%
\begin{equation}
n_{i,\infty}^{\text{n-rel}}\sim\frac{H(T_{i}^{f})}{\left\langle \sigma
_{i}v_{i}\right\rangle }\sim g_{\ast}^{1/2}(T_{i}^{f})\frac{m_{i}^{2}}{M_{PL}%
}\frac{(x_{i}^{f})^{2}}{\left\langle \sigma_{i}v_{i}\right\rangle }\text{.}%
\end{equation}
Dividing by the entropy density at the time of freeze out, the yield is
therefore:%
\begin{equation}
Y_{i,\infty}^{\text{n-rel}}=\frac{n_{i,\infty}^{\text{n-rel}}}{s(T_{i}^{f}%
)}\sim\frac{1}{M_{PL}}\frac{g_{\ast}^{1/2}(T_{i}^{f})}{g_{\ast S}(T_{i}^{f}%
)}\frac{x_{i}^{f}}{m_{i}\left\langle \sigma_{i}v_{i}\right\rangle }\text{.}%
\end{equation}
Plugging back into equation (\ref{relicyield}), and restoring all numerical
factors, the relic abundance is then given by:%
\begin{align}
\Omega_{i}^{\text{rel}}h^{2}  &  \sim8\times10^{-2}\cdot\frac{g_{\text{eff}}%
}{g_{\ast S}(x_{i}^{f})}\left(  \frac{m_{i}}{\text{eV}}\right)
\label{omegrel}\\
\Omega_{i}^{\text{n-rel}}h^{2}  &  \sim10^{9}\cdot\frac{\text{GeV}^{-1}%
}{M_{PL}}\frac{g_{\ast}^{1/2}(x_{i}^{f})}{g_{\ast S}(x_{i}^{f})}\frac
{x_{i}^{f}}{\left\langle \sigma_{i}v_{i}\right\rangle }\text{.}
\label{omegnrel}%
\end{align}
As a point of terminology, relics which decouple when they are relativistic
are often called \textquotedblleft hot\textquotedblright, whereas
relics which decouple when they are non-relativistic are
called \textquotedblleft cold\textquotedblright. An important feature of the cold relic density is that
it is inversely proportional to the thermally averaged cross section
$\left\langle \sigma_{i}v_{i}\right\rangle $. Assuming that $\left\langle
\sigma_{i}v_{i}\right\rangle \sim\alpha^{2}/M^{2}$ for $\alpha$ a fine
structure constant on the order of $\sim1/50$, in a model such as the MSSM\ or
Standard Model where $g_{\ast}\sim100$, a suggestive feature of the above
formula is that $\Omega_{i}^{\text{rel}}h^{2}\sim0.1$ when:
\begin{equation}
M\sim1\text{ TeV.}%
\end{equation}

An important caveat to the above computations is that it implicitly assumes
that the Universe starts at a high enough temperature that the species in
question is in thermal equilibrium, and then falls out of equilibrium. Indeed,
it is in principle also possible to consider scenarios where the production of
a given species is truncated because of $T_{RH}^{0}$ being lower than the
freeze out temperature of the species. Although somewhat ad hoc, this is one
mechanism which has often been invoked to avoid over-production of gravitinos
in models where this particle is the LSP.

\subsection{The Gravitino and its Consequences}

In the context of F-theory GUTs, the gravitino corresponds to the LSP. Due to
the fact that R-parity is typically preserved in such models, the gravitino is
stable, and can potentially correspond to a cosmological relic. For example,
in the context of high scale gauge mediation scenarios, the gravitino can have
a mass as high as $1$ GeV, although in the specific context of F-theory GUTs,
this value is somewhat lower at $10-100$ MeV. In this subsection we review the
fact that in many models, the gravitino relic abundance can overclose the
Universe. Indeed, one of the aims of the present paper is to explain how
F-theory GUTs naturally solve this \textquotedblleft problem\textquotedblright.

\subsubsection{Freeze Out of the Gravitino}

Because it interacts so weakly with other particles, the freeze out
temperature $T_{3/2}^{f}$ of the gravitino is typically quite high. To
estimate the value of $T_{3/2}^{f}$, we first clarify how the gravitino
interacts with the thermal bath of MSSM\ particles. Following the discussion
in for example \cite{Moroi:1993mb}, after supersymmetry is broken, the
longitudinal mode of the gravitino $\psi_{3/2}^{\mu}$ eats the spin $1/2$
Goldstino mode, $\psi$ associated with supersymmetry breaking so that:%
\begin{equation}
\psi_{3/2}^{\mu}\sim\frac{1}{m_{3/2}}\partial^{\mu}\psi\text{.}%
\end{equation}
Labeling the bosonic component first, given a chiral multiplet $(\phi,\chi)$
or vector multiplet $(A_{\mu},\lambda)$ the Goldstino mode couples to these
fields through the associated supercurrent so that the gravitino Lagrangian
density contains the terms \cite{Fayet:1986zc}:%
\begin{equation}
L_{3/2}\supset\frac{im_{\lambda}}{m_{3/2}\cdot M_{PL}}\left[  \gamma^{\mu
},\gamma^{\nu}\right]  \overline{\psi}\lambda F_{\mu\nu}+\frac{m_{\chi}%
^{2}-m_{\phi}^{2}}{m_{3/2}\cdot M_{PL}}\overline{\psi}\lambda\phi^{\ast
}+h.c.\text{,} \label{gravuniversal}%
\end{equation}
where in the above, the $m$'s denote the masses of various particles, the
$\gamma$'s are the usual Dirac matrices, and we have dropped various constants
which are not crucial for the discussion to follow. Letting $m$ denote the
characteristic mass scale associated with the mass splitting between members
of a given supermultiplet, it follows that the relevant cross section for the
gravitino is of the form:%
\begin{equation}
\sigma_{3/2}\sim\frac{1}{M_{PL}^{2}}\left(  \frac{m}{m_{3/2}}\right)
^{2}\text{.} \label{gravscatt}%
\end{equation}

When in equilibrium, the primary thermal production mechanism for gravitinos
is given by the conversion of particles of supersymmetric QCD into gravitinos
via processes of the form:%
\begin{equation}
AB\rightarrow C\psi_{3/2}\text{,}%
\end{equation}
where here, $A,B,C$ are shorthand for quarks, squarks, gluinos and gluons so
that:%
\begin{equation}
\sigma_{3/2}\sim\frac{1}{M_{PL}^{2}}\left(  \frac{m_{\widetilde{g}}}{m_{3/2}%
}\right)  ^{2}\text{,} \label{crosssec}%
\end{equation}
where $m_{\widetilde{g}}$ is the mass of the gluino. We refer the interested
reader to \cite{Ellis:1984eq,Moroi:1993mb,Bolz:2000fu} for a complete list of
interactions and the detailed form of the corresponding amplitudes.

Returning to cosmological considerations, the gravitino roughly freezes out at
the temperature $T_{3/2}^{f}$ defined by:%
\begin{equation}
H(T_{3/2}^{f})\sim n_{3/2}\left\langle \sigma_{3/2}v_{3/2}\right\rangle
\text{.}%
\end{equation}
Precisely because the gravitino only interacts quite weakly with the
background thermal bath, it decouples when it is still relativistic. For this
reason, it is appropriate to use the relation $n_{3/2}\sim T^{3}$ for a
relativistic species. Furthermore, because the decoupling of the gravitinos
happens at high temperatures, this decoupling occurs in an era when radiation
dominates the energy density of the Universe. We note in passing that after
the gravitino decouples, there could be a transition to a more exotic epoch
where matter, or the coherent oscillation of a field dominates the energy
density of the Universe. Using the relation between temperature and the Hubble
parameter in an era of radiation domination provided by equation
(\ref{raddom}):%
\begin{equation}
H^{2}\sim g_{\ast}\frac{T^{4}}{M_{PL}^{2}}%
\end{equation}
with $g_{\ast}$ the total number of relativistic degrees of freedom, it
follows that the freeze out temperature satisfies:%
\begin{equation}
T_{3/2}^{f}\sim g_{\ast}^{1/2}M_{PL}\left(  \frac{m_{3/2}}{m_{\widetilde{g}}%
}\right)  ^{2}\text{,} \label{TFREEZE}%
\end{equation}
where in the above, we have set $v_{3/2}\sim1$, as appropriate for a
relativistic species. Including all appropriate numerical factors
and performing a more precise estimate based on integrating the
Boltzmann equations yields a value $T_{3/2}^{f}\sim 10^{10} - 10^{11}$ GeV
for a gravitino of mass $m_{3/2} \sim 10$ MeV \cite{Steffen:2006hw,Pradler:2006hh}. Indeed,
in comparing the overall gravitino relic abundance obtained for $T^{0}_{RH} < T^{f}_{3/2}$
(a computation we will shortly review) with the value in the opposite
regime where $T^{0}_{RH} > T^{f}_{3/2}$, continuity of the gravitino
relic abundance across this interpolation yields:
\begin{equation}
T_{3/2}^{f}\sim2\times10^{10}\text{ GeV}\cdot\left(  \frac{m_{3/2}}{10\text{
MeV}}\right)  ^{2}\left(  \frac{1\text{ TeV}}{m_{\widetilde{g}}}\right)
^{2}\text{.} \label{TFREEZEGRAV}%
\end{equation}

\subsubsection{Gravitino Relic Abundance}

Having specified the temperature at which the gravitino freezes out, in this
subsection we determine the corresponding relic abundance. As in previous
subsections, at this point we will not assume that any late decaying field
dilutes the total entropy of the Universe. Assuming that the initial reheating
temperature $T_{RH}^{0}$ is greater than the freeze out temperature, the
formulae for the relic abundance of a \textquotedblleft hot\textquotedblright%
\ species given by equation (\ref{omegrel}) is applicable so that:%
\begin{equation}
T_{RH}^{0}>T_{3/2}^{f}:\Omega_{3/2}^{T}h^{2}\sim8\times10^{-2}\cdot
\frac{g_{\text{eff}}}{g_{\ast S}(x_{3/2}^{f})}\left(  \frac{m_{3/2}}%
{\text{eV}}\right)  \text{,} \label{HIGHRELICGRAV}%
\end{equation}
where in the above, $\Omega_{3/2}^{T}$ denotes the fact that these gravitinos
are thermally produced. Since the gravitino decouples at such high
temperatures, the value of $g_{\ast S}(x_{i}^{f})$ is given by the total
number of degrees of freedom in the MSSM. Including all relevant numerical
factors, the resulting relic abundance of gravitinos is \cite{Pagels:1981ke}:%
\begin{equation}
\Omega_{3/2}^{T}h^{2}\sim\frac{m_{3/2}}{2\text{ keV}}\text{.}%
\end{equation}
A perhaps distressing feature of this formula is that for a gravitino of mass
$m_{3/2}\geq0.2$ keV, the relic abundance would appear to overclose the Universe!

As the above analysis shows, if the MSSM\ thermal bath starts out at a very
high temperature and only cools to the gravitino freeze out temperature at
some later stage of expansion, there is a risk that the abundance of
gravitinos could overclose the Universe. For this very reason, it is common in
the literature to consider scenarios where thermal production of gravitinos
has been truncated by lowering the initial reheating temperature $T_{RH}^{0}$
below $T_{3/2}^{f}$.

To determine how low $T_{RH}^{0}$ must be in order to avoid an over-production
of gravitinos, we next repeat our analysis of the freeze out temperature of a
species detailed in section \ref{SUBSEC:RELDEC}, but now in the more general
case where the start of thermal production commences either above or below the
freeze out temperature of the gravitinos. The main change from our previous
analysis is that the relic abundance is now determined by the yield at the
temperature $T_{RH}^{0}$, rather than the freeze out temperature. As
throughout this review section, our aim here is to give a rough derivation of
this formula. More precise derivations may be found for example in
\cite{Ellis:1984eq,Moroi:1993mb,Bolz:2000fu}.

To estimate the abundance of thermally produced gravitinos, we again use the
fact that $Y_{3/2}^{T}=n_{3/2}^{T}/s$ is roughly constant after the initial
production of gravitinos. Here, it is important to note that in principle, the
initial reheating temperature $T_{RH}^{0}$ can either be greater than, or less
than the temperature at which gravitinos freeze out. In the latter case, the
thermal bath of MSSM\ particles will begin producing gravitinos up until the
temperature at which they freeze out. On the other hand, in the latter
scenario, the scattering processes described above will convert
MSSM\ particles into gravitinos at a temperature, $T_{RH}^{0}$ which
immediately freeze out. In this case, the yield of gravitinos is given as:%
\begin{equation}
Y_{3/2}^{T}\left(  T_{RH}^{0}\right)  =\frac{n_{3/2}}{s}\sim\frac
{C_{3/2}(T_{RH}^{0})}{H(T_{RH}^{0})s(T_{RH}^{0})}\text{,}%
\end{equation}
where the final estimate follows from the Boltzmann equation
(\ref{Boltzequation}). In this case, the equilibrium number density appearing
in $C_{3/2}$ as in equation is given by the number density of the background
radiation $n_{r}(T_{RH}^{0})$ so that:%
\begin{equation}
C_{3/2}(T_{RH}^{0})\sim\left\langle \sigma_{3/2}v_{3/2}\right\rangle n_{r}%
^{2}(T_{RH}^{0})\text{.}%
\end{equation}
Using the relation $n_{r}(T_{RH}^{0})=s(T_{RH}^{0})$, the yield is:%
\begin{equation}
Y_{3/2}^{T}\left(  T_{RH}^{0}\right)  \sim g_{\ast}^{1/2}(T_{RH}%
^{0})\left\langle \sigma_{3/2}v_{3/2}\right\rangle M_{PL}T_{RH}^{0}\text{.}%
\end{equation}
Utilizing our expression for the cross section in equation (\ref{crosssec}),
we find:%
\begin{equation}
Y_{3/2}^{T}\left(  T_{RH}^{0}\right)  \sim g_{\ast}^{1/2}(T_{RH}^{0}%
)\frac{T_{RH}^{0}}{M_{PL}}\left(  \frac{m_{\tilde{g}}}{m_{3/2}}\right)
^{2}\text{.}%
\end{equation}
As a consequence, the relic abundance when $T_{RH}^{0}<T_{3/2}^{f}$ is:%
\begin{equation}
T_{RH}^{0}<T_{3/2}^{f}:\Omega_{3/2}^{T}h^{2}\sim\left(  \frac{s_{0}}%
{\rho_{c,0}}h^{2}\right)  g_{\ast}^{1/2}(T_{RH}^{0})\frac{T_{RH}^{0}}{M_{PL}%
}\frac{m_{\tilde{g}}^{2}}{m_{3/2}}\text{.}%
\end{equation}
Restoring all numerical factors as in \cite{Bolz:2000fu}:%
\begin{equation}
T_{RH}^{0}<T_{3/2}^{f}:\Omega_{3/2}^{T}h^{2}\sim2.7\times10^{3}\cdot\left(
\frac{T_{RH}^{0}}{10^{10}\text{ GeV}}\right)  \left(  \frac{10\text{ MeV}%
}{m_{3/2}}\right)  \left(  \frac{m_{\widetilde{g}}}{1\text{ TeV}}\right)
^{2}\text{.}%
\end{equation}
This is to be contrasted with the relic abundance of gravitinos obtained in
the case where $T_{RH}^{0}>T_{3/2}^{f}$, where essentially the same formula
applies with $T_{RH}^{0}$ replaced by the freeze out temperature $T_{3/2}^{f}$.
Defining:%
\begin{equation}
T_{3/2}^{\min}\equiv\min(T_{RH}^{0},T_{3/2}^{f})\text{,}%
\end{equation}
the gravitino relic abundance in either case is given by:%
\begin{equation}
\Omega_{3/2}^{T}h^{2}\sim2.7\times10^{3}\cdot\left(  \frac{T_{3/2}^{\min}%
}{10^{10}\text{ GeV}}\right)  \left(  \frac{10\text{ MeV}}{m_{3/2}}\right)
\left(  \frac{m_{\widetilde{g}}}{1\text{ TeV}}\right)  ^{2}\text{.}
\label{finalthermalrelic}%
\end{equation}
Note that when $T_{RH}^{0}>T_{3/2}^{f}$, substituting $T_{3/2}^{f}$ of
equation (\ref{TFREEZE}) into the above relation reproduces the earlier form
of the gravitino relic abundance given by equation (\ref{HIGHRELICGRAV}):%
\begin{equation}
T_{RH}^{0}>T_{3/2}^{f}:\Omega_{3/2}^{T}h^{2}\sim\frac{m_{3/2}}{2\text{ keV}%
}\text{.}%
\end{equation}

Equation (\ref{finalthermalrelic}) illustrates the general puzzling feature of
models with a light, stable gravitino. On the one hand, particle physics
considerations have a priori nothing at all to do with the value of
$T_{RH}^{0}$. On the other, the overclosure constraint:%
\begin{equation}
\Omega_{3/2}h^{2}\leq0.1
\end{equation}
is badly violated for $m_{3/2}\geq2$ keV when~$T_{RH}^{0}>T_{3/2}^{f}$, and
when $T_{RH}^{0}<T_{3/2}^{f}$ imposes the stringent condition:%
\begin{equation}
T_{RH}^{0}\leq10^{6}\text{ GeV}\cdot\left(  \frac{m_{3/2}}{10\text{ MeV}%
}\right)  \text{,}%
\end{equation}
when the gluino has mass $m_{\widetilde{g}}\sim1$ TeV.

One of the remarkable features which we shall find in section \ref{FCOSMO} is
that these sharp bounds are significantly weaker in F-theory GUT\ models. In
fact, taking the most natural range of parameters dictated from purely
particle physics considerations, \textit{we find that this constraint is
completely absent in F-theory GUTs}! This is due to the fact that the decay of
the saxion can release a significant amount of entropy, diluting the relic
abundance of the gravitino.

\subsection{Cosmological Moduli and Their Consequences}

In the previous subsection we alluded to the crucial role which the decay of the
saxion can play in F-theory GUT scenarios. Here, we review the main effects of
late-decaying moduli for cosmology. To begin, we clarify our nomenclature for
\textquotedblleft moduli\textquotedblright. In the cosmology literature, it is
common to refer to any scalar field which has a nearly flat effective
potential as a \textquotedblleft modulus\textquotedblright. Prominent examples
in the context of supersymmetric models are field directions which are
massless in the limit where supersymmetry is restored. In the context of
string constructions, however, moduli fields typically refer to deformations
of the metric or a given vector bundle which can in principle be stabilized by
high scale supersymmetric dynamics. To distinguish these two notions of
\textquotedblleft moduli\textquotedblright\ we shall always refer to fields
which develop a mass due to supersymmetry breaking effects as
\textquotedblleft cosmological moduli\textquotedblright. In this section we
show that the coherent oscillation and subsequent decay of cosmological moduli
can have important consequences for cosmology.

Cosmological moduli are commonly thought to pose significant problems for
cosmology. Indeed, as we now review, such late decaying fields if present can
come to dominate the energy density of the Universe. If such fields decay too
late, they will disrupt BBN. On the other hand, there is also the potential
for such moduli to resolve various problems via their decays. This fact in
particular will prove important when we turn to the cosmology of F-theory GUTs.

By definition, the effective potential for cosmological moduli are quite flat.
After the initial reheating of the Universe ends at a temperature $T_{RH}^{0}%
$, the corresponding scalar modes will have a non-zero amplitude, which we
denote by $\phi_{0}$. As the Universe cools, these cosmological moduli develop
an effective potential, and begin to oscillate about their minima. In
particular, whereas in the standard cosmology, radiation dominates in the
range of temperatures between $T_{RH}^{0}$ and the start of BBN $T_{BBN}%
\sim10~$MeV, when the initial amplitude of a modulus field is large enough,
this contribution can come to dominate the energy density of the Universe. We
now review the estimate for determining when this can occur, and also
elaborate on some consequences associated with the subsequent decay of such a
modulus field.

Letting $V(\phi)$ denote the effective potential for the modulus, the equation
of motion for the $\phi$ field is given by:%
\begin{equation}
\ddot{\phi}+3H\dot{\phi}+V^{\prime}(\phi)=0
\end{equation}
where the dots above $\phi$ denote derivatives with respect to time and the
prime denotes the derivative with respect to $\phi$. For simplicity, we shall
consider the special case where $V(\phi)=m_{\phi}\phi^{2}/2$. In this case,
the energy density stored in the field is given by:%
\begin{equation}
\rho_{\phi}=\frac{1}{2}\dot{\phi}^{2}+\frac{1}{2}m_{\phi}^{2}\phi^{2}\text{,}%
\end{equation}
where in the above $m_{\phi}$ denotes the mass of the modulus in question. In
the following analysis we shall always assume that the mass $m_{\phi}$ is
roughly constant as a function of temperature. When we discuss the cosmology
of oscillating axions, we will see that temperature effects play a more
significant role.

There are in principle two possibilities for when a modulus will begin to
oscillate. The first possibility is that below the initial reheating
temperature, the effective potential may develop below the initial temperature
of reheating, in which case the modulus begins to oscillate during an era of
radiation domination. On the other hand, it is also possible that the modulus
could begin oscillating during an era above that set by the initial reheating temperature.

Assuming that the modulus begins to oscillate during an era of radiation
domination, the temperature at which the Hubble parameter becomes comparable
to the mass scale determines the oscillation temperature $T_{osc}^{\phi}$ so
that:%
\begin{equation}
H(T_{osc}^{\phi})\sim m_{\phi}\text{.}%
\end{equation}
Using the radiation domination relation $H\sim g_{\ast}^{1/2}T^{2}/M_{PL}$,
the resulting oscillation temperature is:%
\begin{equation}
T_{osc}^{\phi}\sim g_{\ast}^{-1/4}\sqrt{m_{\phi}M_{PL}}\text{.}
\label{ToscPhiDef}%
\end{equation}
Restoring all numerical factors, a more exact analysis yields:%
\begin{equation}
T_{osc}^{\phi}\sim0.3\sqrt{m_{\phi}M_{PL}}\text{.}%
\end{equation}
We shall frequently refer to this temperature as the \textquotedblleft
oscillation temperature\textquotedblright\ of the modulus field.

At temperatures where the modulus field begins to oscillate, the energy
density $\rho_{\phi}$ stored in the modulus field is governed by the initial
amplitude of the modulus and the mass of the modulus field so that:%
\begin{equation}
\rho_{\phi}\sim\frac{1}{2}m_{\phi}^{2}\phi_{0}^{2}\text{.}%
\end{equation}
On the other hand, the energy density stored in the background radiation
$\rho_{r}\sim T^{4}$, so that at the era when the modulus field begins to
oscillate, we have:%
\begin{equation}
\left(  \rho_{r}\right)  _{osc}\sim(T_{osc}^{\phi})^{4}\text{.}%
\end{equation}
Combining this relation with equation (\ref{ToscPhiDef}), it follows that the
ratio of $\rho_{\phi}$ and $\rho_{r}$ at the temperature $T_{osc}^{\phi}$ is:%
\begin{equation}
\left(  \frac{\rho_{\phi}}{\rho_{r}}\right)  _{\phi,osc}\sim\frac{\phi_{0}%
^{2}}{M_{PL}^{2}}\text{.} \label{rhorat}%
\end{equation}
An interesting feature of this formula is that $\phi_{0}$, rather than
$m_{\phi}$ appears.

Following for example \cite{Hashimoto:1998ua}, we now determine the conditions
required for a modulus field to dominate the energy density. For simplicity,
we consider a scenario where a single modulus field undergoes coherent
oscillation. This scenario can be generalized to situations where multiple
fields oscillate and can all contribute a substantial portion of the overall
energy density of the Universe.

Assuming that the Universe is in an era of radiation domination, the
transition to a modulus dominated energy density can occur provided the energy
density stored in the modulus, $\rho_{\phi}$ becomes comparable to the
background radiation $\rho_{r}$. In other words, an era of modulus domination
commences at a temperature $T_{dom}^{\phi}$ where:%
\begin{equation}
\left(  \frac{\rho_{\phi}}{\rho_{r}}\right)  _{\phi,dom}\sim1\text{.}%
\end{equation}
Assuming that the oscillation of the modulus eventually dominates the energy
density at a temperature $T_{dom}^{\phi}$, the scaling behavior $\rho_{\phi
}\varpropto T^{3}$ and $\rho_{r}\propto T^{4}$ translates into the condition:%
\begin{equation}
T_{dom}^{\phi}\left(  \frac{\rho_{\phi}}{\rho_{r}}\right)  _{\phi,dom}%
=T_{osc}^{\phi}\left(  \frac{\rho_{\phi}}{\rho_{r}}\right)  _{\phi
,osc}\text{.}%
\end{equation}
Combined with equation (\ref{rhorat}), the temperature $T_{dom}^{\phi}$ is
given by:%
\begin{equation}
T_{dom}^{\phi}\sim\frac{\phi_{0}^{2}}{M_{PL}^{2}}T_{osc}^{\phi}\sim\frac
{\phi_{0}^{2}}{M_{PL}^{2}}\sqrt{m_{\phi}M_{PL}}\text{.} \label{tphidom}%
\end{equation}
More generally, in scenarios where $T_{osc}^{\phi}$ is either greater than or
less than $T_{RH}^{0}$, $T_{dom}^{\phi}$ is given by:%
\begin{equation}
T_{dom}^{\phi}\sim\frac{\phi_{0}^{2}}{M_{PL}^{2}}\min(T_{osc}^{\phi}%
,T_{RH}^{0})\text{.}%
\end{equation}

In the above estimate, we have implicitly assumed that the modulus field is
sufficiently long lived that it can dominate the energy density. It is also
possible, however, that the modulus field may decay before, or after an era of
modulus domination is able to commence. For example, when $\phi_{0}/M_{PL}$ is
sufficiently low, the resulting value of $T_{dom}^{\phi}$ may correspond to a
timescale which is longer than the lifetime of $\phi$. In such a scenario, the
decay of $\phi$ will occur in an era of radiation domination.

The timescale for the decay of the $\phi$ particle is given by the inverse of
$\Gamma_{\phi}$, the decay rate for $\phi$. The modulus $\phi$ will either
decay during an era of radiation domination, or at the end of $\phi$
domination. In the latter case, we shall assume that the Universe then
transitions back to an era of radiation domination. In both cases, the
corresponding temperature $T_{decay}^{\phi}$ at the time of decay therefore
scales with $t_{decay}^{\phi}=\Gamma_{\phi}^{-1}$ as:%
\begin{equation}
\sqrt{t_{decay}^{\phi}}\propto(T_{decay}^{\phi})^{-1}\text{,}%
\end{equation}
which holds when radiation dominates the energy density of the Universe. The
relation:%
\begin{equation}
H(T_{decay}^{\phi})\sim\Gamma_{\phi}%
\end{equation}
therefore implies:%
\begin{equation}
T_{decay}^{\phi}\sim g_{\ast}^{-1/4}\sqrt{\Gamma_{\phi}M_{PL}}\text{.}%
\end{equation}
Restoring all numerical factors, a more exact analysis yields:%
\begin{equation}
T_{decay}^{\phi}\sim0.5\sqrt{\Gamma_{\phi}M_{PL}}\text{.} \label{tphidecay}%
\end{equation}
We note that if $\Gamma_{\phi}$ is sufficiently small, this can disrupt the
standard predictions of BBN, significantly altering standard cosmology. As we
now explain, in situations where $\phi$ dominates the energy density of the
Universe, it is also common to refer to this temperature as the
\textquotedblleft reheating temperature\textquotedblright\ of $\phi$, and we
shall therefore also use the notation:%
\begin{equation}
T_{RH}^{\phi}\equiv T_{decay}^{\phi}\sim0.5\sqrt{\Gamma_{\phi}M_{PL}}\text{.}%
\end{equation}

In order for the Universe to enter an era of modulus domination, the three
temperatures $T_{decay}^{\phi}$, $T_{dom}^{\phi}$ and $T_{RH}^{0}$ must
satisfy the system of inequalities:%
\begin{equation}
T_{decay}^{\phi}<T_{dom}^{\phi}<\min(T_{osc}^{\phi},T_{RH}^{0})\text{.}%
\end{equation}
Returning to equations (\ref{tphidom}) and (\ref{tphidecay}), this amounts to
a condition on the properties of the modulus:%
\begin{equation}
\sqrt{\Gamma_{\phi}M_{PL}}\lesssim\frac{\phi_{0}^{2}}{M_{PL}^{2}}\sqrt
{m_{\phi}M_{PL}}\text{,}%
\end{equation}
or:%
\begin{equation}
\sqrt{\frac{\Gamma_{\phi}}{m_{\phi}}}\lesssim\frac{\phi_{0}^{2}}{M_{PL}^{2}%
}\text{.}%
\end{equation}

The decay products of the modulus field can also have significant consequences
for the cosmology of the Universe. Again, the immediate consequence naturally
separates into scenarios where the modulus decays in an era of radiation
domination or modulus domination. In the former case, the primary constraint
is that the resulting decay products must satisfy the bound
(which we shall review later) on the total number of relativistic
species, in order to remain in accord with BBN. While a
similar constraint also holds in the case where the modulus comes to dominate
the energy density of the Universe, another significant effect is the release
of entropy into the Universe as a result of the decay of the modulus. This
increase in entropy effectively dilutes the total relic abundance of all
species produced prior to this decay.

Letting $\Omega_{z}^{(0)}h^{2}$ denote the relic abundance of a species $z$
prior to the decay of $\phi$, the relic abundance after dilution is then given
by:%
\begin{equation}
\Omega_{z}h^{2}=D_{\phi}\Omega_{z}^{(0)}h^{2}\equiv\frac{s_{before}}%
{s_{after}}\Omega_{z}^{(0)}h^{2}\text{,}%
\end{equation}
where in the above, $s$ denotes the entropy density, and we have introduced
the \textquotedblleft dilution factor\textquotedblright:%
\begin{equation}
D_{\phi}\equiv\frac{s_{before}}{s_{after}}\text{.} \label{dilutiondef}%
\end{equation}
A similar analysis to that utilized in estimating the ratio of energy
densities in equation (\ref{rhorat}) yields the following estimate for the
dilution factor, which can be found, for example, in
\cite{KT,BanksDineGraesser,Ibe:2006rc,Kawasaki:2008jc}:%
\begin{equation}
D_{\phi}\sim\frac{T_{RH}^{\phi}}{T_{dom}^{\phi}}\sim\frac{M_{PL}^{2}}{\phi_{0}^{2}%
}\frac{T_{RH}^{\phi}}{\min\left(  T_{RH}^{0},T_{osc}^{\phi}\right)  }\text{.}
\label{dilfingen}%
\end{equation}
In the above, the minimum of $T_{RH}^{0}$ and $T_{osc}^{\phi}$ enters because
there are in principal two possible scenarios, where either the temperature of
oscillation is larger, or smaller than $T_{RH}^{0}$. The essential point is
that the lower temperature more strongly determines the resulting entropy
density of the Universe denoted by $s_{before}$, so that it is always
appropriate to take the minimum of these two parameters. The release of this
large entropy into the Universe effectively \textquotedblleft
reheats\textquotedblright\ the Universe.

An implicit assumption of equation (\ref{dilfingen}) is that the dilution
factor $D_{\phi}$ is less than one. Indeed, in scenarios where $\phi$ decays during
an era of radiation domination, the only effect on the history of the Universe
may be an increase in the relic abundance of a given species. When $D_{\phi}$ is
formally greater than one, it follows from equation (\ref{dilfingen}) that the
temperature $T_{dom}^{\phi}$ is less than $T_{RH}^{\phi}$, violating one of
the implicit assumptions used to derive our expression for the dilution factor.

\subsubsection{The Saxion as a Cosmological Modulus}

In supersymmetric models which also solve the strong CP\ problem via an axion,
the other real bosonic degree of freedom of the corresponding chiral
supermultiplet, which we shall refer to as the saxion provides an important
example of a cosmological modulus. The essential point is that in the limit
where supersymmetry is restored, the axion and saxion have the same mass.
Thus, just as for any other cosmological modulus, supersymmetry breaking
effects provide the dominant contribution to the effective potential of the saxion.

The presence of the saxion is especially significant in models where the scale
of supersymmetry breaking is lower than in gravity mediation models. On
general grounds, many of the moduli of the string compactification in such
cases will have large masses determined by high scale supersymmetric dynamics.
Indeed, in such a scenario essentially the only remaining cosmological moduli
are those required from particle physics considerations, the saxion being the
primary example of such a cosmological modulus. The dynamics of this field in
particular will play an important role in the cosmology of F-theory GUT\ models.

\subsection{Oscillations of the Axion}

In the previous subsection we reviewed the fact that the oscillation of a
modulus can alter the evolution of the Universe, leading to an era of modulus
domination, as well as an overall dilution of all relic abundances due to the
release of entropy into the Universe. In particular, we also reviewed the fact
that the saxion, as a component of the axion supermultiplet can play the role
of such a cosmological modulus. In this subsection we review the cosmology
associated with the axion, focussing in particular on the effects of its oscillation.

We first begin by reviewing some details of the axion. By definition, the QCD
axion $a$ couples to the QCD\ instanton density so that the Lagrangian density
for $a$ contains the terms:%
\begin{equation}
L_{axion}=\frac{f_{a}^{2}}{2}\left(  \partial_{\mu}a\right)  ^{2}+\frac
{a}{32\pi^{2}}\varepsilon^{\mu\nu\rho\sigma}Tr_{SU(3)_{C}}F_{\mu\nu}F_{\rho\sigma}%
\end{equation}
where here, $F_{\mu\nu}$ denotes the field strength of $SU(3)_{C}$, and
$f_{a}$ denotes the axion decay constant. The field $a$ corresponds to the
Goldstone mode associated with spontaneous breaking of an anomalous global
$U(1)$ symmetry and takes values in the interval:%
\begin{equation}
-\pi<a<\pi\text{.}%
\end{equation}

Constraints from supernova cooling impose a lower bound on $f_{a}$ so that:%
\begin{equation}
f_{a}>10^{9}\text{ GeV.}%
\end{equation}
The upper bound on $f_{a}$ is based on cosmological considerations, and is
general more flexible. One of the purposes of this subsection is to review the
derivation of this upper bound.

The effective potential for the axion is generated by QCD instanton effects
and can be approximated using the pion Lagrangian, as for example in section
23.6 of \cite{WeinbergII}:%
\begin{equation}
V_{axion}(a)=m_{\pi}^{2}f_{\pi}^{2}\left(  1-\cos a\right)  \text{,}%
\end{equation}
where $m_{\pi}\sim130$ MeV is the mass of the pion, and $f_{\pi}\sim90$ MeV is
the pion decay constant. The mass of the canonically normalized axion is
therefore:%
\begin{equation}
m_{a}\sim\frac{m_{\pi}f_{\pi}}{f_{a}}\sim6\times10^{-5}\text{ eV}\cdot\left(
\frac{10^{12}\text{ GeV}}{f_{a}}\right)  \text{.}%
\end{equation}
At temperatures $T\gg\Lambda_{QCD}\sim0.2$ GeV, the mass of the axion depends
non-trivially on $T$, and is given by:
\begin{equation}
m_{a}(T)=m_{a}\cdot\left(  \frac{\Lambda_{QCD}}{T}\right)  ^{4}\text{.}
\label{axfinitetemp}%
\end{equation}

The axion is a very long lived particle, and can therefore have consequences
for cosmology. Returning to the pion Lagrangian, it can be shown that the
primary decay channel of the axion is into two photons. Following section 23.6
of \cite{WeinbergII}, the relative decay rates between $a\rightarrow
\gamma\gamma$ and~$\pi^{0}\rightarrow\gamma\gamma$ is given by the square of
the ratios of the two decay constants, $f_{a}$ and $f_{\pi}$ multiplied by an
overall phase space factor proportional to $m_{a}^{3}/m_{\pi}^{3}$. The
relative decay rates are therefore:%
\begin{equation}
\frac{\Gamma_{a\rightarrow\gamma\gamma}}{\Gamma_{\pi^{0}\rightarrow
\gamma\gamma}}\sim\left(  \frac{f_{\pi}}{f_{a}}\right)  ^{2}\left(
\frac{m_{a}}{m_{\pi}}\right)  ^{3}\sim\left(  \frac{f_{\pi}}{f_{a}}\right)
^{5}\text{.}%
\end{equation}
Using the lifetime of $\pi^{0}$ given by $\tau_{\pi^{0}}\sim8.4\times10^{-17}$
sec, it follows that the lifetime of the axion is:%
\begin{equation}
\tau_{a}\sim10^{-16}\sec\cdot\left(  \frac{f_{\pi}}{f_{a}}\right)  ^{-5}%
\sim10^{49}\sec\cdot\left(  \frac{10^{12}\text{ GeV}}{f_{a}}\right)
^{-5}\text{,}%
\end{equation}
which is far greater than the current lifetime of the Universe $\sim
10^{18}\sec$.

Because the axion is quite long lived in comparison to cosmological
timescales, it can in principle play an important role in cosmology. Even so,
due to its small mass and the fact that all couplings of the axion to the
Standard Model and MSSM\ degrees of freedom are suppressed by powers of
$1/f_{a}$, the total relic abundance of axions produced from thermal processes
is typically quite small. We refer the interested reader to \cite{KT}, for
example, for further details on such estimates.

The primary cosmological issue connected to the axion is the fact that at high
temperatures, the potential for the axion is nearly flat, and the field can
easily be displaced from its minimum. Much as for cosmological moduli, the
oscillation of the axion can then have consequences for cosmology, appearing
as a zero momentum condensate of non-relativistic particles. The corresponding
contribution to the overall energy density of the Universe can in principle
overclose the Universe, or provide a significant component of the overall dark
matter. To this end, we now review the associated relic abundance from
coherent oscillation of the axion.\footnote{Although it is beyond the scope of
the present paper to present speculations on the evolution of the Universe at
temperatures greater than $T_{RH}^{0}$, in the context of inflationary models
where $T_{RH}^{0}<f_{a}$, quantum fluctuations in the oscillation of the axion
of the form $a=\left\langle a\right\rangle +\delta a$ associated with
oscillation of the axion can in principle induce density perturbations,
leading to small variations in the CMBR. These \textquotedblleft isocurvature
perturbations\textquotedblright\ occur in models where the fluctuation mode
exits the horizon during the expected de Sitter phase and remains frozen in
until some time after inflation ends, at which point the mode re-enters the
horizon. In the context of models in which inflation occurs at a temperature
$T_{RH}^{0}<f_{a}$, this leads to a bound on the reheating temperature of the
form: $T_{RH}^{0}\lesssim10^{13}$ $\mathrm{GeV}\cdot\left(  \Omega
_{\mathrm{ax}}h^{2}\right)  ^{-1/4}\cdot(f_{a}/10^{12}$ GeV$)^{5/24}$. See for
example \cite{KT} for further details of isocurvature perturbations.}

\subsubsection{Axionic Dark Matter}

As indicated previously, the axion is nearly massless at high temperatures. By
inspection of equation (\ref{omegrel}), the thermal production of axionic
relics is very small owing to the small mass of the axion $m_{a}\sim10^{-5}$
eV. For this reason, axionic dark matter is only a viable candidate when
produced through some non-thermal mechanism. Precisely because of its small mass,
the axion can be displaced from its minimum to a value $a_{0}$ such that
$-\pi<a_{0}<\pi$. Throughout our discussion, we shall assume that this
initial displacement is given by roughly the same value in causally
disconnected patches of the Universe. We note that in scenarios where
$T_{RH}^{0}<f_{a}$, whatever mechanism solves the homogeneity problem will
also translate into a uniform value for the initial amplitude in the entire
causal patch of the Universe. At higher temperatures where $T_{RH}^{0}>f_{a}$,
the axion is still not well-defined, so that once the Universe cools to a
temperature below $f_{a}$, the initial amplitude of the axion may be different
in distinct patches of the Universe. Nevertheless, this simply amounts to
replacing the uniform value of the initial amplitude by a rough average over
various causal patches.

Once the Universe cools sufficiently, the axion will begin to oscillate,
creating a condensate of zero momentum particles. We now proceed to estimate
the effective number density of this condensate, and compute the associated
relic abundance of axions. Just as in our review of general cosmological
moduli, under the assumption that the mass term dominates the effective
potential, the equation of motion for the axion is given by:%
\begin{equation}
\ddot{a}+3Ha+m_{a}^{2}(T)a=0
\end{equation}
where here, we have included the explicit $T$ dependence of $m_{a}$ given by
equation (\ref{axfinitetemp}).

A priori, the axion may begin to oscillate during an era of either radiation
or modulus domination. In the latter case, the decay of the modulus can dilute
the relic abundance of the axion, so that in principle, the value of the decay
constant can be increased \cite{BanksDineGraesser}. However, when this is not
done, a comparison of the relic abundances obtained from oscillation in an era
of radiation domination and the \textquotedblleft undiluted\textquotedblright%
\ relic abundance of the axion obtained from an era of modulus domination are
numerically quite similar. For this reason, the diluted relic abundance is
typically negligible when the axion starts oscillating before the modulus
decays. Since we are interested in the case where the undiluted relic
abundances are numerically quite similar anyway, for our present purposes it
is sufficient to review the relic abundance computation in the case where the
axion begins oscillating during an era of radiation domination.

The axion begins to oscillate at a temperature $T_{osc}^{a}$ where the mass
$m_{a}(T)$ is comparable to the overall Hubble parameter:%
\begin{equation}
H\sim m_{a}\left(  T_{osc}^{a}\right)  \text{.}%
\end{equation}
Assuming that it begins oscillating during an era of radiation domination so
that $H\sim g_{\ast}^{1/2}T^{2}/M_{PL}$ it follows that $T_{osc}^{a}$ is given
by:%
\begin{equation}
\frac{T_{osc}^{a}}{\Lambda_{QCD}}\sim\left(  \frac{M_{PL}m_{a}}{g_{\ast}%
^{1/2}\Lambda_{QCD}^{2}}\right)  ^{1/6}\sim10\cdot\left(  \frac{10^{12}\text{
GeV}}{f_{a}}\right)  ^{1/6}\text{.} \label{Toscrad}%
\end{equation}
Dropping the weak dependence on $f_{a}$, the temperature at which the axion
begins to oscillate is therefore given by:%
\begin{equation}
T_{osc}^{a}\sim10\cdot\Lambda_{QCD}\sim1\text{ GeV,}%
\end{equation}
which is only a few orders of magnitude away from the start of BBN, with
$T_{BBN}\sim 2$ MeV $\sim10^{-2}\cdot\Lambda_{QCD}$.

We now proceed to determine the relic abundance of the axion. As in our
analysis of the cosmological modulus, the energy density stored in the axion
when it commences oscillation is given by:%
\begin{equation}
\rho_{a}(T_{osc}^{a})\sim\frac{1}{2}m_{a}^{2}(T_{osc}^{a})(f_{a}a_{0})^{2}\text{.}%
\end{equation}
Treating the field condensate as a collection of non-relativistic particles at
zero momentum, the initial number density is:%
\begin{equation}
n_{a}(T_{osc}^{a})\sim\frac{\rho_{a}(T_{osc}^{a})}{m_{a}(T_{osc}^{a})}%
\sim\frac{1}{2}m_{a}(T_{osc}^{a})(f_{a}a_{0})^{2}\sim\frac{1}{2}m_{a}(f_{a}a_{0})^{2}%
\cdot\left(  \frac{\Lambda_{QCD}}{T_{osc}^{a}}\right)  ^{4}\text{.}%
\end{equation}
The yield of axions is therefore:%
\begin{equation}
Y_{a}=\frac{n_{a}(T_{osc}^{a})}{s(T_{osc}^{a})}\sim m_{a}(f_{a}a_{0})^{2}\cdot\left(
T_{osc}^{a}\right)  ^{-3}\left(  \frac{\Lambda_{QCD}}{T_{osc}^{a}}\right)
^{4}\text{.}%
\end{equation}
Including all relevant numerical factors, a similar analysis to the one
already presented yields the final estimate for the axion relic abundance
\cite{Mukhanov}:%
\begin{equation}
\Omega_{\mathrm{ax}}h^{2}\sim a_{0}^{2} \left(  \frac{f_{a}}{10^{12}\;\mathrm{GeV}%
}\right)  ^{7/6} \text{.}%
\end{equation}
Thus, under circumstances where $a_{0}$ is roughly an order one number,
it follows that overclosure constraints impose the condition $f_{a}%
\lesssim10^{12}$ GeV, so that axions can in principle comprise a component of
dark matter. Nevertheless, this is quite sensitive to the actual value of
$a_{0}$ so that even when $a_{0}\sim10^{-1}$, the corresponding
relic abundance will be negligible.\ Further, in the event that the axion
begins oscillating during an era of modulus domination, the numerical
similarity of the two relic abundances for $f_{a}\sim10^{12}$ GeV implies that
once the effects of dilution are taken into account, in the latter case the
relic abundance is always negligible.

\subsection{Constraints from BBN}

While many extensions of the Standard Model come equipped with potential dark
matter candidates, it is also quite important to check that any such extension
does not introduce additional elements which conflict with well-established
features of the standard cosmology. In this regard, the standard cosmology
prediction for the abundances of the light nuclei $H^{+}$, $D^{+}$ are in
excellent agreement with observation, and are in reasonable accord with
$^{3}He^{++}$, $^{4}He^{++}$. The predicted abundance of $^{7}Li$ derived from
the standard cosmology appears to reveal a discrepancy between theory and
observation. Optimistically speaking, this can be viewed as a potential window
into the physics beyond the Standard Model which could potentially alter some
of the reaction rates present in standard nucleosynthesis.

One of the most remarkable features of BBN\ is that the resulting abundances
of light elements essentially depend on only the expansion rate of the
Universe, and the overall baryon asymmetry:%
\begin{equation}
\eta_{B}\equiv\frac{n_{B}-n_{\overline{B}}}{n_{\gamma}}\text{,}%
\end{equation}
where in the above, $n_{B}$, $n_{\overline{B}}$ and $n_{\gamma}$ respectively
denote the number density of baryons, anti-baryons and
photons. Quite remarkably, although the resulting abundances span
approximately \textit{nine} orders of magnitude, they are all correctly
accounted for when the baryon asymmetry falls within the narrow window:%
\begin{equation}
4.7\times10^{-10}\;\lesssim\eta_{B}\;\lesssim6.5\times10^{-10}\text{.}%
\end{equation}

Extensions of the standard cosmology can potentially threaten this result in
one of two ways. As we will shortly review, BBN\ imposes significant limits on
the overall expansion rate, and as such, effectively constrains the total
number of relativistic species present at the start of BBN. The presence of
late-decaying particles can also alter the results of BBN by either
destroying, or producing too much of a given light element. For example, in
the context of the MSSM, in scenarios with a bino-like NLSP, the decay of this
particle into a photon and gravitino can potentially disrupt the production of
certain elements through the presence of additional background photons. In
certain cases, however, such decays can in fact \textit{improve} the agreement
between theory and observation. An important example of this type is the
overall abundance of $^{7}Li$, which in the standard cosmology turns out to be
a factor of $2-5$ larger than is observed. In the remainder of this
subsection, we provide additional details on these two central constraints.

\subsubsection{BBN Bounds on Relativistic
Species\label{RelSpecCon}}

As reviewed for example in \cite{KT}, increasing the expansion rate of the
Universe leads to an increase in the total amount of $^{4}He$ produced by BBN.
Due to the connection between the Hubble parameter and the number of
relativistic degrees of freedom coupled to the thermal bath:%
\begin{equation}
H=\sqrt{\frac{g_{\ast}(T)\pi^{2}}{90M_{PL}^{2}}}T^{2}\text{,}%
\end{equation}
a constraint on the expansion rate translates into a direct bound on the total
number of relativistic species. Although it is beyond the scope of the present
paper to review the derivation of how production of $^{4}He$ translates into a
bound on the expansion rate of the Universe, the end result of this
calculation is that at the start of BBN, the total number of relativistic
species must satisfy:%
\begin{equation}
g_{\ast}(T\sim\text{MeV})\leq12.5\text{.} \label{BBNlightbound}%
\end{equation}
where $g_{\ast}(T)$ as a function of $T$ is given by equation (\ref{gstdef}).
The relativistic degrees of freedom of the Standard Model already nearly
saturate this upper bound, and are given by the photon $(g_{\gamma}=+2)$,
three species of neutrinos $(g_{\nu}=6)$ and the electrons and positrons
$(g_{e^{-}}=g_{e^{+}}=2)$. Because these particles are all in thermal
equilibrium, equation (\ref{gstdef}) simplifies to the special case where
$T_{i}=T$ for all species so that:%
\begin{equation}
g_{\ast}^{SM}(T\sim1\text{ MeV})=2+\frac{7}{8}\left(  6+2+2\right)
=10.75\text{.}%
\end{equation}
Combined with the strong upper bound provided by (\ref{BBNlightbound}), the
general temperature dependence of $g_{\ast}$ provided by equation
(\ref{gstdef}) leads to the inequality:%
\begin{equation}
\underset{\text{new bose}}{\sum}g_{i}\left(  \frac{T_{i}}{1\text{ MeV}%
}\right)  ^{4}+\frac{7}{8}\underset{\text{new fermi}}{\sum}g_{i}\left(
\frac{T_{i}}{1\text{ MeV}}\right)  ^{4}<1.75\text{.} \label{grelbound}%
\end{equation}
Note in particular that even one fermionic species in thermal equilibrium
already completely saturates this upper bound. For example, this bound implies
that only one additional species of interacting relativistic neutrinos can be
included in an extension of the Standard Model, and that in this case,
absolutely no additional degrees of freedom can be added without disrupting
BBN! This same condition can also be stated as a bound on the total energy
density contributed by an additional relativistic species:%
\begin{equation}
\left(  \frac{\rho_{extra}}{\rho_{r}}\right)  _{BBN}\leq\frac{7}{43}\text{,}%
\end{equation}
where $\rho_{extra}$ denotes the energy density stored in the extra
relativistic species.

It is important to qualify that the bound on the number of relativistic
species is most stringent in the case of additional degrees of freedom which
directly couple to the background thermal bath. For example, this might appear
to contradict the possibility of Dirac-like neutrinos, because \textit{if}
additional light states happened to be in thermal equilibrium at the start of
BBN, inequality (\ref{grelbound}) would be violated.

Additional relativistic species could be present if they decouple at a
sufficiently high temperature, which we denote by $T_{D}$. Comparing the scale
factor dependence in equations (\ref{radscale}) and (\ref{mattscale}) with the
temperature dependence in equations (\ref{rhorel}) and (\ref{rhonrel}), it
follows that the temperature of a decoupled species $i$ obeys the relation:%
\begin{equation}
\label{ati}\text{Decoupled and Relativistic: }a\propto\frac{1}{T_{i}}\text{.}%
\end{equation}
On the other hand, species which remain coupled to the thermal bath are more
directly sensitive to changes in the number of relativistic species. Indeed,
equation (\ref{Tbathevolve}) implies that the overall scaling of the thermal
bath evolves as:%
\begin{equation}
aT\propto g_{\ast S}^{-1/3}(T)\text{.} \label{at}%
\end{equation}
Comparing equations (\ref{ati}) and (\ref{at}), the resulting temperature
ratio entering inequality (\ref{grelbound}) is:%
\begin{equation}
\frac{T_{i}}{T}\sim\left(  \frac{10.75}{g_{\ast}(T_{D})}\right)
^{1/3}\text{.}%
\end{equation}
In other words, if a species decouples at a sufficiently high temperature, the
actual contribution to $g_{\ast}(T\sim1$ MeV$)$ will be significantly suppressed.

Returning to the case of right-handed neutrinos mentioned previously, we note
that such particles decouple at a temperature $T_{D}$ such that:%
\begin{equation}
g_{\ast}(T_{D})\geq106.75\text{,}%
\end{equation}
where this lower bound corresponds to the number of degrees of freedom of the
Standard Model. As a consequence, we obtain the relation:%
\begin{equation}
6\cdot\frac{7}{8}\left(  \frac{T_{\nu_{R}}}{T}\right)  ^{4}\lesssim0.2\text{,}%
\end{equation}
so that inequality (\ref{grelbound}) remains intact.

\subsubsection{BBN\ and Late Decaying Particles}

Given the fact that even crude bounds from BBN tied to the expansion rate of
the Universe translate into detailed constraints on the number of relativistic
species, it is perhaps not surprising that the reaction rates necessary for
generating the correct abundance of light elements from BBN\ are also quite
sensitive to the presence of late decaying particles. On the one hand, this
imposes important constraints on potential extensions of the Standard Model,
because the abundances of the light nuclei $H^{+}$, $D^{+}$, $T^{+}$,
$^{3}He^{++}$, $^{4}He^{++}$ are all in reasonable accord with observation. On
the other hand, this also provides a window into new physics, because the
standard cosmology appears to predict an abundance of $^{7}Li$ which is too
large by a factor of $2-5$ when compared with observation. We refer the
interested reader to \cite{Bailly:2008yy} and references therein for a very
recent account of the current bounds on various abundances.

As briefly mentioned above, late decaying particles are possible in certain
supersymmetric extensions of the Standard Model. For example, in the context
of the MSSM\ where the effects of supersymmetry breaking are communicated via
gauge mediation, the gravitino is the lightest superpartner of the MSSM, and
either the bino or stau corresponds to the next to lightest superpartner
(NLSP). This is the case of primary interest for F-theory GUTs, and so in the
remainder of this subsection we shall therefore restrict attention to this case.

The decay rate of the NLSP\ into a gravitino and its Standard Model
counterpart is determined by the universal coupling of the gravitino to matter
provided by equation (\ref{gravuniversal}). The calculation of the lifetime of
the NLSP\ is reviewed for example, in \cite{MartinPrimer}, and leads to the
well known result:%
\begin{equation}
\tau_{NLSP}\sim\frac{6\times10^{-2}\text{ sec}}{\kappa}\cdot\left(
\frac{m_{NLSP}}{100\text{ GeV}}\right)  ^{-5}\left(  \frac{m_{3/2}}{10\text{
MeV}}\right)  ^{2}\text{,} \label{NLSPLIFE}%
\end{equation}
where $m_{NLSP}$ denotes the mass of the NLSP$\ $and $\kappa$ is a model
dependent factor which is unity for the case of the stau NLSP, and measures
the photino content of the bino in the case of the bino NLSP. The particular
normalization for the two masses has been chosen to conform with natural
values in the range expected in the specific context of a high-scale gauge
mediation model, as is the case in the context of F-theory GUTs. The lifetime
of the NLSP\ is to be compared with the timescale of BBN, which roughly
commences at a temperature of $T_{BBN}\sim1$ MeV, corresponding to the
timescale $t\sim0.2$ s. By inspection of equation (\ref{NLSPLIFE}), it follows
that in certain situations, the NLSP\ could potentially decay just before the
start of, or even during BBN!

At the most conservative level, the usual results of the standard
BBN\ cosmology can typically be retained if the NLSP decays prior to the start
of BBN. Returning to equation (\ref{NLSPLIFE}), decreasing $m_{3/2}$ or
increasing $m_{NLSP}$ will both decrease the value of $\tau_{NLSP}$. In
particular, the fact that the fifth power of $m_{NLSP}$ appears in
$\tau_{NLSP}$ implies that even very mild adjustments in this value can
significantly decrease the lifetime of the NLSP.

Assuming that the NLSP decays during BBN, its decay products could potentially
jeopardize the production of the light element abundances, or could bring the
abundance of light elements such as $^{7}Li$ into \textit{better} accord with
observation. The precise effect of the NLSP depends on whether it decays to a
photon and gravitino as for a bino-like NLSP; or whether it decays to a tau
and gravitino, as in the case of a stau NLSP. The impact from a late decaying
NLSP depends on details of a particular model, such as the overall abundance of the
NLSP\ prior to the start of BBN. Nevertheless, under reasonable assumptions
such that the relative abundance of the NLSP to the overall baryon density is
not distorted by dilution effects, it is possible to estimate the production
of the light elements due to BBN.

In the case of a bino-like NLSP, the analysis reflected in figure 7 of
\cite{Kawasaki:2008qe} indicates that when the gravitino has a mass of
$10-100$ MeV, the main predictions of BBN\ remain intact. At larger values of
the gravitino mass in the range of $\gtrsim1$ GeV, only somewhat large values
of the bino mass remain in accord with BBN. Similarly, in the case of a stau
NLSP, figure 12 of \cite{Kawasaki:2008qe} illustrates that the same range for
the gravitino mass remains in accord with BBN with similar constraints for the
mass of the stau in the range of larger values of the gravitino mass.

Interestingly, the very recent analysis of \cite{Bailly:2008yy} also indicates
that a gravitino in this same mass range appears to also decrease the overall
abundance of $^{7}Li$, bringing the resulting abundance into better agreement
with observation. Although a complete analysis of BBN\ in the context of
F-theory GUTs is beyond the scope of the present paper, quite auspiciously,
the natural range of parameters of F-theory GUTs suggests a gravitino mass in
the range of $10-100$ MeV!

\section{F-theory GUTs and the Axion Supermultiplet\label{FGUTAXION}}

In this section we study the axion supermultiplet in F-theory GUT\ models.
\ To this end, we first briefly review the main features of F-theory GUTs
\cite{BHVI,BHVII,HVGMSB}.\ This is followed by a discussion of the axion
supermultiplet, and in particular, the interactions of the saxion with the MSSM.

In F-theory GUT\ models, the singularity type of an elliptically fibered
Calabi-Yau fourfold with section over subspaces of the threefold base of
complex codimension one, two and three respectively determine the resulting
gauge symmetry, matter content and interactions terms of the low energy
effective theory. \ Singularity type enhancements along complex surfaces are
interpreted as seven-branes with $ADE$\ gauge group. \ These seven-branes then
intersect over Riemann surfaces or \textquotedblleft matter
curves\textquotedblright\ where the singularity type enhances by at least one
rank. \ Finally, this enhancement can increase further at points of the
compactification, corresponding to the intersection of at least three
seven-branes at a single point.

Although perhaps contrary to previous experience with string
compactifications, local F-theory GUTs provide a somewhat rigid framework for
string based model building. \ In these models, the existence of a limit where
gravity can in principle decouple requires that the GUT\ group seven-brane
must wrap a del Pezzo surface. \ Moreover, breaking the GUT\ group to the
Standard Model gauge group is remarkably constrained in such models
\cite{BHVII,DonagiWijnholtBreak}.

Next consider the supersymmetry breaking sector of F-theory GUTs.
For the purposes of this paper we shall assume that the primary
features of the deformation of a minimal gauge mediation scenario
developed in \cite{HVGMSB} are satisfied. The existence of a gauge
mediation scenario assumes that most moduli present are stabilized
due to high scale supersymmetric dynamics. Much as in \cite{HVGMSB}, our attitude will be that phenomenological constraints on
the particle physics content of this class of models should be viewed as imposing interesting restrictions on possible global completions which satisfy these conditions.

We now review some further details of the supersymmetry breaking sector discussed in \cite{HVGMSB}.
Consistent electroweak symmetry breaking requires that the
parameter $\mu$ of the MSSM\ superpotential:%
\begin{equation}
L_{MSSM}\supset\int d^{2}\theta\mu\cdot H_{u}H_{d}%
\end{equation}
must not be significantly different from the weak scale. If supersymmetry
breaking indeed stabilizes the hierarchy between the weak scale and the
GUT\ scale, this naturally suggests that the value of $\mu$ should be
correlated with the scale of supersymmetry breaking. The effects of
supersymmetry breaking can be parameterized in terms of the vev of a
GUT\ group singlet chiral superfield $X$ such that:%
\begin{equation}
\left\langle X\right\rangle =x+\theta^{2}F\text{.} \label{XVEV}%
\end{equation}
In \cite{HVGMSB}, some explicit solutions to the $\mu$ problem were obtained under the
assumption that $X$ localizes on a matter curve which intersects the
GUT\ seven-brane at a point. \ Integrating out the Kaluza-Klein modes on the
$X$ field curve generates a higher dimension operator in the effective theory
of the form:%
\begin{equation}
L_{MSSM}\supset\gamma\cdot\int d^{4}\theta\frac{X^{\dag}H_{u}H_{d}}{M_{X}}
\label{GiudiceMasiero}%
\end{equation}
where, as estimated in \cite{BHVII}, $M_{X}\sim10^{15.5\pm0.5}$ is the
Kaluza-Klein scale associated with the curve supporting the $X$ field and
$\gamma\sim O(10)$. In order to generate the correct value of the $\mu$ term,
$F$ must attain the value $\sim10^{17\pm0.5}$ GeV$^{2}$.\ As explained in
\cite{HVGMSB}, this value is naturally attained via instanton effects
associated with Euclidean three-branes wrapping other complex surfaces of the geometry.

Following the model of \cite{HVGMSB}, the supersymmetry breaking sector is closely
tied to the seven-brane associated with the $U(1)_{PQ}$ gauge symmetry. \ This
gauge theory is anomalous, and as such, instanton contributions can generate
contributions to the superpotential of the form:%
\begin{equation}
W_{inst}\supset M_{PQ}^{2}\cdot q\cdot X
\end{equation}
where:%
\begin{equation}
q \sim e^{-Vol_{PQ}}\text{.}%
\end{equation}
Assuming a fixed value for $q$, this determines the F-term component of $X$. As explained in \cite{HVGMSB}, to properly
analyze the PQ symmetry breaking sector, it is necessary to treat both $q$ and $X$ as dynamical fields. In that context, it was shown that an appropriate tuning in the K\"ahler potential for $q$ and the flux-induced FI parameter for the PQ gauge theory is compatible with stabilizing the vev of $X$ at the scale $10^{12}$ GeV. The Goldstone mode associated with the breaking of the accidental global $U(1)_{PQ}$ symmetry is parameterized by the phase of the gauge invariant operator $q \cdot X$. In particular, the axion is therefore primarily given by the phase of $X$, with a small contribution from $q$. The corresponding axion decay constant is then given as:
\begin{equation}
f_{a}=\sqrt{2}x\sim10^{12}\text{ GeV.}%
\end{equation}

To a large extent, the allowed mediation mechanism which communicates the
effects of supersymmetry breaking to the visible sector is dictated by the
fact that $F/M_{PL}$ is far below the weak scale so that Planck suppressed
operators, and therefore gravity mediated supersymmetry breaking cannot
generate viable soft mass terms. \ By contrast, geometric realizations of
minimal gauge mediation scenarios are far more viable in this scenario, and
explicit models based on minimal gauge mediation have been discussed in
\cite{HVGMSB} (see also \cite{MarsanoGMSB}). In minimal gauge
mediation (MGMSB), the soft mass terms are completely fixed by the gauge couplings of
the MSSM\ and the ratio $F_{X}/x\equiv\Lambda$ as:%
\begin{equation}
m_{soft}\sim\frac{\alpha}{4\pi}\frac{F}{x}=\frac{\alpha}{4\pi}\cdot\Lambda
\end{equation}
up to numerical factors associated with the representation content of a given
field. Here, $\alpha$ is shorthand for the fine structure constants of the
various gauge groups of the Standard Model under which a given superfield may
be charged.

As explained in \cite{HVGMSB}, the F-theory GUT\ actually corresponds to a
deformation of minimal gauge mediation.\ Indeed, precisely because the $X$
field localizes on a matter curve, it will be charged under additional
seven-branes of the compactification. \ These seven-branes endow the low
energy effective theory with additional, generically anomalous $U(1)$ gauge
group factors. \ The generalized Green-Schwarz mechanism cancels this anomaly,
but the presence of the requisite coupling of the gauge field to an axion-like
field in the four-dimensional effective theory generates a large mass for the
gauge boson via the St\"{u}ckelberg mechanism. \ Below the mass scale of this
gauge boson, the theory will therefore retain an anomalous global $U(1)$
symmetry, which in appropriate circumstances can be identified with a $U(1)$
Peccei-Quinn symmetry which we denote as $U(1)_{PQ}$.

The fields of the MSSM\ are charged under $U(1)_{PQ}$ with charges $-2$, $+1$
and $-4$ for the respective Higgs fields, chiral matter and $X$ field
of the F-theory GUT\ model. \ Integrating out the heavy $U(1)_{PQ}$ gauge
fields also generates higher dimension operators in the low energy effective
theory of the form \cite{ArkaniHamedANOM}:%
\begin{equation}
L_{eff}\supset-\frac{4\pi\alpha_{PQ}}{M_{U(1)_{PQ}}^{2}}%
e_{X}e_{\Phi}\cdot\int d^{4}\theta X^{\dag}X\Phi^{\dag}\Phi
\end{equation}
where $M_{U(1)_{PQ}}$ denotes the mass of the heavy $U(1)_{PQ}$ gauge boson
and the $e$'s denote the respective charges of $X$ and MSSM field $\Phi$ under
$U(1)_{PQ}$, and $\alpha_{PQ}$\ is the fine structure constant of this gauge
theory. Once $X$ develops a vev as in equation (\ref{XVEV}), this generates an
additional contribution to the masses squared of the scalar component of
$\Phi$ at the messenger scale:%
\begin{equation}
m_{\Phi,soft}^{2}=m_{\Phi,MGMSB}^{2}+\frac{e_{\Phi}\cdot e_{X}}{\left\vert
e_{X}\right\vert }\cdot\Delta_{PQ}^{2} \label{PQDEF}%
\end{equation}
where we have introduced the PQ\ deformation parameter:%
\begin{equation}
\Delta_{PQ}^{2}\equiv4\pi\alpha_{PQ}\left\vert e_{X}\right\vert \left\vert
\frac{F}{M_{U(1)_{PQ}}}\right\vert ^{2}\text{.}%
\end{equation}
In terms of the anomalous $U(1)_{PQ}$ gauge theory, this contribution can be
interpreted as a supersymmetry breaking D-term. Insofar as the PQ\ deformation
corrects the soft mass terms of the MSSM, the value of $\Delta_{PQ}$ from
prior considerations is on the order of $\sim100$ GeV. As we show later, this
estimate is borne out by the cosmology of the F-theory GUT\ scenario.

By inspection of equation (\ref{PQDEF}), the PQ\ deformation decreases the
soft masses squared when $e_{\Phi}\cdot e_{X}<0$. \ As a consequence, there is
a limit to the size of $\Delta_{PQ}$ before the PQ deformation induces a
tachyon in the squark/slepton sector of the MSSM. \ The precise value of this
vev depends on the value of $\Lambda$. \ For example, in a model with a single
vector-like pair of messenger fields in the $5\oplus\overline{5}$ of $SU(5)$,
the minimal value of $\Lambda$ consistent with the Higgs mass bound $m_{h^{0}%
}\geq114.5$ GeV is $\Lambda\sim1.3\times10^{5}$ GeV, and the maximal value of
$\Delta_{PQ}$ allowed before the mass squared of the lightest stau becomes
tachyonic is $\Delta_{PQ}\sim290$ GeV. \ For vanishing PQ deformation, a
bino-like lightest neutralino is the NLSP. \ On the other hand, for large
PQ\ deformation, the NLSP\ can instead correspond to the lightest stau. \ The
PQ\ deformation also plays a significant role in the dynamics of the $X$ field
which we analyze in detail in subsection \ref{AXIONInteractions}. \ This is
particularly relevant for cosmological considerations because as we explain in
subsection \ref{AXMULT}, the axion and gravitino are both closely tied to the
dynamics of $X$ which is in turn controlled by the value of $\Delta_{PQ}$.

\subsection{Axion Supermultiplet \label{AXMULT}}

In the axion solution to the strong CP problem, an anomalous global
$U(1)_{PQ}$ symmetry is spontaneously broken at an energy scale $f_{a}$. \ The
associated Goldstone mode then corresponds to the axion field, which we denote
by $a$. \ In a supersymmetric theory, $a$ fits into a complete supermultiplet
given by one additional real bosonic degree of freedom $s$, a fermionic
component $\psi_{\bot}$, and an auxiliary field $F_{\bot}$, which we assemble
into the chiral superfield:%
\begin{equation}
\mathcal{A}=a+is+\sqrt{2}\theta\psi_{\bot}+\theta^{2}F_{\bot}\text{.}%
\end{equation}
The field $s$ corresponds to the \textquotedblleft saxion\textquotedblright%
\ and the field $\psi_{\bot}$ corresponds to the \textquotedblleft
axino\textquotedblright. \ By definition, $\mathcal{A}$ couples to the QCD
superfield strength through the coupling:%
\begin{equation}
L\supset\operatorname{Re}\int d^{2}\theta\frac{\mathcal{A}}{16\pi^{2}%
}Tr_{SU(3)}W^{\alpha}W_{\alpha}\text{.}%
\end{equation}

In this section we review and slightly extend the analysis of \cite{HVGMSB} by
showing that the components of $\mathcal{A}$ are to leading order given by the
components of the chiral superfield $X$. In addition to its role in PQ
symmetry breaking, the $X$ field also plays a key role in supersymmetry
breaking. \ In the field theory limit, this leads to an exactly massless
Goldstino mode in the low energy theory. \ Precisely because $X$ is the
primary source of supersymmetry breaking, a linear combination given
predominantly by the fermionic component of the $X$ superfield with smaller
contributions from other fermionic modes corresponds to the Goldstino. \ Away
from the strict field theory limit, the Goldstino is eaten by the gravitino
via the super-Higgs mechanism. \ To leading order, the axino therefore
corresponds to the longitudinal components of the gravitino.

\subsubsection{$U(1)_{PQ}$ Goldstone Mode\label{Goldstone}}

We now describe the Goldstone mode associated with the breaking of the
anomalous $U(1)_{PQ}$ symmetry. \ As explained in \cite{HVGMSB}, to leading order, $\arg X$
corresponds to the axion field. \ Strictly speaking, however, this
identification is not completely accurate because the axion superfield
corresponds to a linear combination of $X$ with subleading contributions from
other chiral multiplets. The reason for this can be traced back to the way in which the PQ gauge boson develops a mass.

In general terms, the anomalous $U(1)_{PQ}$ gauge theory consists of $n$ chiral
superfields $X_{i}$ with charges $q_{i}$ such that:%
\begin{equation}
Q\equiv\underset{i=1}{\overset{n}{\sum}}q_{i}\neq0\text{.}%
\end{equation}
The anomaly for the corresponding $U(1)$ gauge boson is then canceled via the
Green-Schwarz mechanism. This corresponds to introducing an additional
axion-like superfield $\mathcal{C}$ such that the chiral superfield
$e^{\mathcal{C}}$ has charge $-Q$. \ $\mathcal{C}$ couples to the PQ vector
multiplet via the D-term:%
\begin{equation}
L\supset\int d^{4}\theta K(\mathcal{C}+\mathcal{C}^{\dag}-Q\cdot
V_{PQ})\text{,}%
\end{equation}
where here, $K$ denotes an appropriate K\"{a}hler potential. The corresponding
gauge field develops a mass via the St\"{u}ckelberg mechanism, leaving a
nearly exact global symmetry in the low energy effective theory. \ Once some
combination of the $X_{i}$'s develop a vev, this global $U(1)_{PQ}$ will be
broken, leaving behind a Goldstone mode.

The axion supermultiplet is given by a linear combination of $\mathcal{C}$ and
the associated \textquotedblleft phases\textquotedblright\ of the $X_{i}$'s. Another linear combination of these fields is
eaten by the vector multiplet. Assuming a canonical normalization for all of the $X_i$'s, the direction in field space fixed by the
D-term potential of the PQ seven-brane gauge theory is:
\begin{equation}
\underset{i=1}{\overset{n}{\sum}}q_{i} |X_i|^2 - Q K^{\prime} = \xi_{flux} \text{,}%
\end{equation}
where here $\xi_{flux}$ denotes a flux induced FI parameter. The directions unfixed by the D-term potential are conveniently parameterized in terms of chiral superfields $\Theta_i$ defined as:
\begin{equation}
X_j = |x_j| \exp(i \Theta_j)
\end{equation}
The linear combinations of $\Theta_j$ neutral under $U(1)_{PQ}$ correspond to possible flat directions not fixed by the D-term potential. For example,
in the context of the Fayet-Polonyi model, the $U(1)_{PQ}$ invariant combination:
\begin{equation}
\widehat{X} \equiv q \cdot X
\end{equation}
develops a supersymmetry and global PQ symmetry breaking vev. This vev is stabilized by contributions to the K\"ahler potential for
$X$ and $q$ \cite{HVGMSB}. The bosonic component of $\widehat{\Theta}$ is
given by:%
\begin{equation}
\widehat{\Theta}=a+is+...
\end{equation}
where $a$ is the axion and $s$ is the saxion. The mode $a$ takes values in the
interval $-\pi<a<\pi$.

\subsubsection{Supersymmetry Breaking and the Goldstino \label{Goldstino}}

In the above analysis, we assumed that supersymmetry was unbroken. \ At energy
scales below $x\sim10^{12}$ GeV, supersymmetry is broken via the Fayet-Polonyi
model described in \cite{HVGMSB}. \ In the strict field theory limit,
spontaneous supersymmetry breaking generates a massless fermion corresponding
to the Goldstino mode. \ The explicit form of the Goldstino mode depends on
the details of F- and D-term breaking of a particular model. \ Nevertheless,
certain features of the Goldstino mode and how it couples to fields of a
particular model are universal, and our analysis will for the most part only
rely on these well known features. \ As reviewed, for example in
\cite{MartinPrimer}, the explicit form of the Goldstino mode is given as a
linear combination of the fermions $\lambda_{a}$ and $\chi_{i}$ respectively
from the vector and chiral multiplets:%
\begin{equation}
\widetilde{G}=\frac{iD^{a}}{\sqrt{2}}\lambda_{a}+F^{i}\chi_{i}%
\end{equation}
where the effects of supersymmetry breaking are encoded in non-zero vevs for
some subset of the $D^{a}$ and $F^{i}$. \ Taking into account the fact that
$M_{PL}$ is not infinite, spontaneous supersymmetry breaking implies that the
Goldstino mode is eaten by the gravitino. \ The mass of the gravitino is:
\begin{equation}
m_{3/2}^{2}=\frac{1}{3M_{PL}^{2}}\left(  \underset{i}{%
{\displaystyle\sum}
}\left\vert F^{i}\right\vert ^{2}+\frac{1}{2}\underset{a}{%
{\displaystyle\sum}
}\left\vert D^{a}\right\vert ^{2}\right)  \text{.}%
\end{equation}

In the explicit Fayet-Polonyi model presented in \cite{HVGMSB}, there will in
general be contributions to both the F- and D-term components of the
gravitino. \ In addition to contributions to the D-term potential, the
Fayet-Polonyi model contains an instanton induced contribution to the
superpotential of the form:%

\begin{equation}
W=M_{PQ}^{2}e^{\mathcal{C}}X
\end{equation}
where the mass scale $M_{PQ}\sim M_{GUT}$ which as explained in \cite{HVGMSB}
can in general be different from the mass of the PQ\ gauge boson. \ For this
range of mass scales, it turns out that the vev of $e^{\mathcal{C}}$ can
naturally attain the value $\sim10^{-17}$, as required to achieve $F%
\sim10^{17}$ GeV$^{2}$. \ With the identification $e^{C} \propto X_{n+1}$,
we can alternatively view this Fayet-Polonyi model as a Fayet model of supersymmetry
breaking with superpotential:%
\begin{equation}
W=mX_{1}X_{n+1}\text{.}%
\end{equation}
As explained in \cite{HVGMSB}, the full sector requires a non-trivial K\"ahler potential for $X_1$ and $X_{n+1}$. The F-term components of the various superfields are therefore determined as:%
\begin{align}
\overline{F}_{1}  &  =-mx_{n+1}\sim10^{17}\text{ GeV}^{2}\\
\overline{F}_{n+1}  &  =-mx_{1}=-mx_{n+1}\frac{x_{1}}{x_{n+1}}\sim
10^{13}\text{ GeV}^{2}\text{,}%
\end{align}
where in the above we have plugged in rough representative values of
$x_{n+1}\sim M_{U(1)_{PQ}}\sim M_{GUT}$ and $x_{1}\sim10^{12}$ GeV consistent
with the estimates obtained in \cite{HVGMSB}.

In addition to these F-term contributions to supersymmetry breaking, as
explained in \cite{ArkaniHamedANOM}, we should also expect a contribution from
D-term breaking, which in the present class of models is given as:%
\begin{equation}
D=-\frac{4\pi\alpha_{PQ}}{M_{U(1)_{PQ}}^{2}}\underset{i}{\sum}q_{i}\left\vert
F_{i}\right\vert ^{2}\simeq-\Delta_{PQ}^{2}\text{,}%
\end{equation}
which is far smaller than the F-term breaking components.

To summarize the discussion presented above, the gravitino is predominantly
given by the fermionic component of the $X$ field. \ Nevertheless, for certain
purposes, it is important that the gravitino corresponds to a linear
combination of the Goldstino which contains additional contributions from the
fermionic components of $e^{\mathcal{C}}$. \ For example, in the low energy
effective theory for the $X$ field, integrating out the heavy $U(1)_{PQ}$
gauge boson generates the higher dimension operator:%
\begin{equation}
L_{X}\supset-\frac{4\pi\alpha_{PQ}}{M_{U(1)}^{2}}\int d^{4}\theta X^{\dag
}XX^{\dag}X\text{.}%
\end{equation}
Once $x$ and $F_{X}$ develop non-zero values, this would appear to induce a
mass term for the gravitino of order $\left\vert x\Delta_{PQ}^{2}%
/F_{X}\right\vert $. \ In a more complete analysis, the fermionic mode
$\psi_{X}$ mixes with the fermionic components associated with $e^{\mathcal{C}%
}$. \ That this must be the case follows from the fact that in the limit
$M_{PL}\rightarrow\infty$, the gravitino is exactly massless. Indeed, a naive
analysis suggests that the fermionic component $\psi_{\mathcal{C}}$ has a mass term
of order $\left\vert x_{n+1} \Delta_{PQ}^{2}/F_{X}\right\vert $ which
is significantly larger. \ For most purposes,
however, this distinction will not play any crucial role in many of the
cosmological considerations discussed here.

\subsection{Axion Supermultiplet Interaction Terms \label{AXIONInteractions}}

\ For the purposes of cosmological considerations, it is also important to
determine the couplings between the axion supermultiplet and the matter
content of the rest of the F-theory GUT\ model. \ Indeed, these interaction
terms determine the lifetime of the saxion, which can have important
consequences if it is sufficiently long-lived. \ In particular, the decay of
the saxion can reheat the universe. \ The specific details of this reheating
depends on the dominant mode of decay, as well as the primary decay rates.

As derived in\cite{HVGMSB}, the effective action for the $X$ field in 4d
$\mathcal{N}=1$ superspace contains the terms:\footnote{It is important to note that this effective action is only
valid a temperatures far below the messenger scale. Indeed, in the context of models where the dilaton directly couples to
the QCD field strength, high temperature effects can potentially destabilize the value of the dilaton. Here, this problem
is avoided because the actual coupling is only present in the low energy effective theory. See \cite{Buchmuller:2003is,Buchmuller:2004xr} for further details on destabilization of the dilaton in other contexts. Nevertheless, it would be interesting to study in greater detail the high temperature phase of this system.}
\begin{align}
L_{X}  &  =\int d^{4}\theta X^{\dag}X^{\dag}+\operatorname{Re}\int d^{2}%
\theta\frac{C_{W}\log X}{32\pi}TrW^{\alpha}W_{\alpha}-\int d^{4}\theta
C_{\Phi}\left(  \log\left\vert X\right\vert ^{2}\right)  ^{2}\Phi^{\dag}\Phi\\
&  -\frac{4\pi\alpha_{PQ}e_{\Phi}e_{X}}{M_{U(1)_{PQ}}^{2}}\int
d^{4}\theta\Phi^{\dag}\Phi X^{\dag}X - m_{sax}^2 |X - \langle X \rangle|^2
\end{align}
where in the above, $\Phi$ is shorthand for any MSSM\ chiral superfield,
$W^{\alpha}$ denotes any gauge superfield strength, the multiplicative factors
$C_{W}$ and $C_{\Phi}$ are determined by the gauge couplings constants of the
MSSM, as well as the quadratic Casimirs for the representations of the
messenger fields, and $e_{\Phi}$ and $e_{X}$ denote the $U(1)_{PQ}$ charges of
$\Phi$ and $X$. Here, the specific form of the saxion mass squared term is fixed by details
of the K\"ahler potential for $X$, and the axion-like field
$\mathcal{C}$ which enters the Green-Schwarz mechanism \cite{HVGMSB}. The explicit
coupling of the axion supermultiplet is then given by performing the substitution:%
\begin{equation}
X\rightarrow xe^{i\mathcal{A}}+\theta^{2}F \label{XSUB}%
\end{equation}
in the Lagrangian density $L_{X}$.
Although the PQ deformation dependence of the saxion mass is subject to order one
tunings depending on the K\"ahler potentials of $X$ and $\mathcal{C}$ \cite{HVGMSB}, to make our discussion more concrete, we shall estimate the resulting
mass using the higher dimension operator:
\begin{equation}
-\frac{4\pi\alpha_{PQ}e_{X}e_{X}}{M_{U(1)_{PQ}}^{2}}\int
d^{4}\theta X^{\dag}X X^{\dag}X.
\end{equation}
Substituting in our expression from equation (\ref{XSUB}), we shall take as a rough estimate the saxion mass squared as:%
\begin{equation}
m_{sax}^{2}=4\left\vert e_{X}\right\vert \Delta_{PQ}^{2}%
\end{equation}
or:%
\begin{equation}
m_{sax}=4\Delta_{PQ}, \label{msax}%
\end{equation}
where in the above we have used the fact that $\left\vert e_{X}\right\vert =4$.

In the present context, however, the most important interactions are those
which determine the decay modes of the saxion. The dominant interaction terms
originate from the self-interactions of the $X$ field chiral supermultiplet,
and the soft mass terms associated with gauge mediation effects. Expanding the kinetic term $\left\vert \partial_{\mu}X\right\vert ^{2}$ in
terms of the axion multiplet yields the model independent coupling between the
axion and saxion:%
\begin{equation}
L_{X}\supset\left(  f_{a}^{2}+f_{a}\cdot s\right)  \left(  \frac{1}{2}\left(
\partial_{\mu}a\right)  ^{2}+\frac{1}{2}\left(  \partial_{\mu}s\right)
^{2}\right)  \text{.}%
\end{equation}
In particular, the decay rate $s\rightarrow aa$ is given by:%
\begin{equation}
\Gamma_{s\rightarrow aa}\sim\frac{1}{64\pi}\frac{m_{sax}^{3}}{f_{a}^{2}}%
=\frac{1}{\pi}\frac{\Delta_{PQ}^{3}}{f_{a}^{2}}\text{.} \label{GAA}%
\end{equation}
Further details on universal couplings of the axion supermultiplet can be
found in \cite{Chun:1995hc}.

\begin{figure}
[ptb]
\begin{center}
\includegraphics[
height=3.7369in,
width=6.3875in
]%
{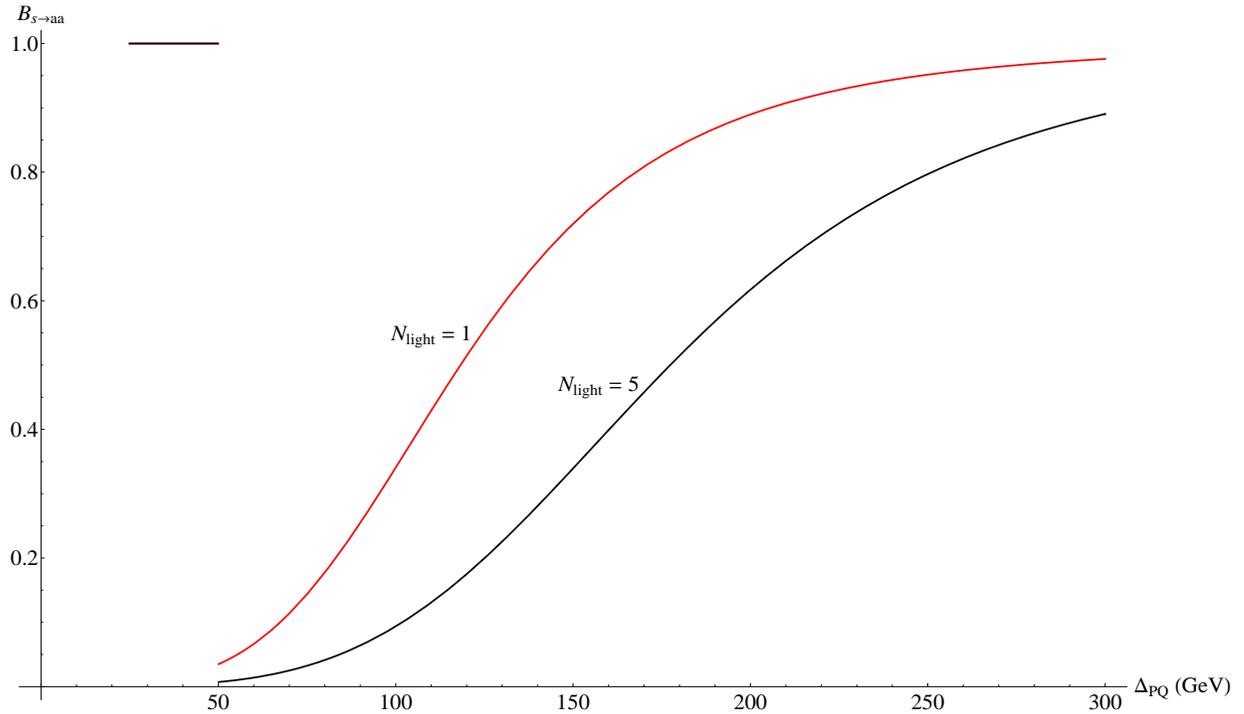}%
\caption{Plot of the \textquotedblleft toy model\textquotedblright\ branching
ratio of the saxion to axions and one (red) to five (black) species of
MSSM\ fields as a function of $\Delta_{PQ}$ for fixed values of $f_{a}%
=10^{12}$ GeV and $m_{soft}=200$ GeV. \ For the decay to a representative
MSSM\ field, we have used the crude estimate provided by equation
(\ref{GMSSM}). This situation is somewhat idealized, because as $\Delta_{PQ}$
increases, the number of decay channels will increase, decreasing the overall
branching ratio to axions.}%
\label{brtwohund}%
\end{center}
\end{figure}

\begin{figure}
[ptb]
\begin{center}
\includegraphics[
height=3.7369in,
width=6.3875in
]%
{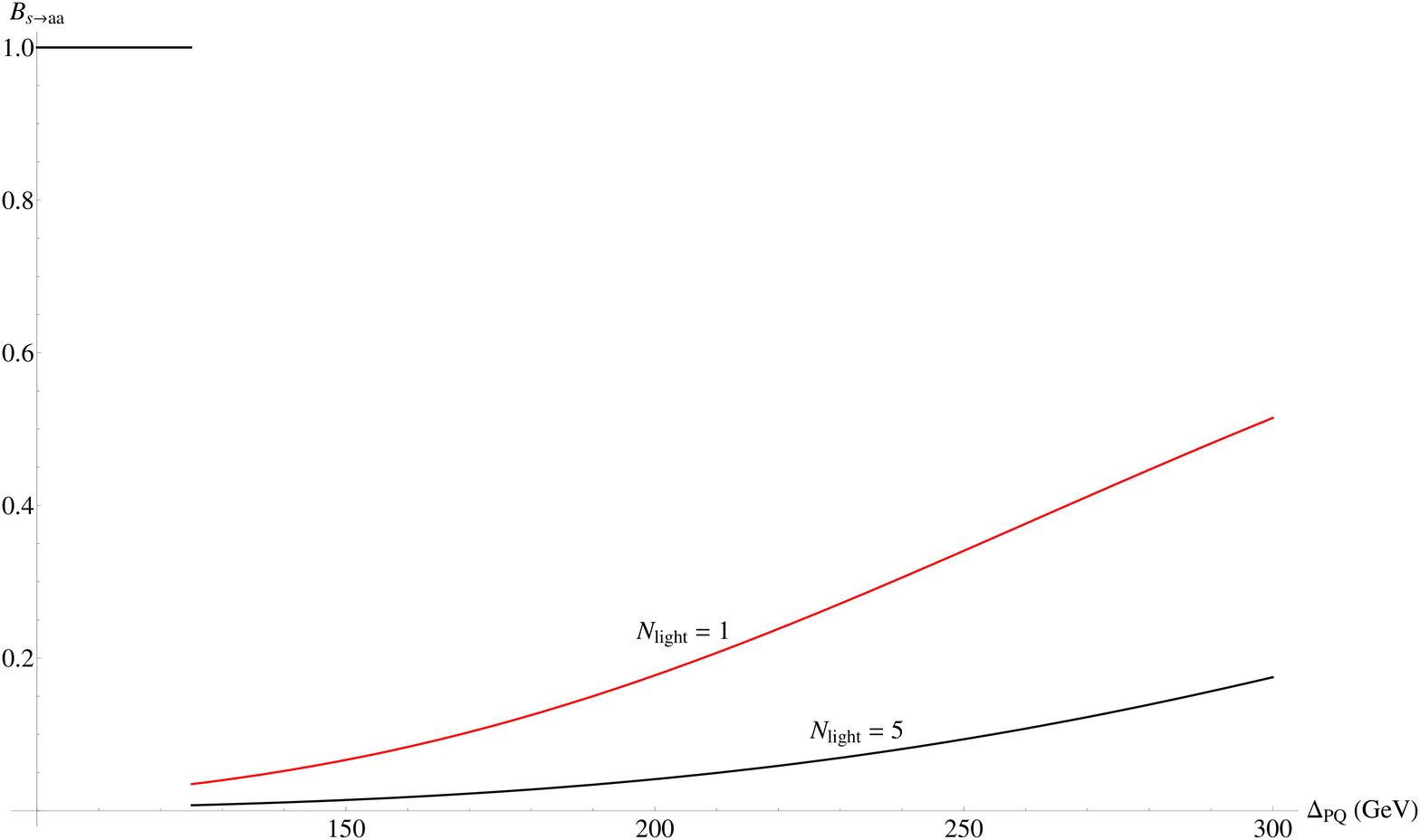}%
\caption{Plot of the branching ratio of the saxion to axions and one (red) to
five (black) species of MSSM\ fields as a function of $\Delta_{PQ}$ for fixed
values of $f_{a}=10^{12}$ GeV and $m_{soft}=500$ GeV. \ For the decay to a
representative MSSM\ field, we have used the crude estimate provided by
equation (\ref{GMSSM}). This situation is somewhat idealized, because as
$\Delta_{PQ}$ increases, the number of decay channels will increase,
decreasing the overall branching ratio to axions.}%
\label{brfivehund}%
\end{center}
\end{figure}

Gauge mediation generates soft mass terms for the scalars and gauginos of the
MSSM, and also induces additional interaction terms for the saxion. The
essential point is that the soft mass terms for a gaugino, $\lambda$, and
scalar, $\phi$ of the MSSM\ are given as:%
\begin{equation}
L_{soft}\supset\frac{1}{2}m_{\lambda,soft}(x)\lambda\lambda+h.c.-m_{\phi
,soft}^{2}\left(  x,\overline{x}\right)  \left\vert \phi\right\vert ^{2}%
\end{equation}
where in the above,
\begin{align}
m_{\lambda,soft}(x)  &  \propto\frac{F}{x}\\
m^{2}_{\phi,soft}(x,\overline{x})  &  \propto\left\vert \frac{F}{x}\right\vert ^{2}\text{.}%
\end{align}
As noted, for example, in \cite{Ibe:2006rc}, the explicit $x$ dependence of
$m_{soft}$ implies that in both cases, $X$, and therefore the saxion directly
couples to $\lambda$ and $\phi$. \ Performing the substitution $x\mapsto x+X$,
it now follows that $X$ couples to these fields as:%
\begin{equation}
L_{soft}\supset-\frac{1}{2}m_{\lambda,soft}(x)\frac{X}{x}\lambda
\lambda+m_{\phi,soft}^{2}\left(  x,\overline{x}\right)  \frac{X}{x}\left\vert
\phi\right\vert ^{2}+h.c.
\end{equation}
Assuming that a given decay is kinematically allowed, the decay rate of the
saxion to MSSM particles is therefore:%
\begin{equation}
\Gamma_{s\rightarrow MSSM}\sim\frac{1}{2\pi m_{sax}}\left(  \frac{m_{soft}%
^{2}}{f_{a}}\right)  ^{2}=\frac{1}{8\pi\Delta_{PQ}}\left(  \frac{m_{soft}^{2}%
}{f_{a}}\right)  ^{2}\text{,} \label{GMSSM}%
\end{equation}
which is to be compared with the decay $s\rightarrow aa$. The ratio
of these decay rates is:
\begin{equation}
\frac{\Gamma_{s\rightarrow aa}}{\Gamma_{s\rightarrow MSSM}}\sim8\left(
\frac{\Delta_{PQ}}{m_{soft}}\right)  ^{4}\text{.}%
\end{equation}
Although a sharp upper bound is somewhat flexible, in order to avoid a
tachyonic mass, it is necessary to assume, for example, that $\Delta
_{PQ}\lesssim m_{soft}\sim100-1000$ GeV. Introducing the branching ratio to
axions%
\begin{equation}
B_{s\rightarrow aa}=\frac{\Gamma_{s\rightarrow aa}}{\Gamma_{sax}}\text{,}%
\end{equation}
where $\Gamma_{sax}$ denotes the total decay rate, we conclude that for
typical values of $\Delta_{PQ}$ and $m_{soft}$, $B_{s\rightarrow aa}$ is in
the range:%
\begin{equation}
10^{-3}<B_{s\rightarrow aa}<10^{-1}\text{.}%
\end{equation}
See figures \ref{brtwohund} and \ref{brfivehund} for plots of the
\textquotedblleft toy model\textquotedblright\ branching ratios $\Gamma
_{s\rightarrow aa}/(N_{light}\Gamma_{s\rightarrow MSSM}+\Gamma_{s\rightarrow
aa})$ as functions of $\Delta_{PQ}$ with $f_{a}$ and $m_{soft}$ fixed to
representative values. \ Here, $N_{light}$ denotes the number of light
MSSM\ species which are sufficiently light to allow a decay channel to an
MSSM\ particle. In realistic models, $N_{light}$ is a non-trivial function of
$\Delta_{PQ}$, although for illustrative purposes, we take it fixed in the
figures.%

When a decay to an MSSM\ particle is kinematically allowed, the branching
fraction to MSSM\ particles will most likely dominate over decays to the
axion. \ This is an important constraint in the context of decays to
relativistic species, such as the axion. \ In a representative case, we can
expect $\Delta_{PQ}\sim100$ GeV and $m_{soft}\sim10^{2.5}$ GeV, in which case:%
\begin{equation}
\frac{\Gamma_{s\rightarrow aa}}{\Gamma_{s\rightarrow MSSM}}\sim10^{-1}\text{.}%
\end{equation}

Restricting to saxion decays to MSSM\ particles, we note that the primary
decay channel through MSSM scalars proceed via the Higgs bosons and
potentially the lightest stau $\widetilde{\tau}_{1}$ for sufficiently large
PQ\ deformation. \ The saxion can also decay to MSSM\ fermions. \ Note,
however, that in the supersymmetric limit, the axion supermultiplet coupling
to the MSSM\ is far weaker. \ Because of this, our expectation is that the
saxion couples more strongly to the fermionic superpartners in comparison to
the fermions of the Standard Model. \ In the context of the MSSM, the gluinos
are nearly $1000$ GeV in mass. \ The primary decay channel through
MSSM\ fermions therefore proceeds via decays to the two lightest neutralinos
$\widetilde{\chi}_{0}^{1}$ and $\widetilde{\chi}_{0}^{2}$, which are primarily
composed of the bino and wino. \ The respective masses of these particles in
the present class of models are $m_{\widetilde{\chi}_{0}^{1}}\sim170$ GeV and
$m_{\widetilde{\chi}_{0}^{2}}\sim350$ GeV. \ In all cases, the relative
branching fractions are roughly determined by the relative masses of two
species $i$ and $j$ to be:%
\begin{equation}
\frac{\Gamma_{s\rightarrow xx}}{\Gamma_{s\rightarrow yy}}\sim\left\vert
\frac{m_{soft}^{x}}{m_{soft}^{y}}\right\vert ^{4}\text{.} \label{SaxRelRates}%
\end{equation}
Here it is important to note that the soft mass $m_{soft}$ may in general
differ from the actual mass of the particle. \ For example, whereas the Higgs
field has mass $m_{h^{o}}\sim115$ GeV, the soft mass term which enters into
the saxion decay rate formula is more accurately approximated by $m_{H_{u}%
}^{(0)}\sim600$ GeV \cite{HVGMSB}.

The saxion also couples to the gravitino. The model independent decay rate
$s\rightarrow\psi_{3/2}\psi_{3/2}$ is given by:
\begin{equation}
\Gamma_{s\rightarrow\psi_{3/2}\psi_{3/2}}\sim\frac{1}{96\pi}\frac{m_{sax}^{3}%
}{M_{PL}^{2}}\left(  \frac{m_{sax}}{m_{3/2}}\right)  ^{2}\text{.}%
\end{equation}
The relative branching fraction to gravitinos versus axions is therefore given
as:
\begin{align}
B_{s\rightarrow\psi_{3/2}\psi_{3/2}}  &  =\frac{\Gamma_{s\rightarrow\psi
_{3/2}\psi_{3/2}}}{\Gamma_{sax}}=B_{s\rightarrow aa}\frac{\Gamma
_{s\rightarrow\psi_{3/2}\psi_{3/2}}}{\Gamma_{s\rightarrow aa}}=\frac{2}%
{3}B_{s\rightarrow aa}\left(  \frac{f_{a}}{M_{PL}}\right)  ^{2}\left(
\frac{m_{sax}}{m_{3/2}}\right)  ^{2}\label{BSPSIPSI}\\
&  \sim2\times10^{-4}\cdot B_{s\rightarrow aa}\left(  \frac{\Delta_{PQ}%
}{100\text{ GeV}}\right)  ^{2}\left(  \frac{10\text{ MeV}}{m_{3/2}}\right)
^{2}\text{.}%
\end{align}

The decay rates obtained above allow us to determine the lifetime of the
saxion.\ The lifetime of the saxion is given by the inverse of its decay rate:%
\begin{equation}
\tau_{sax}=\Gamma_{sax}^{-1}=B_{s\rightarrow aa}\cdot\Gamma_{s\rightarrow
aa}^{-1}\text{,}%
\end{equation}
where as before $B_{s\rightarrow aa}$ denotes the branching ratio of the
saxion to axions. Combined with our expression for the decay rate to two
axions given by equation (\ref{GAA}), the lifetime of the saxion is roughly
given by:%
\begin{equation}
\tau_{sax}=B_{s\rightarrow aa}\Gamma_{s\rightarrow aa}^{-1}\sim B_{s\rightarrow aa}\frac{\pi f_{a}^{2}}{\Delta_{PQ}^{3}}\sim2\times10^{-6}\sec\cdot
B_{s\rightarrow aa}\left(  \frac{100\text{ GeV}}{\Delta_{PQ}}\right)
^{3}\text{,}%
\end{equation}
which shows that the saxion is long-lived.

Summarizing, the primary decay channel of the saxion is either to an
MSSM\ field such as the Higgs, or to a pair of axions. In terms of the
PQ\ deformation parameter $\Delta_{PQ}$, the relevant decay rates are:%
\begin{align}
\Gamma_{s\rightarrow aa}  &  \sim\frac{1}{\pi}\frac{\Delta_{PQ}^{3}}{f_{a}%
^{2}}\\
\Gamma_{s\rightarrow MSSM}  &  \sim\frac{1}{8\pi\Delta_{PQ}}\left(
\frac{m_{soft}^{2}}{f_{a}}\right)  ^{2}\\
\Gamma_{s\rightarrow\psi_{3/2}\psi_{3/2}}  &  \sim\frac{1024}{96\pi}%
\frac{\Delta_{PQ}^{3}}{M_{PL}^{2}}\left(  \frac{\Delta_{PQ}}{m_{3/2}}\right)
^{2}\text{.}%
\end{align}
Having detailed the primary channels of the saxion, we now proceed to the
cosmology of F-theory GUTs.

\section{Cosmology of F-theory GUTs\label{FCOSMO}}

A priori, a seemingly consistent particle physics model could be in severe
conflict with cosmology. In this regard, cosmological considerations translate
into constraints on properties of the particle physics model. In this section we
study the cosmology of F-theory GUTs. Quite remarkably, we find that over the
available range of parameters dictated by purely particle physics
considerations, F-theory GUT scenarios appear to non-trivially satisfy many
cosmological constraints. Further, we find that much of the tension typically
present in gravitino cosmology finds a natural resolution in the context of
F-theory GUTs, which has additional repercussions for issues related to
generating a baryon asymmetry with a sufficiently high value of $T_{RH}^{0}$.
Indeed, many popular mechanisms for generating sufficient baryon asymmetry
hinge on allowing high values for the initial reheating temperature, which is
commonly viewed as being in conflict with the requirements of gravitino
cosmology. This resolution comes about because of a remarkable conspiracy in
the mass of the gravitino and the expected dilution effects associated to the
decay of the saxion. Thus, while one component of the axion supermultiplet
might appear to create a source of tension, the other component completely
relaxes it away.

Cosmological constraints also provide an important window into
UV\ sensitive features of the F-theory GUT\ model. Indeed, the PQ deformation
is directly sensitive to the mass of the anomalous $U(1)_{PQ}$ gauge boson.
This PQ deformation plays a prominent role in the dynamics of the saxion,
which can in turn significantly impact the evolution of the Universe. In this
way, constraints from cosmology on the dynamics of the saxion field directly
translate into seemingly far removed constraints on the compactification and
particle physics content of the F-theory GUT!

Because the ensuing discussion has many interlinked parts, we now summarize
the main features which allow F-theory GUTs to evade the typically stringent
bounds on the initial reheating temperature derived from bounds on the relic
abundance of gravitinos. Starting from the initial temperature $T_{RH}^{0}$,
the Universe begins to cool. At high temperatures, the associated thermal bath
converts MSSM\ particles into gravitinos. Due to the small total cross section
of the gravitinos, these particles fall out of equilibrium at a relatively
high temperature, $T_{3/2}^{f}$, which as before we will denote as the
\textquotedblleft freeze out\textquotedblright\ temperature. While all of this
is occurring, however, the saxion modulus begins to oscillate at a temperature
$T_{osc}^{s}$. The essential point is that because its potential is so flat,
the saxion will naturally be displaced away from its minimum. In fact, for
values of the initial amplitude $s_{0}$ of the saxion field which are quite
reasonable from the perspective of F-theory GUTs, the Universe eventually
enters an era of saxion domination at a temperature $T_{dom}^{s}$. At this
point, the relic abundance of gravitinos is already determined and is given by
the estimates already presented in section \ref{CosmoReview}. The era of
saxion domination continues until the saxion decays, at which point it reheats
the Universe to a temperature $T_{RH}^{s}$, releasing a large amount of
entropy as a consequence of its decay. This dilutes the overall relic
abundance of all species, such as gravitinos.

It remains to say whether the actual relic abundance of gravitinos is
sufficiently low to remain in accord with overclosure bounds. As we will
describe in greater detail in the sections to follow, although a priori, the
oscillation temperature of the saxion and the gravitino mass are unrelated, in
the context of F-theory GUTs, typical values for the saxion and gravitino mass
lead to the relation:%
\begin{equation}
\text{F-theory GUT}:T_{osc}^{s}\sim T_{3/2}^{f}\text{.}%
\end{equation}
This turns out to have the remarkable consequence that\textit{\ the overall
relic abundance of gravitinos is independent of }$T_{3/2}^{f}$, $T_{osc}^{s}$
and $T_{RH}^{0}$! Moreover, the relic abundance of gravitinos thus obtained
does not overclose the Universe, and could potentially correspond to a large
component of the total dark matter. As reviewed in section \ref{CosmoReview},
the axion can potentially also contribute to the overall dark matter content
when it begins to oscillate at temperatures less than the reheating
temperature of the saxion. This analysis only depends on the crude details of
the saxion reheating temperature, which depending on the choice of inputs can
sometimes be above, or below the axion temperature. As such, we will not dwell
on this possibility in any great detail.

In addition, although far less likely, it is also possible to consider
scenarios where the saxion does not dominate the energy density. In such
cases, the usual very stringent overclosure bounds on the gravitino relic
abundance apply, effectively requiring $T_{RH}^{0}<10^{6}$ GeV for a gravitino
of mass $m_{3/2}\sim10-100$ MeV. This can occur, for example, when the initial
amplitude of the saxion is far smaller, at around the scale set by the decay
constant of the axion. Nevertheless, because this requires a significant
calibration of various parameters such as the initial reheating temperature,
in this paper we study the most natural situation where the saxion comes to
dominate the energy density of the Universe.

While the saxion neatly resolves some puzzling features typically found in
gravitino cosmology, it also has the potential to introduce additional issues.
One such issue is that its decay products must not disrupt BBN. Indeed, as
reviewed in section \ref{CosmoReview}, the tight restrictions on the number of
additional relativistic species which can be present translate into a
constraint on the decay channels of the saxion. We find that this either
requires introducing some additional weakly interacting particle into which
the saxion can decay, or that the mass of the saxion must be sufficiently high
so that other decay channels to MSSM\ particles become kinematically available.

Because the decay of the saxion indiscriminately dilutes various relics, it is
important to check whether an appropriate baryon asymmetry can be generated in
F-theory GUT models. Rather than presenting an impediment, the decay of the
saxion appears to open up more possibilities for generating a suitable baryon
asymmetry! This is due to two key features. First, because the relic abundance
of gravitinos can quite naturally fall in a viable range, the usually very tight
prohibition on increasing the initial reheating temperature $T_{RH}^{0}$ is no
longer present. Rather importantly, many mechanisms for generating a large
baryon asymmetry require a high value for $T_{RH}^{0}$, which are commonly
thought to be in conflict with the requirements of gravitino cosmology. Upon
dispensing with the \textquotedblleft gravitino problem\textquotedblright,
these options are once again available! Having said this, in typical models,
the generated baryon asymmetry expected from such mechanisms is sometimes too
large. In the present context, the same dilution effect discussed earlier
turns out to retain an appropriate baryon asymmetry from scenarios such as standard
leptogenesis in a natural range of parameters for F-theory GUT models.

The rest of this section is organized as follows. Because it will play a very
central role in much of the analysis to follow, we begin by analyzing the
history of the saxion, and determine the precise conditions necessary for
saxion domination. After this, we study the relic abundance of the gravitino.
To this end, we first frame the discussion in general terms, asking under
which circumstances we can expect the decay of a cosmological modulus to
render the gravitino relic abundance independent of the temperatures
$T_{3/2}^{f}$, $T_{osc}^{s}$ and $T_{RH}^{0}$. We find that there is a
remarkably small range of parameters available, which are in fact consistent
with the most naive expectations from F-theory GUTs! After this motivation, we
present further details of how the saxion of F-theory GUTs satisfies all of
these requirements. Next, we discuss the decay products generated by the decay
of the saxion. After analyzing constraints from BBN, we show that standard
leptogenesis can generate a suitable baryon asymmetry in the present class
of models, with no fine tuning.

\subsection{Cosmology of the F-theory GUT\ Saxion}

Having outlined the general cosmology of F-theory GUTs, in this subsection we
detail the primary role that the saxion plays as a cosmological modulus. In
order for the saxion to dilute the relic abundance of a species such as the
gravitino, it must be sufficiently long-lived in order for the coherent
oscillations of the saxion field to dominate the energy density of the
Universe. As obtained in section \ref{FGUTAXION}, the lifetime of the saxion
is given by:%
\begin{equation}
\tau_{sax}\sim2\times10^{-6}\sec\cdot B_{s\rightarrow aa}\left(
\frac{100\text{ GeV}}{\Delta_{PQ}}\right)  ^{3}\text{.}%
\end{equation}
Due to its long lifetime, the initial amplitude of the saxion field and its
subsequent coherent oscillation can potentially dominate over other
contributions to the energy density of the Universe.

Because the saxion is a cosmological modulus, much of the general discussion
reviewed in section \ref{CosmoReview} can now be applied to this special case
of interest. Letting $s_{0}$ denote the initial amplitude of the saxion,
the temperature at which the saxion begins to oscillate is given by
equation (\ref{ToscPhiDef}) so that:%
\begin{equation}
T_{osc}^{s}\sim0.3\cdot\sqrt{m_{sax} M_{PL}}=0.6\cdot\sqrt{\Delta
_{PQ} M_{PL}}\sim10^{10}\text{ GeV}\cdot\left(  \frac{\Delta_{PQ}%
}{100\text{ GeV}}\right)  ^{1/2}\text{,} \label{Tsosc}%
\end{equation}
where in the second equality we have used the relation between the mass of the
saxion and $\Delta_{PQ}$ provided by equation (\ref{msax}). As anticipated, we
note that this provides a link between cosmology and the PQ deformation.

Once the saxion begins to oscillate, it will continue to do so until it
decays. Using the general formula for the decay of a modulus, the associated
temperature of decay is given by:
\begin{equation}
T_{RH}^{s}=T_{decay}^{s}\sim0.5\cdot\sqrt{\Gamma_{sax} M_{PL}}%
\sim0.4\text{ GeV}\cdot B_{s\rightarrow aa}^{-1/2}\left(  \frac{\Delta_{PQ}%
}{100\text{ GeV}}\right)  ^{3/2}\text{,} \label{TSRH}%
\end{equation}
where by a small abuse of notation, we have identified the decay temperature
with a \textquotedblleft reheating temperature\textquotedblright, which is
strictly speaking only correct if the saxion comes to dominate the energy
density. Note in particular that $T_{decay}^{s}$ is far lower than
$T_{osc}^{s}$. In addition, we also observe that the decay temperature of the
saxion falls above the temperature required for BBN. Taking the maximal allowed
branching ratio to axions consistent with BBN (which we will discuss later)
so that $B_{s\rightarrow aa} \sim 1/6$, and for a representative value
of $\Delta_{PQ} \sim 100$ GeV, we obtain the crude estimate:\footnote{Although
the saxion is a cosmological modulus, it is interesting to note that in scenarios
where it dominates the energy density of the Universe, it avoids the
usual \textquotedblleft moduli problem \textquotedblright because
$f_{a} \ll M_{GUT}$. Indeed, for a typical modulus of mass $m_{\phi} \sim m_{sax}$, the analogous decay rate
is of the form $\Gamma_{\phi} \sim m^{3}_{\phi}/\Lambda^{2}$, where $\Lambda$ is of the GUT,
or Planck scale. As a consequence, the corresponding reheating temperature
would then be much lower, jeopardizing BBN.}
\begin{equation}
T^{s}_{RH} \sim 1 \text{ GeV}\text{.}
\end{equation}

In between the temperature at which it begins to oscillate and decay, the
saxion may come to dominate the energy density of the Universe. Letting
$T_{dom}^{s}$ denote this temperature, this amounts to the condition:%
\begin{equation}
T_{osc}^{s}>T_{dom}^{s}>T_{decay}^{s}\text{,} \label{Tinequ}%
\end{equation}
where as reviewed in section \ref{CosmoReview} for a general cosmological
modulus, the temperature $T_{dom}^{s}$ is given by:%
\begin{equation}
T_{dom}^{s} \sim \frac{s_{0}^{2}}{M^2_{PL}}\min(T_{RH}^{0},T_{osc}^{s})\text{.}%
\end{equation}

There are a priori two natural scales associated to the initial amplitude of
$s_{0}$. Because the saxion localizes on a matter curve with
characteristic mass scale $M_{X}\sim10^{15.5}$ GeV, it is natural to take:%
\begin{equation}
s_{0}\sim M_{X}\sim10^{15.5}\text{ GeV.}%
\end{equation}
Comparing our expressions for $T_{osc}^{s}$, $T_{dom}^{s}$ and $T_{decay}^{s}
$, it follows that inequality (\ref{Tinequ}) indeed holds for $s_{0}\sim10^{15.5}$
GeV. We note in passing that from the perspective of the effective field
theory, it is also possible to consider smaller field ranges set by the value
of the axion decay constant so that $s_{0}\sim f_{a}\sim10^{12}$ GeV.\ Note
that in this case, $T_{dom}^{s}$ is smaller than $T_{decay}^{s}$, indicating
that the saxion in this case never comes to dominate the energy density of the Universe.

Restricting to the most natural scenario where the saxion does indeed come to
dominate the energy density, it will release a large amount of entropy when it
decays. The associated dilution factor for any cosmological modulus again
applies to the special case of the saxion, with the result:%
\begin{equation}
D\sim\frac{M_{PL}^{2}}{s_{0}^{2}}\frac{T_{RH}^{s}}{\min(T_{osc}^{s},T_{RH}%
^{0})}\text{.}%
\end{equation}
Treating separately the two cases $T_{osc}^{s}>T_{RH}^{0}$ and $T_{osc}%
^{s}<T_{RH}^{0}$, we therefore obtain the following expressions for the
overall dilution factor:%
\begin{align}
D_{T_{osc}^{s}>T_{RH}^{0}}  &  \sim10^{-5}\cdot B_{s\rightarrow aa}%
^{-1/2}\left(  \frac{10^{10}\text{ GeV}}{T_{RH}^{0}}\right)  \left(
\frac{10^{15.5}\text{ GeV}}{s_{0}}\right)  ^{2}\left(  \frac{\Delta_{PQ}%
}{100\text{ GeV}}\right)  ^{3/2}\label{sdtbu}\\
D_{T_{osc}^{s}<T_{RH}^{0}}  &  \sim10^{-5}\cdot B_{s\rightarrow aa}%
^{-1/2} \left(\frac{10^{15.5}\text{ GeV}}{s_{0}}\right)  ^{2}\left(  \frac{\Delta_{PQ}%
}{100\text{ GeV}}\right)  \text{.}%
\end{align}
An important feature of these expressions is the overall dependence of the
dilution factor on the initial reheating temperature, $T_{RH}^{0}$. Indeed, we
note that when $T_{RH}^{0}<T_{osc}^{s}$, the dilution factor becomes more
potent as $T_{RH}^{0}$ increases. This continues until $T_{RH}^{0}\sim
T_{osc}^{s}$, at which point, the dilution factor ceases to depend on
$T_{RH}^{0}$. This situation closely parallels the $T_{RH}^{0}$ dependence of
the gravitino, to which we shall shortly turn.

Finally, it is also convenient to introduce the minimal reheating temperature required in
order for saxion dilution to occur. This is given by the value of $T_{RH}^{0}$ at which the
dilution factor first equals one (the case of no dilution). Solving for $T_{RH}^{0}$ in the equality:
\begin{equation}
1 = D_{min} \sim \frac{M_{PL}^{2}}{s_{0}^{2}}\frac{T_{RH}^{s}}{T_{RH}^{0}}%
\end{equation}
yields:
\begin{equation}
T^{s}_{D_{min}} \equiv T_{RH}^{0} \sim \frac{M_{PL}^{2}}{s_{0}^{2}} T_{RH}^{s} \text{.}
\end{equation}
Using the explicit numerical values found in this section, it follows that the typical value of $T^{s}_{D_{min}}$ is
on the order of $10^5$ GeV.

\begin{figure}
[ptb]
\begin{center}
\includegraphics[
height=4.12in,
width=6.5769in
]%
{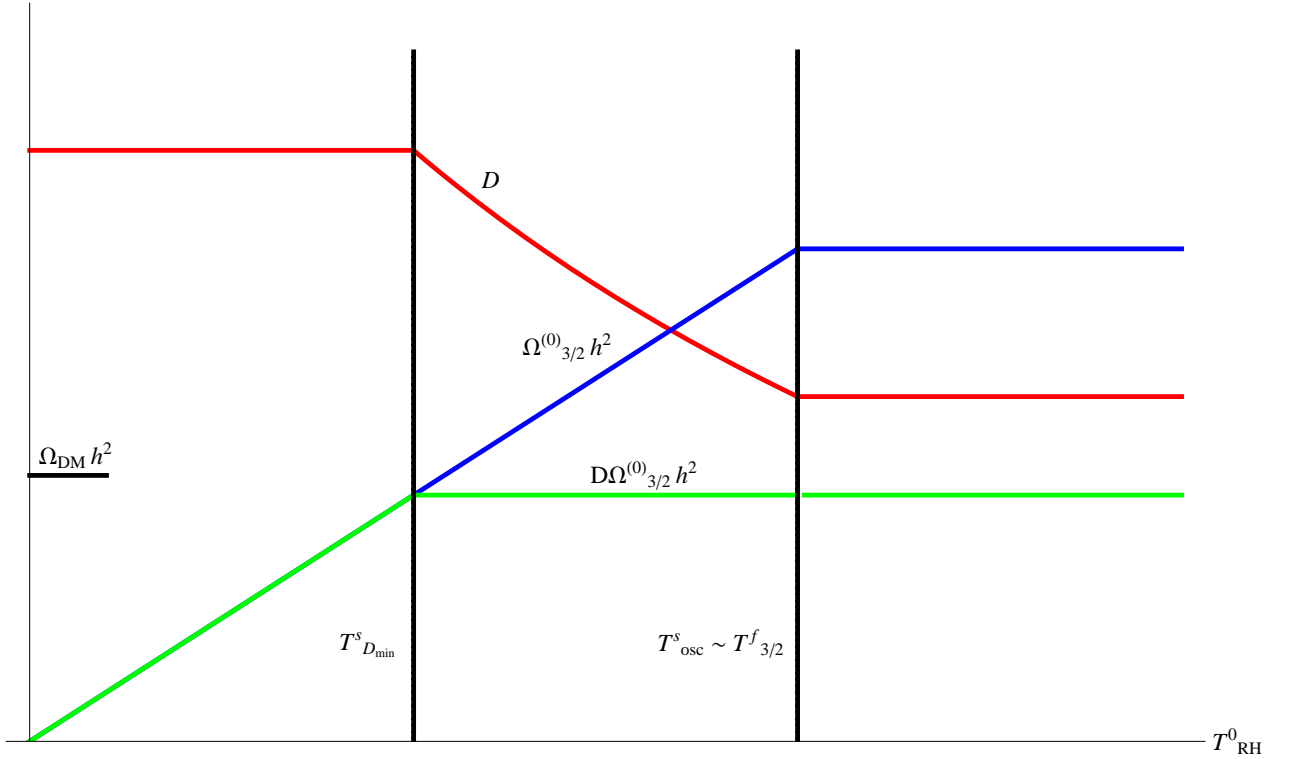}%
\caption{Schematic plot of the dilution factor $D$ of the F-theory GUT saxion,
the gravitino relic abundance in the absence of dilution, $\Omega_{3/2}%
^{(0)}h^{2}$ and the net relic abundance after taking account of the dilution
factor of the saxion as a function of the initial reheating temperature
$T_{RH}^{0}$. The graph of the dilution is given with respect to a scale
distinct from that for the relic abundances. The plot depicts the special case where the freeze out
temperature for the gravitino $T_{3/2}^{f}\sim T_{osc}^{s}$, the temperature
at which the saxion begins to oscillate. When $T_{RH}^{0}$ is below
$T^{s}_{D_{min}}$, there is no dilution factor $(D=1)$. In the special case
where $T_{3/2}^{f}\sim T_{osc}^{s}$, for all values of $T_{RH}^{0}>T^{s}_{D_{min}}$,
the total relic abundance of gravitinos is independent of $T_{RH}^{0}$.
See figures \ref{sketcha} and \ref{sketchb} for schematic plots of scenarios
where $T_{3/2}^{f}\neq T_{osc}^{s}$.}%
\label{specialsketch}%
\end{center}
\end{figure}
\begin{figure}
[ptb]
\begin{center}
\includegraphics[
height=4.1208in,
width=6.5708in
]%
{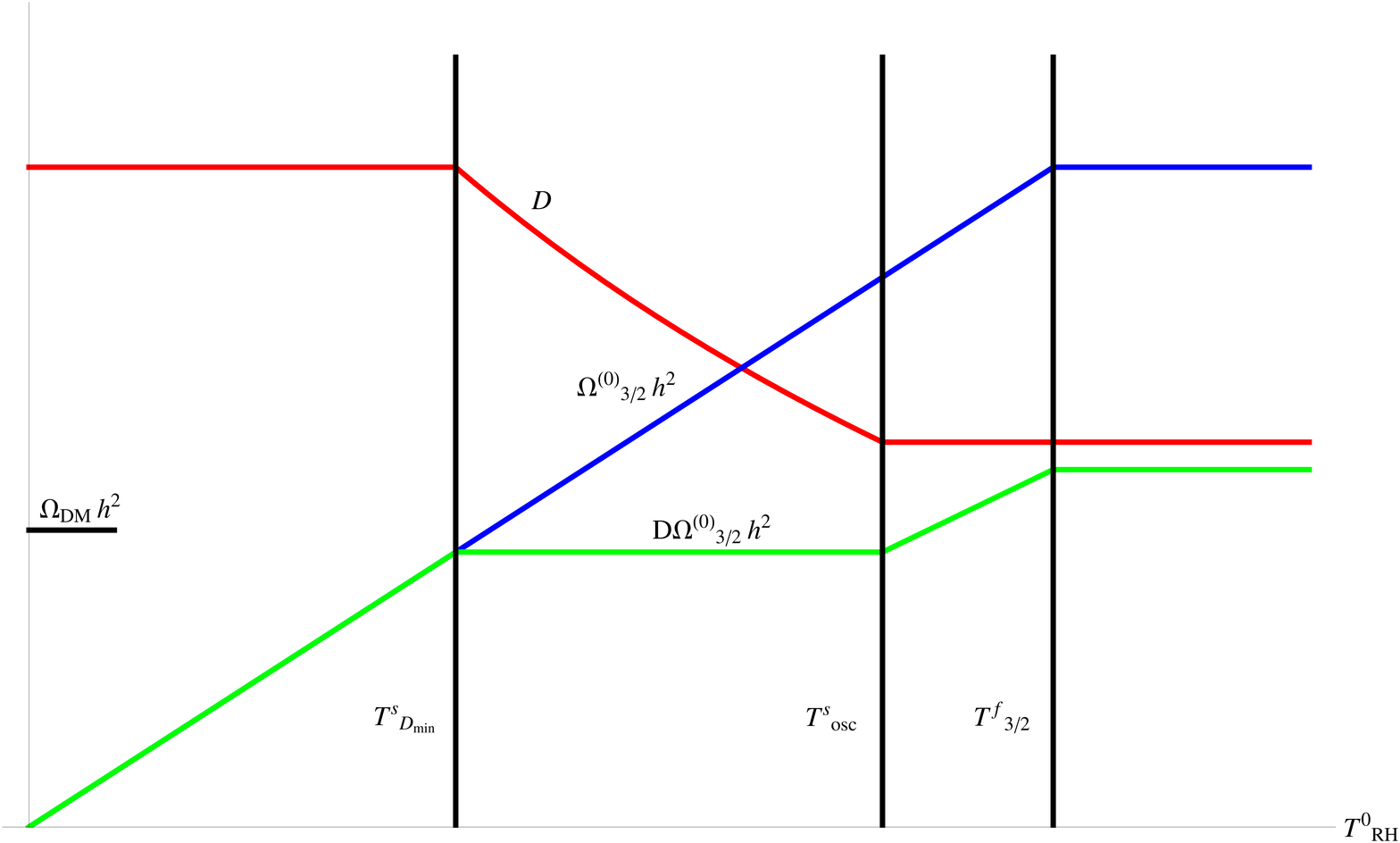}%
\caption{Schematic plot of the dilution factor $D$ of the F-theory GUT saxion,
the gravitino relic abundance in the absence of dilution, $\Omega_{3/2}%
^{(0)}h^{2}$ and the net relic abundance after taking account of the dilution
factor of the saxion as a function of the initial reheating temperature
$T_{RH}^{0}$. The graph of the dilution is given with respect to a scale
distinct from that for the relic abundances. The plot depicts the case where the freeze out temperature for
the gravitino $T_{3/2}^{f}>T_{osc}^{s}$, the temperature at which the saxion
begins to oscillate. When $T_{RH}^{0}<T^{s}_{D_{min}}$, there is no dilution
factor. Note that in this case, the total relic abundance of gravitinos
increases for values of the initial reheating temperature such that
$T_{osc}^{s}<T_{RH}^{0}<T_{3/2}^{f}$.}%
\label{sketcha}%
\end{center}
\end{figure}
\begin{figure}
[ptb]
\begin{center}
\includegraphics[
height=4.1208in,
width=6.5657in
]%
{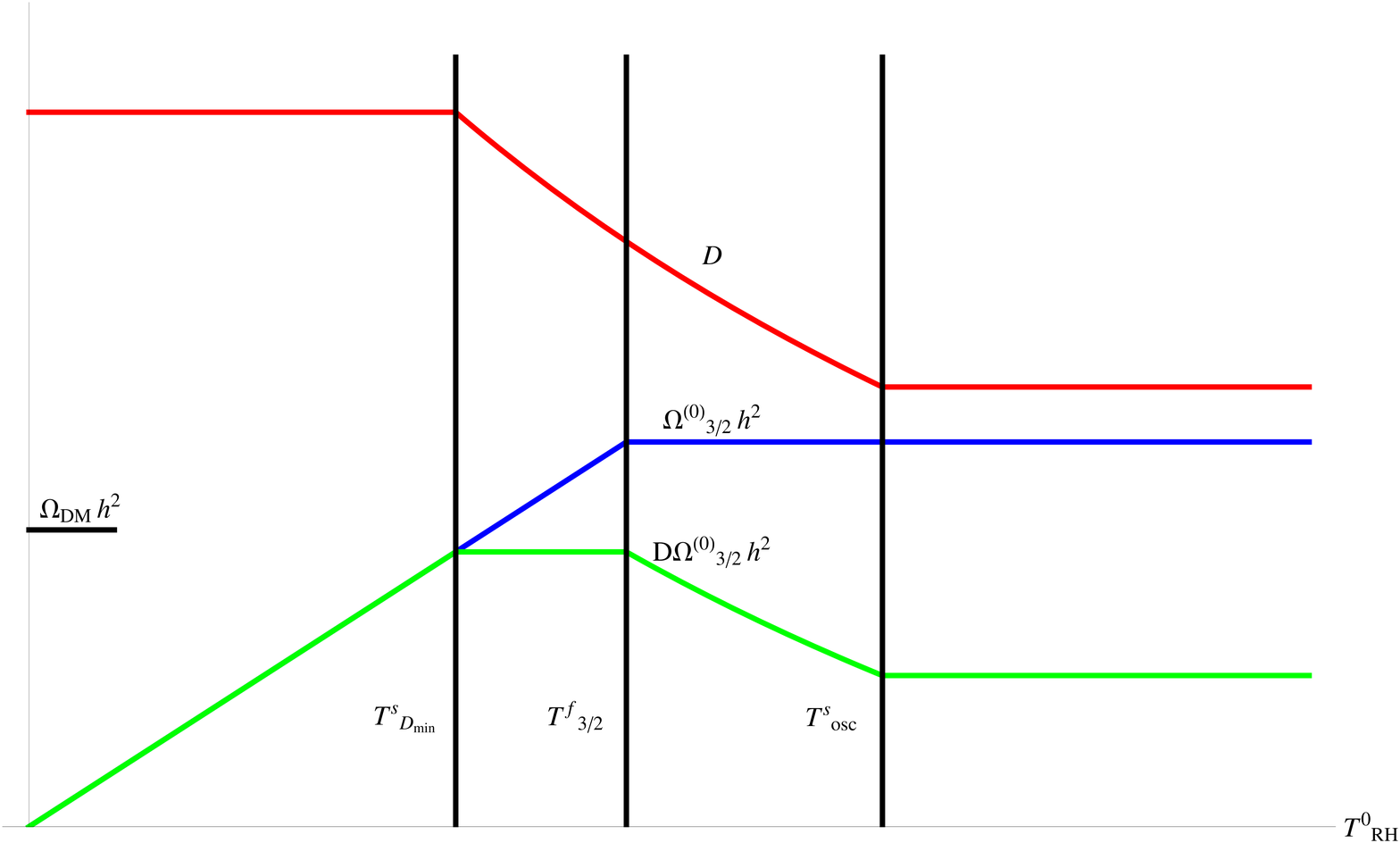}%
\caption{Schematic plot of the dilution factor $D$ of the F-theory GUT saxion,
the gravitino relic abundance in the absence of dilution, $\Omega_{3/2}%
^{(0)}h^{2}$ and the net relic abundance after taking account of the dilution
factor of the saxion as a function of the initial reheating temperature
$T_{RH}^{0}$. The graph of the dilution is given with respect to a scale
distinct from that for the relic abundances. The plot depicts the case where the freeze out temperature for
the gravitino $T_{3/2}^{f}<T_{osc}^{s}$, the temperature at which the saxion
begins to oscillate. When $T_{RH}^{0}<T^{s}_{D_{min}}$, there is no dilution
factor. Note that the total relic abundance
of gravitinos decreases as $T_{RH}^{0}$ increases in the range $T_{3/2}%
^{f}<T_{RH}^{0}<T_{osc}^{s}$.}%
\label{sketchb}%
\end{center}
\end{figure}

\subsection{The Saxion-Gravitino Connection}

In the previous section we described the primary features of the saxion. The
main point we found is that for typical values of F-theory GUT\ parameters,
the oscillation of the saxion eventually comes to dominate the energy density
of the Universe. The subsequent decay of the saxion will then dilute the relic
abundance of all particle species. In this subsection we study the effects of
this dilution on the relic abundance of gravitinos. We note that the idea of
solving the gravitino problem due to a late decaying field has appeared for
example in \cite{Fujii:2002fv,Fujii:2003iw}. In this regard, the quite natural way
that these ideas automatically appear in F-theory GUTs provides
strong motivation for this type of cosmological scenario.

Recall that in section \ref{CosmoReview} we reviewed the fact that for a
gravitino of mass $m_{3/2}\sim10-100$ MeV, the resulting relic abundance would
at first appear to violate the usual overclosure bound:%
\begin{equation}
\Omega_{3/2}^{T}h^{2}\leq0.1\text{.}%
\end{equation}
We find that the dilution of the saxion naturally resolves this issue because
of a remarkable confluence of elements involving the masses of the gravitino
and saxion, and the initial amplitude of the saxion. The basic point is that
the relic abundance of gravitinos is governed by equation (\ref{finalthermalrelic}):
\begin{equation}
\Omega_{3/2}^{T}h^{2}\sim D\cdot2.7\times10^{3}\cdot\left(  \frac{\min
(T_{3/2}^{f},T_{RH}^{0})}{10^{10}\text{ GeV}}\right)  \left(  \frac{10\text{
MeV}}{m_{3/2}}\right)  \left(  \frac{m_{\widetilde{g}}}{1\text{ TeV}}\right)
^{2}%
\end{equation}
where here, we have included explicitly the dilution factor associated with
the decay of the saxion. In addition, the gravitino freeze out temperature is
given by equation (\ref{TFREEZEGRAV}) as:%
\begin{equation}
T_{3/2}^{f}\sim2\times10^{10}\text{ GeV}\cdot\left(  \frac{m_{3/2}}{10\text{
MeV}}\right)  ^{2}\left(  \frac{1\text{ TeV}}{m_{\widetilde{g}}}\right)
^{2}\text{.} \label{ourTFREEZE}%
\end{equation}
Note that the overall relic abundance is given by the minimum of the freeze
out temperature $T_{3/2}^{f}$ and the initial reheating temperature
$T_{RH}^{0}$. Indeed, recall that for $m_{3/2}\sim10$ MeV and $m_{\widetilde
{g}}\sim1$ TeV, consistency with the overclosure bound would appear to require
$T_{RH}^{0}\lesssim10^{6}$ GeV. Quite curiously, the amount of dilution
associated with the saxion is also limited by the minimum of its oscillation
temperature, and $T_{RH}^{0}$ so that:%
\begin{equation}
D\sim\frac{M_{PL}^{2}}{s_{0}^{2}}\frac{T_{RH}^{s}}{\min(T_{osc}^{s},T_{RH}%
^{0})}\text{.}%
\end{equation}
In other words, the thermally produced relic abundance of gravitinos is:%
\begin{equation}
\Omega_{3/2}^{T}h^{2}\sim2.7\times10^{3}\cdot\frac{M_{PL}^{2}}{s_{0}^{2}%
}\left(  \frac{\min(T_{3/2}^{f},T_{RH}^{0})}{\min(T_{osc}^{s},T_{RH}^{0}%
)}\right)  \left(  \frac{T_{RH}^{s}}{10^{10}\text{ GeV}}\right)  \left(
\frac{10\text{ MeV}}{m_{3/2}}\right)  \left(  \frac{m_{\widetilde{g}}}{1\text{
TeV}}\right)  ^{2}\text{.} \label{omegagrav}%
\end{equation}
While a priori, the oscillation temperature of the saxion and the gravitino
mass are unrelated, in the context of F-theory GUTs, comparing the oscillation
temperature $T_{osc}^{s}$ of equation (\ref{Tsosc}) with the freeze out
temperature of the gravitino provided by equation (\ref{ourTFREEZE}), we have:%
\begin{align}
T_{osc}^{s}  &  \sim10^{10}\text{ GeV}\cdot\left(  \frac{\Delta_{PQ}%
}{100\text{ GeV}}\right)  ^{1/2}\\
T_{3/2}^{f}  &  \sim2\times10^{10}\text{ GeV}\cdot\left(  \frac{m_{3/2}%
}{10\text{ MeV}}\right)  ^{2}\left(  \frac{1\text{ TeV}}{m_{\widetilde{g}}%
}\right)  ^{2}\text{.} \label{TFREEZEAGAIN}%
\end{align}
In other words for typical values of $m_{3/2}\sim10-100$ MeV, $m_{\widetilde
{g}}\sim1$ TeV and $\Delta_{PQ}\sim100$ GeV, we obtain the remarkable
relation:%
\begin{equation}
T_{osc}^{s}\sim T_{3/2}^{f}\text{,} \label{temprelation}%
\end{equation}
which should hold as an order of magnitude estimate. Returning to the
gravitino relic abundance of equation (\ref{omegagrav}), we therefore find:%
\begin{equation}
\Omega_{3/2}^{T}h^{2}\sim 0.1 \cdot \left(  \frac{10^{15.5}\text{ GeV}}{s_{0}}\right)^{2}
\left(  \frac{T_{RH}^{s}}{1\text{ GeV}}\right)  \left(  \frac{10\text{
MeV}}{m_{3/2}}\right)  \left(  \frac{m_{\widetilde{g}}}{1\text{ TeV}}\right)
^{2}\text{.}%
\end{equation}
\textit{As a consequence, the overall relic abundance of gravitinos is
independent of }$T_{3/2}^{f}$, $T_{osc}^{s}$ and $T_{RH}^{0}$!

Using our expression for $T_{RH}^{s}$ given by equation (\ref{TSRH}) finally
yields:%
\begin{equation}
\Omega_{3/2}^{T}h^{2}\sim0.07\cdot B_{s\rightarrow aa}^{-1/2}\left(
\frac{10^{15.5}\text{ GeV}}{s_{0}}\right)  ^{2}\left(  \frac{\Delta_{PQ}%
}{100\text{ GeV}}\right)  ^{3/2}\left(  \frac{10\text{ MeV}}{m_{3/2}}\right)
\left(  \frac{m_{\widetilde{g}}}{1\text{ TeV}}\right)  ^{2}\text{,}
\label{TGRAV}%
\end{equation}
which without any fine tuning satisfies the overclosure bound, and in certain
circumstances could account for the observed dark matter density!

Turning the discussion around, the relation $T_{osc}^{s}\sim T_{3/2}^{f}$ may
be viewed as a type of \textquotedblleft resonance condition\textquotedblright%
\ which preferentially selects a window of values for the mass of the
gravitino. Setting $T_{osc}^{s}\sim T_{3/2}^{f}$ and solving for $m_{3/2}$ in
terms of $\Delta_{PQ}$, we find:%
\begin{equation}
m_{3/2}\sim10\text{ MeV}\cdot\left(  \frac{\Delta_{PQ}}{100\text{ GeV}%
}\right)  ^{1/4}\left(  \frac{m_{\widetilde{g}}}{1\text{ TeV}}\right)
\label{coincidencegrav}%
\end{equation}
as an order of magnitude estimate. An intriguing consequence of this formula
is that the mass of the gravitino is fairly insensitive to the value of the PQ
deformation parameter.

It is important to note that in the above estimate we have neglected potential
temperature dependent corrections to the mass of the gravitino. In general,
one would expect these corrections to decrease the mass of the gravitino above
$\sqrt{F}\sim10^{8.5}$ GeV. Our estimate for $T_{3/2}^{f}\sim10^{10}$
GeV is close to this scale, which suggests that there should be a mild
decrease in the actual mass of the gravitino at this temperature. Thus, equation
(\ref{coincidencegrav}) should be viewed as a lower estimate for the gravitino
mass, which is reassuring in that $10$ MeV is on the lower end of the mass
range for the gravitino in F-theory GUT models. As a consequence, the
coincidence window is roughly in the range $m_{3/2}\sim10-100$ MeV, which is
intriguingly in the natural range expected for F-theory GUTs.

\subsubsection{F-theory and a Confluence of Parameters}

The result of the previous section suggests that some of the most distressing
features of gravitino cosmology find a natural resolution in the context of
F-theory GUTs. The crucial feature of this analysis is that although the
gravitino relic abundance has non-trivial dependence on $T_{RH}^{0}$, this is
exactly counterbalanced by the $T_{RH}^{0}$ dependence of the dilution factor
derived from the saxion.

Given this remarkable conspiracy, it is natural to ask whether more general
models with a late decaying modulus $\phi$ could potentially exhibit similar
properties. In fact, as we now explain, the required relations between the
mass of this modulus, its associated decay width, and the mass of the
gravitino are only satisfied in a small window of values.

The condition that the dilution factor exactly cancel the $T_{RH}^{0}$
dependence of $\Omega_{3/2}^{T}h^{2}$ requires that the analogue of equation
(\ref{temprelation}) must hold:%
\begin{equation}
T_{osc}^{\phi}\sim T_{3/2}^{f}\text{.} \label{phifreezecomp}%
\end{equation}
Using the explicit expression for the oscillation temperature of a modulus as
well as the value of $T_{3/2}^{f}$ given by equation (\ref{TFREEZEAGAIN}), we
obtain:%
\begin{equation}
0.5\sqrt{m_{\phi}M_{PL}}\sim T_{osc}^{\phi}\sim T_{3/2}^{f}\sim2\times
10^{10}\text{ GeV}\cdot\left(  \frac{m_{3/2}}{10\text{ MeV}}\right)
^{2}\text{,}%
\end{equation}
where for simplicity, we have set $m_{\widetilde{g}}\sim1$ TeV. The mass of
the gravitino is therefore given by:%
\begin{equation}
\frac{m_{3/2}}{10\text{ MeV}}\sim0.6\cdot\left(  \frac{m_{\phi}}{100\text{
GeV}}\right)  ^{1/4}\text{,} \label{gravtophi}%
\end{equation}
In other words, a cosmological modulus with mass on the order of $100-1000$
GeV will lead to a gravitino abundance which is independent of $T_{RH}%
^{0},T_{osc}^{\phi}$ and $T_{3/2}^{f}$ when the mass of the gravitino is in
the range $10-100$ MeV!

We can also deduce further properties of the modulus by examining the overall
relic abundance of gravitinos. Under conditions where equation
(\ref{phifreezecomp}) is satisfied, it is enough to consider the special case
where $T_{RH}^{0}>T_{osc}^{\phi},T_{3/2}^{f}$. In this case, the constraint
$\Omega_{3/2}h^{2}\leq0.1$ translates to the condition%
\begin{equation}
D_{\phi}\left(  \frac{m_{3/2}}{2\text{ keV}}\right)  \leq0.1\text{,}%
\end{equation}
where in the above, $D_{\phi}$ denotes the dilution factor of the cosmological
modulus. Using the explicit expression for $D_{\phi}$ reviewed in section
\ref{CosmoReview}, this becomes:%
\begin{equation}
\Omega_{3/2}^{T}h^{2}\sim\frac{M_{PL}^{2}}{\phi_{0}^{2}}\sqrt{\frac
{\Gamma_{\phi}}{m_{\phi}}}\left(  \frac{m_{3/2}}{2\text{ keV}}\right)
\leq0.1\text{,}%
\end{equation}
where as before, $\Gamma_{\phi}$ denotes the decay rate of the modulus and
$\phi_{0}$ denotes the initial amplitude of this field. Parameterizing the
decay rate as:%
\begin{equation}
\Gamma_{\phi}\sim\frac{1}{64\pi}\frac{m_{\phi}^{3}}{\Lambda^{2}}\text{,}%
\end{equation}
where $\Lambda$ is some characteristic scale associated with the dynamics of
the $\phi$ modulus, we obtain the bound:%
\begin{equation}
\Omega_{3/2}^{T}h^{2}\sim0.02\cdot\left(  \frac{10^{15.5}\text{ GeV}}{\phi
_{0}}\right)  ^{2}\left(  \frac{10^{12}\text{ GeV}}{\Lambda}\right)  \left(
\frac{m_{\phi}}{100\text{ GeV}}\right)  ^{5/4}\leq0.1\text{,}%
\end{equation}
which suggests a range of parameters similar to those of F-theory GUTs.

\subsection{Decay Products of the Saxion\label{DecayProducts}}

In the previous subsection we observed that the decay of the saxion and the
associated release of entropy modifies the expected relic abundance of
gravitinos, finding a remarkable confluence between the saxion oscillation
temperature and the freeze out temperature for the gravitino. In addition we
also found that the resulting relic abundance of thermally produced gravitinos
is quite close to saturating the total amount of dark matter.

But while the decay of the saxion effectively dilutes previously generated
thermal relics, the end products of its decay can re-introduce another source
for these same particles! Due to their overall longevity, decays to gravitinos
and axions comprise the main source of additional potential relics. In
this subsection we compute the relic abundance from
such \textquotedblleft non-thermally produced\textquotedblright%
\ gravitinos and axions. Whereas the amount of non-thermally produced axions
is entirely negligible, depending on the mass of the gravitino, non-thermal
production of gravitinos can also contribute towards the total dark matter.\footnote{This is to be contrasted with
for example, the result of \cite{Ibe:2006rc}, where in that case the
production of gravitinos from the decay of the field responsible for
supersymmetry breaking leads to essentially all of the gravitino relic
abundance because the branching ratio to gravitinos is somewhat higher in the
\textquotedblleft sweet spot\textquotedblright\ scenario.}

For simplicity, in this subsection we restrict our attention to scenarios
where the saxion comes to dominate the energy density of the Universe. Besides
being the primary case of interest for F-theory GUTs, in the other more
specialized case where the saxion is a subdominant component of the overall
energy density, the resulting relic abundance will on general grounds generate
a smaller component of the total matter content in comparison with a scenario
with an era of saxion domination.

In a scenario where the saxion dominates the energy density of the Universe,
the fraction of the energy density transferred to the $i^{th}$ decay product
is dictated by the branching ratio:%
\begin{equation}
B_{s\rightarrow ii}=\frac{\Gamma_{s\rightarrow ii}}{\Gamma_{sax}}%
\end{equation}
where in the above $\Gamma_{s\rightarrow ii}$ the decay rate of the saxion to
the $i^{th}$ species, and as before, $\Gamma_{sax}$ denotes the total decay
rate of the saxion. In this case, the overall yield of the $i^{th}$ species is
given by:%
\begin{equation}
Y_{i}^{NT}=\frac{n_{i,after}^{NT}}{s_{after}}=B_{s\rightarrow ii}\cdot
\frac{s_{before}}{s_{after}}\frac{n_{sax,before}}{s_{before}}=\frac{3}%
{2}B_{s\rightarrow ii}\cdot\frac{T_{RH}^{s}}{m_{s}}\text{.}%
\end{equation}
where here, \textquotedblleft before\textquotedblright\ and \textquotedblleft
after\textquotedblright\ are in reference to times close to the decay of the
saxion. Note that as usual, $Y_{i}$ is constant as the Universe subsequently
evolves. The non-thermally produced relic abundance
from the $i^{th}$ species is therefore given by:
\begin{equation}
\Omega_{i}^{NT}h^{2}=\left(  \frac{s_{0}}{\rho_{c,0}}h^{2}\right)  \cdot
m_{i}Y_{i}^{NT}=\left(  \frac{s_{0}}{\rho_{c,0}}h^{2}\right)  \cdot\frac{3}%
{2}m_{i}B_{s\rightarrow ii}\frac{T_{RH}^{s}}{m_{sax}}\text{.}%
\end{equation}

We now compute the value of $\Omega_{i}^{NT}h^{2}$ in terms of the branching
ratio $B_{s\rightarrow aa}$. Using the explicit expression for
$B_{s\rightarrow\psi_{3/2}\psi_{3/2}}$ in terms of $B_{s\rightarrow aa}$ given
by equation (\ref{BSPSIPSI}), plugging in the present values of $\rho_{c,0}$ and
$s_{0}$ reviewed in section \ref{CosmoReview} as well as the value of
$T_{RH}^{s}$ obtained in equation (\ref{TSRH}), the non-thermally produced
relic abundance of gravitinos and axions are respectively given by:%
\begin{align}
\Omega_{3/2}^{NT}h^{2}  &  \sim0.9\cdot B_{s\rightarrow aa}^{1/2}\left(
\frac{10\text{ MeV}}{m_{3/2}}\right)  \left(  \frac{\Delta_{PQ}}{100\text{
GeV}}\right)  ^{5/2}\label{NTGRAVI}\\
\Omega_{ax}^{NT}h^{2}  &  \sim5\times10^{-9}\cdot B_{s\rightarrow aa}%
^{1/2}\left(  \frac{m_{a}}{10^{-5}\text{ eV}}\right)  \left(  \frac
{\Delta_{PQ}}{100\text{ GeV}}\right)  ^{1/2}\text{.}%
\end{align}
In the natural range of masses for F-theory GUTs, it follows that the relic
abundance from axions is completely negligible. On the other hand, the relic
abundance of non-thermally produced gravitinos can potentially play a more
prominent role. For example, using the representative values $B\sim10^{-2}$,
$m_{3/2}\sim20$ MeV and $\Delta_{PQ}\sim100$ GeV, the resulting relic
abundance of non-thermally produced gravitinos is $\sim0.05$. On the other
hand, when the mass of the gravitino is closer to the end range of F-theory
GUT values at $\sim100$ MeV, the relic abundance is at most $\sim10\%$ of the
dark matter, so that such gravitinos could comprise an additional component of
the dark matter.\footnote{Due to the fact that non-thermally produced
gravitinos are created at temperatures only a few orders of magnitude
different from BBN, it is interesting to ask whether the resulting relics are
indeed \textquotedblleft cold\textquotedblright\ or \textquotedblleft
warm\textquotedblright\ at the time of matter recombination. In principle,
\textquotedblleft hot\textquotedblright\ dark matter can disrupt structure
formation. As dark matter candidates, both gravitinos and axions are typically
both sufficiently non-relativistic at the time of matter recombination to
constitute cold dark matter candidates. We refer the interested reader to
\cite{Taoso:2007qk} for further discussion on the distinctions between hot,
warm and cold dark matter.} This is to be contrasted with the case of the
\textquotedblleft sweet spot\textquotedblright\ model of supersymmetry
breaking \cite{Ibe:2006rc,KitanoIbeSweetSpot}, where the late decay of the
field responsible for supersymmetry breaking generates most of the dark matter abundance.

\subsection{BBN and F-theory GUTs}

As reviewed in section \ref{CosmoReview}, it is important to ascertain whether
a given extension of the Standard Model disrupts the successful predictions of
BBN. In this regard, we have already seen that the saxion can significantly
alter the evolution of the Universe at temperatures above the start of BBN. In
the most typical F-theory GUT\ scenario with an era of saxion domination, we
find that the resulting reheating temperature is somewhat higher than
$T_{BBN}$. Indeed, this imposes only the mild constraint $\Delta_{PQ}\gtrsim1$
GeV. A far more significant constraint from BBN\ originates from the bound on
the number of relativistic species. In the context of models with minimal
matter content, this translates into the condition that the saxion must be
allowed to decay to additional species beyond the axion, which in turn imposes
a lower bound on the mass of the saxion. In terms of the PQ\ deformation, this
amounts to the condition:%
\begin{equation}
\Delta_{PQ}\gtrsim60\text{ GeV.}%
\end{equation}
Note that this bound is indeed in accord with the crude expectation that the
value of $\Delta_{PQ}$ is most naturally near the weak scale.

A late-decaying NLSP can also potentially alter the expected abundances of
light elements generated by BBN. While a full study of BBN\ is beyond the
scope of the present paper, a cursory inspection of the recent literature
suggests that in comparison with standard cosmology, in a range of
parameters for the gravitino and NLSP\ favored by
F-theory GUTs, the expected abundance of light elements, such as the
typically problematic $^{7}Li$ is in somewhat \textit{better} accord with observation. This
is due to the fact that the late decay of the NLSP\ in such scenarios can alter
the reaction rates of BBN.

\subsubsection{Decay Channels of the Saxion}

In subsection (\ref{RelSpecCon}) we reviewed the general constraint on how many
additional relativistic species can contribute such that the predictions of
BBN\ remain in accord with observation. In this subsection we consider the
special case associated with the decay products of the saxion, closely
following the quite general analysis for saxion decays presented in
\cite{Hashimoto:1998ua}. There are in principle two possibilities, depending
on whether or not the oscillation of the saxion comes to dominate the energy
density of the Universe. As throughout, we shall specialize to the case where
the saxion dominates the energy density of the Universe. Although the latter
scenario is indeed more typical in the context of F-theory GUTs, for the sake
completeness, in this subsection we study both possibilities in turn. Using
the general constraint on the total number of allowed relativistic species
reviewed in subsection \ref{RelSpecCon}, the contribution to the energy
density from saxion generated axions at the time of BBN\ must satisfy the
inequality:%
\begin{equation}
\left(  \frac{\rho_{a}}{\rho_{r}}\right)  _{BBN}\leq\frac{7}{43}\text{.}
\label{rhoatorhor}%
\end{equation}

Assuming the saxion comes to dominate the energy density of the Universe,
the energy density stored in the saxion is directly transferred into the
background radiation, as well as the axions. As such, we obtain the estimate:%
\begin{equation}
\left(  \frac{\rho_{a}}{\rho_{r}}\right)  _{BBN}\sim B_{s\rightarrow
aa}\text{,}%
\end{equation}
with $B_{s\rightarrow aa}$ the branching ratio of the saxion to two axions.
Inequality (\ref{rhoatorhor}) therefore provides a bound on the branching
fraction to axions:%
\begin{equation}
B_{s\rightarrow aa}\leq\frac{7}{43}\text{.}%
\end{equation}
In the minimal F-theory GUT, the primary decay channel at small $\Delta_{PQ}$
is given by the saxion to two axions. In order to satisfy this bound, the
F-theory GUT must include some additional mode into which the saxion can
decay. While it is certainly possible to posit the existence of some
additional mode beyond the ones already present in the MSSM, in a minimal
scenario, the saxion will decay to MSSM\ degrees of freedom. As found in
section \ref{FGUTAXION}, when available, the primary decay channel is for the
saxion to decay to two Higgs bosons. This decay is kinematically allowed
provided the mass of the saxion is sufficiently large:%
\begin{equation}
m_{sax}\geq2m_{h^{0}}\text{.}%
\end{equation}
Using the expected mass of the Higgs of $\sim115$ GeV, this translates into
the following lower bound on the PQ\ deformation:%
\begin{equation}
\Delta_{PQ}\gtrsim60\text{ GeV.}%
\end{equation}
We note that this crude kinematic constraint is fairly independent of the
details of a particular F-theory GUT, such as the number of messengers, and in
this regard is likely to be quite robust. Although it is also possible to
derive a similar bound in the case where the saxion does not come to dominate
the energy density of the Universe, in the context of F-theory GUTs, this is a
far less likely scenario. We refer the interested reader to
\cite{Hashimoto:1998ua} for further details on this special case. We note that
in general, the bound from the decay products of the saxion provides a
somewhat tighter lower bound on $\Delta_{PQ}$ in comparison to the bound
obtained from the requirement $T_{RH}^{s}>T_{BBN}$.

\subsubsection{Comments on the Abundance of $^{7}Li$}

While the abundance of the light nuclei $H^{+}$, $D^{+}$, $T^{+}$,
$^{3}He^{++}$, $^{4}He^{++}$ expected from BBN are all in reasonable accord
with observation, there is also some tension between the abundance of $^{7}Li$
expected based on the Standard Cosmology, and the observed abundance, which is
typically a factor of $0.2-0.5$ smaller. In fact, as mentioned in section
\ref{CosmoReview}, recent studies of the MSSM\ in scenarios with a gravitino
in the mass range $10-100$ MeV, with a bino or stau NLSP\ have recently been
studied in \cite{Kawasaki:2008qe} and even more recently in
\cite{Bailly:2008yy}. In this regard, it is interesting to note that the range
of mass parameters expected in F-theory GUTs based on crude particle physics
considerations are in rough accord with these studies. Although it is beyond
the scope of this paper to address such detailed properties of BBN, it is quite
encouraging that in \cite{Bailly:2008yy}, in the context of a gauge mediation
scenario with a gravitino of mass $\sim80$ MeV and a bino NLSP\ of mass
$\sim200$ GeV that the resulting abundance of $^{7}Li$ appears to be in better
agreement with observation. It would be interesting to study this issue in
greater detail.

\subsection{Baryon Asymmetry}

In the previous subsection we explained that the additional particle content
of F-theory GUTs does not appear to disrupt the reaction rates necessary in
BBN. At a more basic level, however, it is important to verify that the
primary input of BBN, namely a sufficient baryon number asymmetry:%
\begin{equation}
\eta_{B}^{\mathrm{obs}}\;\equiv\;\frac{n_{B}-n_{\overline{B}}}{n_{\gamma}%
}=\frac{s}{n_{\gamma}}\frac{n_{B}-n_{\overline{B}}}{s}\sim7.04\cdot Y_{B}%
\sim6\times10^{-10}, \label{obsbara}%
\end{equation}
has in fact been generated! Here, $Y_{B}$ denotes the net yield of baryons.

The creation of a sufficient baryon asymmetry is an especially acute problem
in supersymmetric models with a gravitino LSP. Indeed, as reviewed in section
\ref{CosmoReview}, in models without an era of moduli domination, it is quite
common to lower the initial reheating temperature $T_{RH}^{0}$ to avoid
overclosing the Universe from the thermal production of gravitinos. This in
turn imposes strong constraints on the available mechanisms which can generate
an appropriate baryon asymmetry. For example, in the context of models where
the neutrinos of the Standard Model develop a small mass via the seesaw
mechanism, heavy right-handed neutrinos couple to the Higgs up and lepton
doublet through the superpotential term:%
\begin{equation}
W\supset\lambda_{\nu}^{ij}H_{u}L^{i}N_{R}^{j}+M^{i}_{maj}N^{i}_{R}N^{i}%
_{R}\text{,}%
\end{equation}
where in the above,~$M_{maj}$ denotes the Majorana mass of the right-handed
neutrino, and $\lambda_{\nu}^{ij}$ is the Yukawa matrix in the neutrino sector,
which for simplicity we shall take to be a $3\times3$ matrix. In the specific
context of minimal F-theory GUTs which incorporate a seesaw mechanism, simple
estimates for the mass of the lightest Majorana mass give \cite{BHVII}:%
\begin{equation}
M_{1}\sim3\times10^{12\pm1.5}\text{ GeV.} \label{Mone}%
\end{equation}

In standard leptogenesis, the subsequent decay of the right-handed neutrino to
the Higgs and lepton doublet generates an overall lepton number density which
is converted via sphaleron processes to a baryon asymmetry. In order for this
decay process to generate a sufficient baryon number asymmetry, however, the
initial reheating temperature must be greater than the Majorana mass:%
\begin{equation}
T_{RH}^{0}\gtrsim M_{maj}\text{.} \label{TOBOUND}%
\end{equation}
Indeed, for lower values of the initial reheating temperature, the decay
products of the right-handed neutrinos are too dilute to generate the required
baryon asymmetry.

But while the decay of the saxion dilutes the relic abundance of thermally
produced gravitinos, it will indiscriminately also dilute any pre-existing
baryon asymmetry! In this section we analyze whether a sufficient baryon
asymmetry can be generated once the dilution effects of the saxion are taken
into account, focussing on standard leptogenesis. We find that in the typical
range of parameters for F-theory GUTs, standard leptogenesis typically
generates a surplus baryon asymmetry which is diluted to acceptable values by
the decay of the saxion. Similar studies on the compatibility of standard
leptogenesis with a late decaying field which solves the gravitino problem
have appeared, for example, in \cite{Fujii:2002fv,Fujii:2003iw}. Again, we
find it very reassuring that F-theory GUTs provide a natural setting for
realizing such scenarios.

Although we do not do so here, it is also possible to
consider scenarios based on Dirac leptogenesis. In this case, an analogue of
the seesaw mechanism generates small Dirac masses for the neutrinos, where the
decay of the heavy particle associated with this \textquotedblleft Dirac
seesaw\textquotedblright\ generates light left- and right-handed neutrinos.
Due to the difference in the efficiency of their interactions rates, this
again can generate a lepton asymmetry, which is again converted to a baryon
asymmetry. Insofar as standard leptogenesis can generate a viable level of
baryon asymmetry, models with a similar range of parameters can also generate
a sufficient baryon asymmetry in Dirac leptogenesis scenarios. While it would
be interesting to study other alternative mechanisms for generating a large
baryon asymmetry, it is beyond the scope of the present work to perform such
an analysis. Indeed, the primary aim of this section is to demonstrate that in
F-theory GUTs, standard mechanisms already generate an appropriate baryon
asymmetry, without any additional assumptions.

\subsubsection{Review of Standard Leptogenesis}

We now briefly outline the main points of standard leptogenesis \cite{Fukugita:1986hr},
following the review \cite{Chen:2007fv}.In extensions of the Standard Model which generate a
suitable Majorana mass term for the light neutrinos via the seesaw mechanism,
the decay of heavy right-handed neutrinos into leptons and Higgses can
generate a lepton asymmetry. This process satisfies the Sakharov conditions
reviewed in section \ref{CosmoReview} because by construction, the lepton
number violation present in the Majorana mass term is converted to a baryon
number violation via sphaleron processes. The violation of C and CP$\ $is
somewhat more delicate, and at leading order originates from one loop
contributions to the decay:%
\begin{equation}
\nu_{R}\rightarrow l+h_{u}\text{,}%
\end{equation}
in the obvious notation. The necessity of the one loop contribution for C\ and
CP violation can be established for example, by appealing to the optical
theorem. The amount of CP\ violation can be characterized in terms of the
parameter $\epsilon_{1}$, which roughly measures the overall complex phase of
this one loop contribution. As reviewed in \cite{Chen:2007fv}, the net yield
of leptons produced from this decay is:%
\begin{equation}
Y_{L}=\frac{\kappa}{g_{\ast}}\;\epsilon_{1}\text{,} \label{12}%
\end{equation}
where in the context of supersymmetric models, $g_{\ast}(MSSM)\sim228.75$, and
$\kappa$ is the \textquotedblleft washout\textquotedblright\ factor which
quantifies to what extent the decay of heavy neutrinos occurs out of
equilibrium. The washout factor is given by integrating the Boltzmann
equations, and in the range relevant to us is given by \cite{Chen:2007fv}:
\begin{align}
0  &  \lesssim r\lesssim10:\kappa\left(  r\right)  \sim\frac{1}{2\sqrt
{r^{2}+9}}\label{kapp1}\\
10  &  \lesssim r\lesssim10^{6}:\kappa\left(  r\right)  \sim\frac{0.3}{r(\log
r)^{0.8}}\label{kapp2}\\
10^{6}  &  \lesssim r:\kappa\left(  r\right)  \sim\left(  0.1r\right)
^{1/2}\cdot\exp\left(  -\frac{4}{3}\left(  0.1r\right)  ^{1/4}\right)
\label{kapp3}%
\end{align}
where in the above, $r$ denotes a parameter associated with the efficiency of
the reaction \cite{Chen:2007fv}:%
\begin{equation}
r\equiv\frac{\Gamma_{N_{1}}}{H(M_{1})}\sim\frac{g_{\ast}^{-1/2}\cdot
(\lambda_{\nu}\lambda_{\nu}^{\dagger})_{11}}{1.7\cdot32\pi}\frac{M_{PL}}%
{M_{1}}\text{,} \label{reactionrate}%
\end{equation}
with $\Gamma_{N_{1}}$ the decay rate for the lightest of the heavy
right-handed neutrinos. Sphaleron processes at high temperatures will
automatically convert the lepton number asymmetry into a baryon asymmetry. The
yield of net baryons $Y_{B}$ from this conversion is:%
\begin{equation}
Y_{B}=\frac{10}{31}\cdot Y_{L}=\frac{10}{31}\cdot\frac{\kappa}{g_{\ast}%
(MSSM)}\epsilon_{1}\text{.}%
\end{equation}

The amount of CP\ violation in a given model depends on the size of
the hierarchy in the right-handed neutrino masses. For example, in the case of an
\textquotedblleft extreme\textquotedblright\ hierarchy, $M_{2}/M_{1}$,
$M_{3}/M_{1}\gtrsim10^{3}$, the amount of CP\ violation is essentially
dictated by a single complex number, and $\epsilon_{1}$ satisfies the
Davidson-Iberra bound \cite{DI,Davidson:2008bu}:%
\begin{equation}
\text{Extreme Hierarchical}:|\epsilon_{1}|\;\lesssim\;\frac{3\alpha}{16\pi
}\frac{\delta m\cdot M_{1}}{\left\langle H_{u}\right\rangle ^{2}}%
\equiv\epsilon_{DI}\text{,} \label{dib}%
\end{equation}
where in the above, $\left\langle H_{u}\right\rangle $ denotes the vev of the
Higgs up, and $\delta m\equiv m_{\mathrm{max}}-m_{\mathrm{min}}\sim0.05$ eV is
the mass splitting in the light-neutrino sector. Further, $\alpha$ is an order
one parameter associated with the value of the Yukawa couplings in the
neutrino sector. Finally, in the context of the large $\tan\beta=\left\langle
H_{u}\right\rangle /\left\langle H_{d}\right\rangle $ scenarios studied in
\cite{HVGMSB}, the actual value of $\left\langle H_{u}\right\rangle $ is given
by:%
\begin{equation}
\left\langle H_{u}\right\rangle =v\sin\beta\sim v\sim246\;\mathrm{GeV.}%
\end{equation}

Away from this extreme limit, the mass matrix for the right-handed neutrinos
plays a more essential role, and there can be further sources of
CP\ violation. For example, figure 1 of \cite{Hambye:2003rt} illustrates that
even when $M_{2}/M_{1},M_{3}/M_{1}\sim10$, the value of $\epsilon_{1}$ can
deviate from $\epsilon_{DI}$ by one order of magnitude. More generally, in
the case of less hierarchical masses, the resulting bound on $\epsilon_{1}$ is
\cite{Hambye:2003rt,Davidson:2008bu}:%
\begin{equation}
\text{Less Hierarchical}:|\epsilon_{1}|\;\lesssim\max\left(  \epsilon
_{DI},\frac{M_{1}^{3}}{M_{2}M_{3}^{2}}\right)  \sim O(1)\text{.}
\end{equation}

Plugging in all numerical factors, the resulting upper bound on the baryon
asymmetry in these two situations is:%
\begin{align}
\text{Extreme Hierarchical}  &  :\eta_{B}^{(0)}\lesssim5\times10^{-7}%
\cdot\kappa\left(  \frac{M_{1}}{10^{12}\;\mathrm{GeV}}\right)
\label{hierarchical}\\
\text{Less Hierarchical}  &  :\eta_{B}^{(0)}\lesssim10^{-2}\cdot\kappa,
\label{Nonhierarchical}%
\end{align}
where in the above, the superscript on the baryon asymmetry reflects the fact
that the decay of a cosmological modulus could in principle dilute the total
amount of baryon asymmetry generated. In general, we can expect an
interpolation between the extreme, and less hierarchical scenarios. We now
show that the expected range of Majorana masses in F-theory GUTs quite
comfortably fits with the observed baryon asymmetry.

\subsubsection{Saxion Dilution and Standard Leptogenesis}

In the previous subsection we reviewed the main features of standard
leptogenesis, focussing in particular on the distinction between hierarchical
and non-hierarchical Majorana masses. In this subsection we analyze the effect
of saxion dilution on the net baryon asymmetry. As reviewed near equation
(\ref{Mone}), in F-theory GUTs where the right-handed neutrinos have large
Majorana masses, the natural mass scale associated to such fields is roughly
$\sim3\times10^{12\pm1.5}$ GeV. In the present context then, standard
leptogenesis is most natural in such cases when the initial reheating
temperature $T_{RH}^{0}$ is at or above this range of energy scales.

We now determine whether the dilution factor from the saxion decay is
sufficiently small to avoid overclosure from gravitinos, whilst at the
same time, sufficiently large to avoid completely diluting the necessary baryon
asymmetry generated by standard leptogenesis. In the natural range of
parameters for F-theory GUTs, the Majorana mass of the right-handed neutrinos
$M_{maj}\gtrsim3\times10^{12\pm1.5}$ GeV is greater than the freeze out
temperature of the gravitino $T_{3/2}^{f}\sim10^{10}$ GeV. As a consequence,
the relic abundance of thermally produced gravitinos is given by:%
\begin{equation}
\Omega_{3/2}^{T}h^{2}\sim D\cdot\left(  \frac{m_{3/2}}{2\;\mathrm{keV}%
}\right)  \leq0.1\text{.}%
\end{equation}
In the range $m_{3/2}\sim10-100$ MeV, it follows that the dilution factor is
bounded above by:%
\begin{equation}
D\;\lesssim\;2\times10^{-5}\cdot\left(  \frac{10\;\mathrm{MeV}}{m_{3/2}%
}\right)  \text{.}%
\end{equation}
Multiplying the baryon asymmetry estimated in equations (\ref{hierarchical})
and (\ref{Nonhierarchical}) by the dilution factor, the baryon asymmetry is
therefore bounded above by:%

\begin{align}
\text{Extreme Hierarchical}  &  :\eta_{B}\lesssim10^{-11}\cdot\kappa\left(
\frac{M_{1}}{10^{12}\;\mathrm{Gev}}\right) \label{hiereta}\\
\text{Less Hierarchical}  &  :\eta_{B}\lesssim2\times10^{-7}\cdot
\kappa\text{,} \label{nhiereta}%
\end{align}
where in the above estimate, we have used the fact that $m_{3/2}\gtrsim10$
MeV. To proceed further, we now estimate the overall size of the wash out
factor, $\kappa$ by determining the value of the parameter $r$ in equation
(\ref{reactionrate}). Even without a complete theory of neutrino flavor, for
our present purposes, it is enough to use the order of magnitude estimate for
the Yukawa couplings:%
\begin{equation}
(\lambda_{\nu}\lambda_{\nu}^{\dagger})_{11}\sim\alpha_{GUT}^{3/2}\sim
8\times10^{-3}%
\end{equation}
Combined with the value of $g_{\ast}(MSSM)\sim228.75$, the resulting value of
$r$ is:%
\begin{equation}
r\sim\frac{6\times10^{12}\text{ GeV}}{M_{1}}\sim2\times10^{\pm1.5}\text{,}
\label{ourrr}%
\end{equation}
where in the final estimate we have plugged in the explicit value of $M_{1}$
suggested by F-theory GUTs \cite{BHVII}.

In the extreme hierarchical case, we see that a viable baryon asymmetry is
only possible provided $M_{1}>10^{12}$ GeV and $\kappa$ an order one
parameter. Returning to equation (\ref{ourrr}), it follows that in this range
$r$ is indeed quite small, so that equation (\ref{kapp1}) implies $\kappa
\sim1/6$. Plugging this value of $\kappa$ into (\ref{hiereta}), the baryon
asymmetry is bounded above by:%
\begin{equation}
\text{Extreme Hierarchical}:\eta_{B}\lesssim\;2\times10^{-12}\cdot\left(
\frac{M_{1}}{10^{12}\;\mathrm{GeV}}\right)  \text{.}%
\end{equation}
Generating the observed baryon asymmetry in this extreme case would then
require:%
\begin{equation}
\text{Extreme Hierarchical}:M_{1}\sim10^{14}\text{ GeV,}%
\end{equation}
which is remarkably close to the upper bound on $M_{1}$ expected in F-theory GUTs.

Next consider the more natural case for F-theory GUTs where the Majorana
masses are not \textit{extremely} hierarchical. In this case, inequality
(\ref{nhiereta}) is saturated provided $\kappa\sim10^{-2}-10^{-3}$. Returning
to equations (\ref{kapp1})-(\ref{kapp3}), this range of values requires
$\kappa$ to be in the second range so that $10\leq r\leq10^{6}$. Evaluating
$\kappa\left(  r\right)  $ at some representative values, we have:
\begin{align}
\kappa\left(  r\sim10\right)   &  \sim10^{-2}\\
\text{ }\kappa\left(  r\sim100\right)   &  \sim10^{-3}\text{.}%
\end{align}
In other words, in the range $10\lesssim r\lesssim100$, leptogenesis with less
hierarchical masses generates the requisite baryon asymmetry. In terms of the
Majorana mass $M_{1}$, this corresponds to the range:%
\begin{equation}
\text{Less Hierarchical}:10^{11}\text{ GeV}\lesssim M_{1}\lesssim10^{12}\text{
GeV,}%
\end{equation}
which is in the expected range of Majorana masses estimated in \cite{BHVII}!

To summarize, when the masses of the heavy neutrinos are not extremely
hierarchical, we find that within a natural window of values for $M_{1}$, the
resulting baryon asymmetry matches with the observed value.\ This is due to
the interplay between the dilution due to the saxion, and the natural range of
Majorana masses expected in F-theory GUTs. Indeed, F-theory GUTs elegantly
reconcile the apparent tension between standard leptogenesis and the
\textquotedblleft gravitino problem\textquotedblright.

\subsection{Messenger Relics}

In the above sections, we have seen that the cosmology of F-theory GUTs is
remarkably insensitive to the initial reheating temperature of the Universe,
$T^{0}_{RH}$. In the specific context of high scale gauge mediation scenarios,
though, it is natural to ask whether the relic abundance of messengers will
overclose the Universe. Indeed, if produced from the thermal bath, the large
mass of these particles in F-theory GUTs, $M_{mess} \sim10^{12}$ GeV would
lead to a very large relic abundance which even the decay of the saxion cannot
dilute to an acceptable value. One possibility would be to take $T^{0}_{RH}<
M_{mess} \sim10^{12}$ GeV. In such a scenario the messengers would not have
been produced by this initial condition assumption. However, this would not be
attractive in our scenario, because we have seen that essentially all the
relevant physics is independent of $T^{0}_{RH}$. Moreover this upper bound on
the value of $T^{0}_{RH}$, is potentially in conflict with leptogenesis. A
more natural assumption, in line with the spirit of the present paper is to
assume that the messenger fields can decay to some lighter fields, such as
those present in the MSSM. We are currently investigating explicit models of
F-theory GUTs which take this feature into account \cite{FGUTSDARK}.

\section{Future Directions \label{CONC}}

In this paper we have found that in F-theory GUTs, the gravitino and in some
cases the axion can provide a prominent component of the total dark matter,
which quite remarkably is independent of the initial reheating temperature of
the Universe, $T_{RH}^{0}$. On the other hand, such candidates leave open
the issue of accounting for the recent experimental results
such as PAMELA \cite{Adriani:2008zr}, which could potentially be explained in terms of dark matter physics.
In this regard, it is important to investigate
whether F-theory GUTs provide additional dark matter candidates
\cite{FGUTSDARK}.

In the context of F-theory GUTs, the decay of the saxion which dilutes the
abundance of gravitinos will also dilute the relic abundance of any other dark
matter candidate. Assuming that the dark matter originates from a cold thermal
relic, its abundance scales inversely with its cross section:%
\begin{equation}
\Omega_{\text{DM}}h^{2}\propto\frac{1}{\left\langle \sigma_{\text{DM}%
}v_{\text{DM}}\right\rangle }\text{.}%
\end{equation}
Letting $M$ denote a characteristic mass scale associated with the dark
matter, $\sigma_{\text{DM}}\sim M^{-2}$, which would naturally suggest a mass
scale of order $M\sim1$ TeV. Taking into account the dilution factor
$D\sim10^{-4}-10^{-5}$ from the decay of the saxion would instead suggest that
$M\sim100$ TeV. The downside to this is that this also lowers the cross
section by a factor of $10^{-4}-10^{-5}$. Unless there is a substantial
enhancement either in its density or cross section at small velocities, this
type of dark matter candidate would then be too small to be detectable in
current dark matter searches.

Nevertheless, an exact analysis of potential dark matter candidates depends on
the details of a given model. In this regard, the tight structure of F-theory
GUTs also provides additional candidates associated with degrees of freedom
located near the F-theory GUT\ seven-brane. For example, four-dimensional GUTs
always include a $U(1)_{B-L}$ gauge boson. We are currently investigating the
details of such a scenario \cite{FGUTSDARK}.

\section*{Acknowledgements}

We thank N. Arkani-Hamed, D. Baumann, M. Dine, D. Finkbeiner, P. Kumar, J.
Mason, D. Morrissey, D. Poland and T. Slatyer for helpful discussions. The
work of the authors is supported in part by NSF grant PHY-0244821.

\newpage
\bibliographystyle{ssg}
\bibliography{fgutscosmo}

\end{document}